\documentclass[superscriptaddress, a4paper, 11pt]{revtex4-2}

\usepackage[utf8]{inputenc}
\usepackage[dvipsnames]{xcolor}

\usepackage{amsmath,amssymb,amsthm}
\usepackage{dsfont}
\usepackage{faktor}

\usepackage[export]{adjustbox} 

\usepackage{mathptmx}

\theoremstyle{definition}
\newtheorem{definition}{Definition}
\newtheorem{exmpl}{Example}[section]

\theoremstyle{remark}
\newtheorem*{remark}{Remark}
\usepackage{relsize}

\usepackage{graphicx}
\graphicspath{{./graphics/}}
\usepackage{float}

\usepackage{diagbox}

\usepackage{enumitem}
\usepackage{rotating}

\usepackage{physics}

\usepackage{url}

\newcommand{\cC}{\mathcal{C}}
\newcommand{\cZ}{\mathcal{Z}}
\newcommand{\cB}{\mathcal{B}}
\newcommand{\cM}{\mathcal{M}}

\newcommand{\cL}{\mathcal{L}}
\newcommand{\cH}{\mathcal{H}}
\newcommand{\bZ}{\mathbb{Z}}
\newcommand{\bC}{\mathbb{C}}
\DeclareMathOperator{\Aut}{Aut}

\DeclareMathOperator{\End}{End}
\DeclareMathOperator{\Obj}{Obj}
\DeclareMathOperator{\spn}{span}
\DeclareMathOperator{\vV}{Vec}
\DeclareMathOperator{\Rep}{Rep}
\DeclareMathOperator{\Stab}{Stab}

\DeclareMathOperator{\Id}{Id}

\newcommand{\minv}{\rotatebox[origin=c]{180}{$m$}}

\usepackage{mathtools}
\newcommand{\tildehat}[1]{\tilde{\raisebox{0pt}[0.85\height]{$\hat{#1}$}}}
\newcommand{\hattilde}[1]{\hat{\raisebox{0pt}[0.85\height]{$\tilde{#1}$}}}

\definecolor{darkblue}{RGB}{50,50,150}
\usepackage{hyperref}
\hypersetup{
	colorlinks=true,
	citecolor={darkblue},
	linkcolor={darkblue},
	urlcolor={darkblue},
	pdfencoding=auto,
	psdextra,
	pdfhighlight=/I
}

\begin{document}
\title{Bulk-to-boundary anyon fusion from microscopic models}
\date{November 10, 2023}
\author{Julio C. Magdalena de la Fuente}
\email{julio.magdalena@fu-berlin.de}
\affiliation{Dahlem Center for Complex Quantum Systems, Freie Universit\"at Berlin, 14195 Berlin, Germany}
 \author{Jens Eisert}
\affiliation{Dahlem Center for Complex Quantum Systems, Freie Universit\"at Berlin, 14195 Berlin, Germany}
\affiliation{Helmholtz-Zentrum Berlin f{\"u}r Materialien und Energie, 14109 Berlin, Germany}
\author{Andreas Bauer}
\affiliation{Dahlem Center for Complex Quantum Systems, Freie Universit\"at Berlin, 14195 Berlin, Germany}

\begin{abstract}
    Topological quantum error correction based on the manipulation of the anyonic defects constitutes one of the most promising frameworks towards realizing fault-tolerant quantum devices. Hence, it is crucial to understand how these defects interact with external defects such as boundaries or domain walls.
    Motivated by this line of thought, in this work, we study the fusion events between anyons in the bulk and at the boundary in fixed-point models of 2+1-dimensional non-chiral topological order defined by arbitrary fusion categories. Our construction uses generalized tube algebra techniques to construct a bi-representation of bulk and boundary defects. We explicitly derive a formula to calculate the fusion multiplicities of a bulk-to-boundary fusion event for twisted quantum double models and calculate some exemplary fusion events for Abelian models and the (twisted) quantum double model of $S_3$, the simplest non-Abelian group-theoretical model. Moreover, we use the folding trick to study the anyonic behavior at non-trivial domain walls between twisted $S_3$ and twisted $\bZ_2$ as well as $\bZ_3$ models. A recurring theme in our construction is an isomorphism relating twisted cohomology groups to untwisted ones. The results of this work can directly be applied to study logical operators in two-dimensional topological error correcting codes with boundaries described by a twisted gauge theory of a finite group.
\end{abstract}

\maketitle

\section{Introduction}

Topological phases of matter are intriguing zero-temperature quantum phases that are accompanied by
robust ground state degeneracy and patterns of long-range quantum entanglement. They constitute cornerstones of
modern condensed matter physics~\cite{WenTopological,Levin2004}.
At the same time, they play a central role in notions of topological
quantum error correction that is widely seen as one of the most promising paradigms for scalable quantum computing, a paradigm in which anyonic defects in the code are suitably manipulated
to perform quantum information processing~\cite{kitaev2006anyons,RevModPhys.87.307}.
In more abstract terms, a common high-level description of topologically ordered models is via the codimension-2 defects or ``excitations''. The most famous example for this is the anyons in a 2+1-dimensional model, which are characterized by their fusion and braiding statistics~\cite{kitaev2006anyons}.
Such a description via higher-level invariant data can be extended to boundaries or domain walls, by also considering excitations within the boundary.

A more concrete lower-level description of topological order is via microscopic fixed-point models, which are exactly solvable due to a notion of discrete topological invariance~\cite{bauer2020unified}.
Those models allow for the computation of any higher-level invariants usind only finite-dimensional linear algebra.
In $2+1$ space-time dimensions any non-chiral topological order
can be represented by such a microscopic fixed-point model, with finite-dimensional Hilbert spaces.

There are two major pictures in which those fixed-point models can be formulated, namely the space (or string-net) picture, and the space-time picture.
In the space picture, we start by assigning a Hilbert space to a cellulation of a two-dimensional manifold.
The most well-known examples are string-net models defined on a trivalent cellulation~\cite{Levin2004} or its dual formulation in terms of a triangulation of the same manifold~\cite{Hu2013Twisted}.
To achieve topological invariance, these models are also equipped with partial isometries mapping between the vector spaces of different cellulations.
In fact, arbitrary changes of the cellulation are generated by local moves such as Pachner moves, so the model is defined by the associated local isometries.
A local ground-state projector, or Hamiltonian, can be defined via a local sequence of moves/isometries which have no net effect on the cellulation.

In the space-time picture, a model is given by a discrete state-sum path integral in Euclidean space-time, the most famous one being the Turaev-Viro state-sum~\cite{Turaev1992}.
Such a state-sum is defined on any 3-dimensional cellulation of that space-time, and assigns
a finite label set
to any edge in that cellulation.
The highest-dimensional cells carry weights depending on those labels. The state-sum is evaluated by taking the product of all weights and summing over all label configurations, and it must be invariant under changes in the cellulation.
The cellulations in the space picture can be interpreted as codimension-1 sections of the cellulations in the space-time picture, giving rise to the equivalence of the two pictures.
In this work, we study topological phases on manifolds with boundary using both pictures, since they each have their own up- and downsides.
The space/string-net picture is more illustrative to most physicists since it
is directly related to the usual quantum mechanical language of Hilbert spaces, states, and Hamiltonians.
Hence, we mainly present our constructions in this space picture.
However, the mapping between different cellulations in this picture can be quite tedious to work out, so we fall back on the space-time picture
to evaluate complicated sequences of moves.

Topological fixed-point models on manifolds with boundaries have been studied in various places in the literature to understand properties of the defects on the boundary and how they interact with each other.
Refs.\,\cite{Bridgeman2020computingdatalevin, barter2019domain, bridgeman2019fusing} focus on \textit{corners}, defects between different types of boundaries or domain walls.
In particular, they calculate \textit{vertical} fusion events of codimension 2 defects along the same domain wall and \textit{horizontal} fusion of defects on neighboring domain walls.
Combining their methods allows them to also calculate the associator of these defects.
Ref.\,\cite{kitaev2012models} gives an algebraic description of what happens to bulk defects when approaching the boundary in terms of a forgetful functor on tensor category describing the bulk anyons. However, they do not give a constructive framework to calculate these properties.
Ref.\,\cite{Beigi11} gives a formula to calculate the set of \textit{condensable anyons} in gauge theory models with trivial 3-cocycle.
In general, it is known that the set of condensable anyons have to form a \textit{Lagrangian algebra object} in the modular tensor category that describes the bulk anyons~\cite{Kapustin2011Topological, levin2013protected, davydov2017lagrangian}.

What we have found to be missing is a full description of the fusion of bulk anyons to boundary anyons.
More explicitly, we are interested in the dimensions
$m_{ij}\geq 0$ of the fusion spaces between an (ingoing) bulk anyon $i$ and an (outgoing) boundary anyon $j$.
Importantly, we are interested in constructive formulas to calculate $m_{i,j}$.

Apart from a mathematical
interest, these fusion events find application in topological quantum error correction and computing with boundaries~\cite{cong2016topological}.
For example, the logical operators of a topological code on a manifold with boundary are associated to ribbon operators ~\cite{Kitaev2003fault} of anyons
which connect different boundary segments through the bulk.
Moreover, any lattice-surgery-based computation scheme ultimately relies on deforming non-transparent domain walls between code patches into (partly-)transparent ones in a systematic way~\cite{kesselring2022condensation, thomsen2022lowoverhead, Litinski2019gameofsurfacecodes}.
Understanding how the anyon ribbon operators precisely behave close to these domain walls is therefore essential in the design of novel computational protocols in topological codes. In this sense, the work done here is also expected to provide guidance when devising novel schemes of topological quantum computing involving notions of lattice-surgery with codes beyond untwisted quantum doubles.

With the framework established in this work  we aim to contribute to a further understanding of topological phases with boundaries by formulating a framework to describe bulk-to-boundary anyon fusion events in topological fixed-point models.
We explicitly derive a closed formula for fusion multiplicities of fusion events between bulk and boundary anyons for 2+1-dimensional twisted gauge theory models, also known as \emph{Dijkgraaf-Witten state-sums}~\cite{dijkgraaf1990topological}.
In this case, the anyons in the bulk as well as on the boundary are classified by irreducible sub-spaces of some special type of algebras, which we call \textit{twisted group algebras with action}.
We show how such an algebra is diagonalized and discover that there is an intimate connection to a group cohomological isomorphism that also appears in the classification of topological boundaries of gauge theories.

This manuscript is structured as follows. In Section~\ref{sec:twistedwithaction}, we give a general recipe to find the irreducible sub-spaces of twisted group algebras with action, which characterize both the bulk and boundary anyons in gauge theory models of a finite group.
In Section~\ref{sec:latticemodels}, we give a self-contained introduction into string-net fixed-point models for topological phases with boundaries in two spatial dimensions and illustrate the equivalence to space-time state-sum models.
This section is mainly addressed to readers not yet familiar with these models.
Readers with a background in both formulations of fixed-point models might want to use that section to get familiar with our notation and conventions in the upcoming sections.
In Sections~\ref{sec:tube} and~\ref{sec:RectangleAlgebra}, we classify the anyons in the bulk and boundary. In particular, we focus on gauge theory models of a finite group.
The main result of this paper is presented in Section~\ref{sec:bulkbdryfusion}, where we define a bimodule that allows to calculate the dimensions of bulk-to-boundary fusion events in any fixed-point model and explicitly derive a closed formula for the gauge theory case.
Lastly, we give many examples, that partly already appeared in the literature, but combine them with new calculations to show the wide applicability of our formula to boundaries as well as domain walls.
Finally, we conclude the results and give an outlook into possible continuations of this work in Section~\ref{sec:conslusion}.
For a reader interested in the technical details going into the derivation of the bimodule used in Section~\ref{sec:bulkbdryfusion} and tools to diagonalize the group algebras characterizing point defects, we refer to the appendix for further details.

\section{Diagonalizing twisted group algebras with action}\label{sec:twistedwithaction}
Before we introduce topological fixed-point models, we want to highlight the technical tools used to classify bulk and boundary anyons of gauge theory models in Secs.\,\ref{sec:tube} and~\ref{sec:RectangleAlgebra}.
In both cases topological invariance of the anyonic subspaces naturally defines a finite-dimensional algebra.
In most parts of this paper we focus on gauge theory models, derived from a finite group and a 3-cocycle on it.
For these models both bulk and boundary anyons are classified by a special type of algebra.
In this section, we will present the technical tools used to find the irreducible sub-spaces of, i.e., block-diagonalize, these algebras.

Consider a finite group $G$. Let $X$ be a finite (left) $G$-set, i.e., there exist a map $\triangleright: G\times X\to X$ representing the $G$-multiplication on $X$,
\begin{align}
    g\triangleright(h \triangleright x) = (gh)\triangleright x\;\forall g,h\in G,x\in X.
\end{align}
Given a $G$-set $X$, we define an algebra $A$ over $\bC^{G\times X}$ via the multiplication
\begin{align}\label{eq:mult_algebrawithaction}
    (g,x)\ast(h,y) = \delta_{x, h\triangleright y} \Psi^{y}(g,h) (gh, y),
\end{align}
with $\Psi: G\times G\times X \to U(1)\subset\bC$.
For this multiplication to define an algebra it has to be associative. This imposes a non-trivial  condition on the phase $\Psi$,
\begin{align}\label{eq:twisted2cocyclegeneral}
\Psi^{x}(h,k) \Psi^x(g,hk) = \Psi^x(gh,k)\Psi^{k\triangleright x}(g,h)%
\qcomma \forall g,h,k\in G, x\in X.
\end{align}
This can be seen as a 2-cocycle condition over $U(1)^X$ as a $G$-module with non-trivial action defined via the $G$-action on $X$.
Moreover, we require that $\Psi^x$ is normalized, i.e., $\Psi^x(1_G,h) = \Psi^x(h,1_G) = 1\;\forall g\in G, x\in X$.
We can in fact redefine the basis states of $A$ with $\xi:G\to \bC^\times$ by $(g,x)\mapsto \xi(g)(g,x)$ to effectively map $\Psi^x$ to a normalized 2-cocycle~\cite{brown2012cohomology, chen2013symmetry} so we do not loose generality with this assumption.
We call such an algebra \textit{twisted group algebra with action}.
For more details on $G$-modules and group cohomology, see App.\,\ref{app:cohomology}.

The goal of this section is to find the irreducible sub-spaces of $A$. In particular, we find faithful invariants classifying the sub-spaces, determine their dimension and find the associated central idempotents whose representations project onto the respective sub-spaces. We give a comprehensive summary in terms of a recipe to construct the central idempotents at the end of this section.

First, we note that due to the delta in the multiplication in Eq.\,\eqref{eq:mult_algebrawithaction}, $A$ decomposes over transitive subsets of $X$, or equivalently, over $G$-orbits $\{X_i\subseteq X\}$, so we have an isomorphism
\begin{align}
    A \simeq \bigoplus_{i\in \{X_i\}} A_i.
\end{align}
For any $X_i$ there exists a subgroup $K_i\subseteq G$ such that $X_i$ is isomorphic to $\flatfrac{G}{K_i}$, the set of left $K_i$-cosets. After this isomorphism, $G$ acts via left translation, $g\triangleright hK_i=(gh)K_i$, $g,h\in G$, onto $X_i$.
Since $X_i$ is a transitive $G$-set the \textit{stabilizer groups} of any element $x\in X_i$,
\begin{align}
    \Stab_G(x) := \{g\in G\;|\; g\triangleright x = x\},
\end{align}
are isomorphic.
In particular, for $x\in X,g\in G$,
\begin{align}\label{eq:stabilizerrelation}
    \Stab_G(g\triangleright x) = g \Stab_G(x)g^{-1}.
\end{align}
Hence, we can define an abstract stabilizer group of $X$, $\Stab_G(X)\simeq\Stab_G(x)$ for any $x\in X$.
In fact, $K_i$ is isomorphic to $\Stab_G(X)$.
As a subgroup of $G$, $K_i$ depends on the chosen identification $X_i \leftrightarrow \flatfrac{G}{K_i}$.
We pick a representative $\hat{x}_i\in X_i$ and define
\begin{align}
    K_i:=\Stab_G(\hat{x}_i)\subseteq G.
\end{align}
Note that we can obtain the stabilizer group of any other element in $X_i$ from $K_i$ with Eq.\,\eqref{eq:stabilizerrelation}.

Next, we show how to further decompose each $A_i$ into irreducible components. In fact, we find that they are in one-to-one correspondence with
\emph{irreducible unitary $\Psi^{\hat{x}_i}$-projective representations} (IPRs) of $K_i$.
To see this, we explicitly construct the indecomposable central idempotents in $A_i$ from IPRs of $K_i$.
Let $\{\rho^{\hat{x}_i}\}$ be the irreducible $\Psi^{\hat{x}_i}$-projective representations of $K_i$, defined via a representative $\hat{x}_i\in X_i$. In particular, they fulfill
\begin{align}
    \rho^{\hat{x}_i}(k) \rho^{\hat{x}_i}(k') = \Psi^{\hat{x}_i}(k,k')\rho^{\hat{x}_i}(kk') \qq{}\forall k,k'\in K_i,
\end{align}
and are unitary. If not stated otherwise, any representation considered in this manuscript is assumed to be unitary.
We can construct an equivalent IPR of any other isomorphic stabilizer group $\Stab_G(y)$ for $y\in X_i$.
Concretely, starting from an irrep $\rho^x$ of $\Stab_G(x)$ and $k'\in\Stab_G(x)$ we can define the irreducible representations on all of $A$ via
\begin{align}\label{eq:relation_projective_irreps}
    \rho^{g\triangleright x}(k) = \overline{\Psi^{x}(kgk'^{-1},k')} \Psi^{x}(k,g) \rho^{x}(k')
\end{align}
which by construction fulfill
\begin{align}
    \rho^{y}(k)\rho^{y}(k') = \Psi^{y}(k,k')\rho^{y}(kk')\qq{}\forall y\in X_i; k,k'\in\Stab_G(y),
\end{align}
when the group label is restricted on the respective (isomorphic) stabilizer groups.
Note that we act with the inverse group element on the left-hand-side of Eq.\,\eqref{eq:relation_projective_irreps} because we pull of the $G$-action on $X$ back onto $\rho$.

This allows us to uniquely construct the IPRs of all stabilizer groups $\Stab_G(x)$ for $x\in X_i$ from the IPRs of the stabilizer group of a single representative $\hat{x}_i\in X_i$.

Given all the equivalent IPRs, we can construct the indecomposable central idempotents in $A_i$. They are labeled by IPRs $\{\rho_i\}$ of $K_i$ and given by
\begin{align}
    c_{i,\rho_i} = \frac{\dim(\rho_i)}{\abs{K_i}} \sum_{x\in X_i}\sum_{g\in\Stab_G(x)} \overline{\Tilde{\chi}^{x}_{\rho_i}(g)} (g,x),
\end{align}
where $\Tilde{\chi}^x_{\rho_i} := \Tr(\rho_i^x): G\to \bC$ is the \textit{irreducible $\Psi^x$-projective character} derived from the IPR $\rho_i^x$.
Using unitarity and irreducibility within $A_i$,
\begin{align}
    \sum_g \overline{\rho_i^x(g)_n^m}\rho_i^x(g)_{n'}^{m'} = \delta_{\rho,\rho'}\delta_{n,n'}\delta_{m,m'} \frac{\abs{K_i}}{\dim(\rho_i)} \qcomma \forall x\in X_i,
\end{align}
we can show straight forwardly that the $c_{i,\rho_i}$s are idempotent,
\begin{align}
    c_{i,\rho_i}\ast c_{j,\rho'_j} = \delta_{i,j}\delta_{\rho,\rho'} c_{i,\rho_i},
\end{align}
and central,
\begin{align}
    c_{i,\rho_i}\ast (g,x) = (g,x)\ast c_{i,\rho_i}\qcomma \forall (g, x) \in G\times X.
\end{align}
For now, we have found a set of invariants, $\{i,\rho_i\}$ labeling independent central idempotents.
To show that they are complete, i.e., correspond to irreducible invariant sub-spaces of $A$, we have to show that the $c_{i,\rho_i}$s are not only central and idempotent but also indecomposable.
To see this, consider the following set of smaller idempotents
\begin{align}
    d_{x_{i,\rho_i}} = \frac{\dim(\rho_i)}{\abs{K_i}} \sum_{g\in\Stab_G(x)} \overline{\Tilde{\chi}^{x}_{\rho_i}(g)} (g,x),
\end{align}
that are irreducible (by definition of $\Tilde{\chi}$) but not central.
In fact, we can get to any value of $x\in X_i$ via left- or right-multiplication of an element in $A$ which makes $c_{i,\rho_i}$ irreducible.
Furthermore, the isomorphism discussed in App.\,\ref{app:sec:isomorphism} provides a 1-1 mapping from irreducible representations of the twisted algebra with action in Eq.~\eqref{eq:mult_algebrawithaction} and $\Psi^{\hat x}$-twisted group algebras without action of (a collection of) subgroups.
In App.\,\ref{app:thin_boundary} we give an interpretation of this isomorphism in terms of an invertible domain wall mapping between two different kinds of state-sums with boundaries.

We summarise this section by giving a recipe on how to find the irreducible representations of an algebra of the form of Eq.\,\eqref{eq:mult_algebrawithaction}.
\begin{enumerate}
    \item Decompose $X$ into transitive $G$-orbits $\{X_i\}$.
    \item For each $X_i$:
    \begin{enumerate}
        \item Pick representative $\hat{x}_i\in X_i$ and calculate its stabilizer group $K_i = \Stab_G(\hat{x}_i)$.
        \item Find irreducible $\Psi^{\hat{x}_i}$-projective representations of $K_i$, denote them with $\rho_i$.\label{hardpart}
        \item Use Eq.\,\eqref{eq:relation_projective_irreps} to derive irreducible representations and character functions, $\Tilde{\chi}^x_{\rho_i}(g)  = \Tr(\rho_i^x(g))$ for all $x\in X_i$.
    \end{enumerate}
    \item The indecomposable central idempotents in $A$ are labeled by pairs $(i,\rho_i)\in (G$-orbits of $X$,$\Psi^x$-projective Irreps of $K_i)$ and given by
    \begin{align}
        c_{i,\rho_i} = \frac{\dim(\rho_i)}{\abs{K_i}}\sum_{x\in X_i}\sum_{g\in \Stab_G(x)} \overline{\Tilde{\chi}^{x}_{\rho_i}(g)} (g,x) .
    \end{align}
\end{enumerate}
Note that the above is not a complete algorithm for finding the algebra irreducible representations but merely reduces it to finding the irreducible representations of a much smaller algebra in step~\ref{hardpart}. Those irreducible representations are equivalent to projective group representations which can be found in the literature in many cases.

\section{Models for topological phases with boundary}\label{sec:latticemodels}
Topological phases which possess gapped boundaries can be studied using \emph{fixed-point models} on a discretized space-time. Topological invariance highly restricts the microscopic constituents of these models. 
In the Turaev-Viro state-sum~\cite{Turaev1992, Barrett1993}, or tensor-network path integrals~\cite{bauer2020unified,Bauer2022towards} for $2+1$-dimensional topological order, the topological invariance corresponds to recellulations in a three-dimensional space-time, and the algebraic constraints correspond to the ones defining spherical fusion categories.
A different but equivalent picture are Levin-Wen string-net models~\cite{Levin2004}, where recellulations on a two-dimensional space triangulation are represented by linear operators acting on the local degrees of freedom.
In this picture, the topological invariance in space-time takes the form of coherence axioms between different equivalent space recellulations.
Since these models are equivalent, one can construct the linear operators implementing topological invariance in Levin-Wen models from a state-sum on particular space-time cellulations.

In this section, we introduce microscopic models for gapped topological phases on manifolds with boundaries.
We mainly use the string-net picture, but occasionally refer to the space-time picture where we find it more instructive.
An overview of the space-time picture can be found in App.\,\ref{app:state_sum}.
First, we introduce the bulk degrees of freedom and how states hosting exact topological invariance are constructed based on a spherical fusion category $\cC$.
Secondly, we extend the models to boundaries and show how the model is constrained by the bulk data close to the boundary, leading to a description of the boundary in terms of a $\cC$-module category.

\subsection{Bulk}\label{sec:prelim_bulk}
In the bulk, the microscopic model is defined by a \textit{spherical fusion category} $\cC$. We denote the set of (finitely many) simple objects in $\cC$ by $\Obj(\cC) = \{1_{\cC},i,j, \dots, k\}$ and \textit{fusion multiplicities} $N_{ij}^k\in\bZ_{\geq 0}$. These define a \textit{fusion-operation}
\begin{align}\label{eq:fusion}
    i\times j=\sum_k N_{ij}^k k\qq{with}\quad a\times 1_{\cC} = 1_{\cC}\times a = a \quad\forall a\in\Obj(\cC) .
\end{align}
A fusion category is called \textit{Abelian} if for every pair $(i,j)\in \Obj(\cC)$ there exists only one $k\in\Obj(\cC)$ such that $N_{ij}^k>0$.
In other words, the fusion of $i$ and $j$ is unique.
Note that the term Abelian does not refer to the commutativity of the fusion operation.
Moreover, the fusion above can be equipped with a non-trivial \textit{associator} capturing the isomorphism between objects obtained from fusing in different orders. The associator is part of the input category $\cC$ and will be described in terms of so-called \textit{F-symbols} in microscopic models.

A microscopic (topological) model is defined on a \textit{framed} trivalent graph (tessellating some two-dimensional manifold) with local Hilbert spaces $\cH_i = \spn(\Obj(\cC))$ on each edge.
A framed graph has ``flags" on each edge pointing perpendicular to it. The orientations have to be chosen such that the flags do not point in the same direction around any vertex. This induces a local ordering of the faces around any vertex by the number of flags pointing into the faces. Analogously, one can think of such a framing as a branching structure on the dual triangulation, see for example \eqref{fig:pachner_F}.

The total Hilbert space is the tensor product space of all the local spaces, $\cH_{\rm tot} = \bigotimes_i\cH_i$. $\Obj(\cC)$ defines a natural basis on $\cH_i$ and with that on $\cH_{\rm tot}$.
The fusion multiplicities define a local constraint at each vertex, defining the \textit{physical subspace} $\cH_{\rm phys}\subset\cH_{\rm tot}$. $\cH_{\rm phys}$ is defined by the span of basis states for which the local labels $(i,j,k)$ at every vertex fulfill $N_{ij}^k\neq 0$. For $N_{ij}^k>1$ the vertex itself carries in additional degree of freedom, $\spn(\{0,1, \dots, N_{ij}^k-1\})$.
For the rest of this work, we will work within $\cH_{\rm phys}$ and depict a vertex and its adjacent edge labels $(i,j,k)$ in an allowed basis configuration with
\begin{align}\label{fig:bulk_vertex_general}
    \raisebox{-0.5\height}{\includegraphics[width=0.125\textwidth]{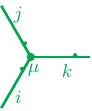}}.
\end{align}
Note that, when going around a vertex, one frame has to point in a different direction than the other two. We call this condition \textit{local acyclicity} and is related to the fact that one has to distinguish the left and the right hand side of Eq.\,\eqref{eq:fusion} when interpreting it as vertex.
Given this prescription, one can tessellate (the bulk of) any two-dimensional manifold with these trivalent vertices, each of which enforce a local constraint on the edge labels. Every allowed configuration defines a basis vector in the state space assigned to the cellulation. To construct a topological fixed-point model, topological invariance is imposed exactly. In particular, one can relate different cellulations (with the same input category $\cC$) via \emph{Pachner moves}, for example an \textit{F-move}
\begin{align}\label{fig:pachner_F}
    \raisebox{-0.5\height}{\includegraphics[width=0.3\textwidth]{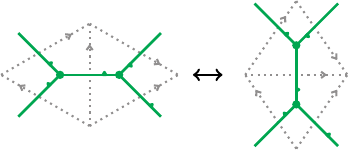}
    }.
\end{align}
The fusion category $\cC$ defines linear transformations that relate vector spaces of different cellulations to each other. The matrix entries of this linear map in the basis of string labels are complex numbers $\{F_{cdf,\nu\rho}^{aabe,\mu\eta}\}$,
\begin{align}\label{eq:fsymbol}
    \raisebox{-0.5\height}{\includegraphics[width=0.4\textwidth]{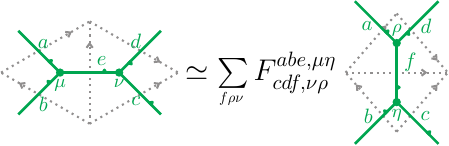}}.
\end{align}
These \textit{F-symbols} are part of the input category $\cC$. Any sequence of topological moves can be represented by a 3-dimensional triangulation derived from the dual cellulation indicated in gray in Eqs.\,\eqref{fig:pachner_F} and \eqref{eq:fsymbol}. In the multiplicity-free case, where $\mu,\nu,\rho,\eta$ are fixed, the F-move above, for example, is represented by a tetrahedron,
\begin{align}
       \mathlarger{F_{cdf}^{abe} =}
       \raisebox{-0.5\height}{\includegraphics[width=0.1\textwidth]{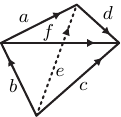}}.
\end{align}
In general, the faces would carry the additional multiplicity labels,
but we will suppress those multiplicity labels for the rest of this paper.
One can view the bottom (dotted) edge as the ``initial", vertical, edge and the top edge as the ``final", horizontal, edge in Eq.\,\eqref{eq:fsymbol}. The remaining four edges correspond to the outgoing/incoming edges in Eq.\,\eqref{eq:fsymbol}. Note that the boundary of the complex is the union of the initial and final cellulation. In a similar way, any sequence of Pachner moves can be represented by a three-dimensional cellulation whose boundary is the union of the initial and the final (two-dimensional) triangulation. The associated amplitude is evaluated by taking the product of numbers associated to the subsimplices. For details on the space-time picture for topological moves in our models, see App.\,\ref{app:state_sum}.

\begin{figure}
    \centering
    \includegraphics[width=0.5\textwidth]{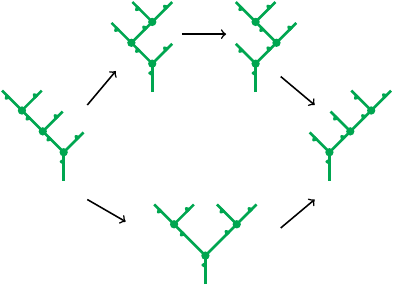}
    \caption{Different sequences of $F$-moves evaluating to the same transformation on of the graph have to compose to the same map on the associated vector spaces. In particular, the above diagram has to commute, i.e., the composite map corresponding to the top path has to evaluate to the same as composing the maps corresponding to the bottom path. The resulting condition on the $F$-symbols is called \textit{pentagon equation}.}
    \label{fig:pentagon}
\end{figure}

The numbers associated to the labeled space-time simplices have to fulfill several consistency conditions to implement exact topological moves. For example, if one combines more than 2 vertices, there are different sequences of Pachner moves that
have the same effect on the tesselation, see, for example, Fig.\,\ref{fig:pentagon}. This results in the non-trivial \textit{pentagon equation} on the $F$-symbols,
\begin{align}
\label{eq:pentagon_equation}
    F_{fdc}^{eba} F_{fgd}^{hai} = \sum_k F_{cgd}^{hek} F_{fgc}^{kbi} F_{ihk}^{eba}
\end{align}
for all $a,b,c,d,e,f,g,h,i$ which can be depicted as
\begin{align}
\label{eq:pentagon_move}
    \raisebox{-0.5\height}{\includegraphics[width=0.3\textwidth]{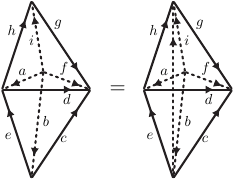}}.
\end{align}
On the left hand side the diamond formed by the 5 vertices is tessellated with two tetrahedra (top and bottom) whereas on the right hand side the same diamond is tessellated with three tetrahedra around the middle axis. Both should give the rise to the same number when summed over labels in the interior.

Strictly speaking, the pentagon equation in Eq.~\eqref{eq:pentagon_equation} is not the only constraint to the $F$-symbols. Firstly, for complete topological invariance, we would need to impose the move in Eq.~\eqref{eq:pentagon_move} for all possible branching structure configurations. Equivalently, one can add simpler auxiliary axioms specifically targeted to change the branching structure~\cite{bauer2020unified}. Secondly, for a physical model we have to impose Hermiticity/unitarity,
which means that all triangulations carry an orientation, and orientation reversal equals complex conjugation.

A spherical fusion category $\cC$~\cite{etingof2016tensor} is the mathematical object giving all the data for a consistent definition of topological moves in the bulk. The topological (ground) space on a given manifold is modeled by the space of all labeled  modulo topological moves. In practice, one chooses a minimal reference cellulation whose labelings form the basis of the associated vector space. One can use topological moves to relate any (basis) state on a given cellulation to an equivalent one on the reference cellulation.

\subsection{Boundaries and domain walls}
At the boundary, the bulk model is terminated along \textit{boundary edges} which can host different degrees of freed. However, the there is some (possibly non-trivial) action of the bulk edges connected to the boundary on these new degrees of freedom. In order to have full topological invarince of the model, the action has to fulfill certain consistency conditions. As we will show in this section, a boundary to a bulk with input category $\cC$ is given by a (left) \textit{$\cC$-module category}~$_\cC \cM$~\cite{etingof2016tensor, kitaev2012models, Bridgeman2020computingdatalevin}.
The module category $_\cC\cM$ is a (semi-simple) category equipped with a (left) $\cC$ action. On the level of the simple objects of $_\cC\cM$, the $\cC$ action is defined via \textit{module fusion multiplicites} $M_{a,\alpha}^\beta\in\bZ_{\geq 0}$,
\begin{align}
    a\triangleright\alpha = \sum_\beta M_{a\alpha}^\beta \beta\qcomma a\in\Obj(\cC), \alpha,\beta\in\Obj(_\cC\cM).
\end{align}
This action has to be compatible with the fusion operation in $\cC$, i.e.,
\begin{align}
    (a\triangleright(b\triangleright \alpha))\simeq& (a\times b)\triangleright\alpha\qq{} \forall a,b,\alpha.
\end{align}
This gives consistency conditions on $\{M_{a\alpha}^\beta\}$ in terms of $\{N_{a,b}^c\}$.

Graphically, the $\cC$-action can be represented by trivalent boundary vertices
\begin{align}
    \raisebox{-.5\height}{\includegraphics[width=0.075\textwidth]{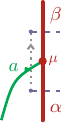}},
\end{align}
where $\mu$ is non-trivial for $M_{a\alpha}^\beta>1$.
We will later resort to the space-time picture where the dual edges of the above cellulation enter. As before, we depict them by dashed edges.
Similar to the $F$-symbols in the bulk, there are linear maps corresponding to deformations of the boundary, whose entries we refer to as \textit{$L$-symbols},
\begin{align}\label{fig:Lsymbol}
    \raisebox{-.5\height}{\includegraphics[width=0.4\textwidth]{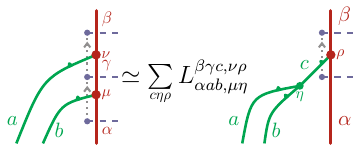}}.
\end{align}
In the space-time picture this move can be represented by a triangle with additional boundary labels at each of its vertices. Omitting the multiplicity labels it can be represented by
\begin{align}
    \mathlarger{L_{\alpha a b}^{\beta\gamma c} =}     \raisebox{-.5\height}{\includegraphics[width=0.1\textwidth]{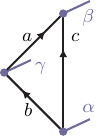}}.
\end{align}

\begin{figure}
    \centering
    \includegraphics[width=0.6\textwidth]{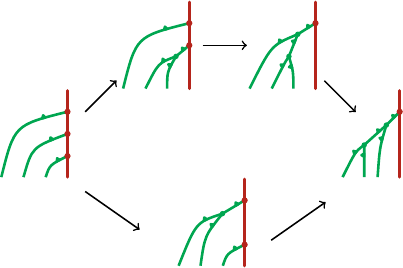}
    \caption{Different sequences of $L$- and $F$-moves evaluating to the same transformation on of the graph have to compose to the same map on the associated vector spaces. In particular, the above diagram has to commute, i.e., the composite map corresponding to the top path has to evaluate to the same as composing the maps corresponding to the bottom path. The resulting condition on the $L$- and $F$-symbols is called \textit{boundary pentagon equation}.}
    \label{fig:bounadry_pentagon}
\end{figure}

The $L$-symbols have to fulfill an associativity condition similar to the bulk pentagon equation. This can be expressed as a commutative diagram in Fig.\,\ref{fig:bounadry_pentagon}. The resulting condition on the $L$- (in relation to the $F$-symbols) reads
\begin{align}
\label{eq:boundary_pentagon}
L_{\alpha e a}^{\delta\beta d} L_{\beta c b}^{\delta\gamma e} = \sum_{f} L_{\alpha b a}^{\gamma\beta f}L_{\alpha c f}^{\delta\gamma d} F_{dce}^{baf} %
\qq{represented as}\raisebox{-0.4\height}{\includegraphics[width=0.3\textwidth]{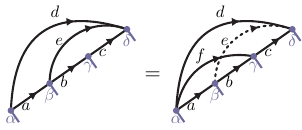}}
\end{align}
The equation on the right shows the corresponding space-time recellulation. On the left, there are two boundary triangles, whereas on the right there is one bulk tetrahedron with two boundary triangles. Note that the edge labeled by $e$ on the right side is not part of a boundary triangle but extends into the bulk.
Again, Eq.~\eqref{eq:boundary_pentagon} does not yield full topological invariance, but we have to add additional axioms as well as Hermiticity. For the models we are going to study, these additional axioms will be fulfilled automatically.

Taken together, the mathematical structure describing the bulk-boundary microscopics is a $\cC$-module category. Given $\cC$ and module categories associated to (distinct) boundaries, we can now describe a topologically ordered ground state on manifolds with boundaries by the space of its cellulations modulo topological moves. The exact topological invariance allows us to work with a minimal reference cellulation and use moves to relate a state on a different cellulation to an equivalent one on the chosen reference cellulation.

\subsubsection{Domain walls and the folding trick}\label{sec:folding}
\begin{figure}
    \centering
    \includegraphics[width=0.6\textwidth]{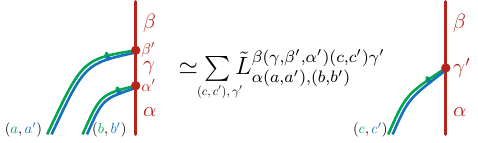}
    \caption{Via the folding trick, a $\cC$-$\cC'$ domain wall, classified by a $\cC$-$\cC'$bimodule, is equivalent to a $\cC\otimes \cC'$ module, defined by $L$ symbols of the above form. In Eq.\,\eqref{eq:folding_Lsymbol} we give the equation relating the data defining the bimodule to the boundary data after the fold.}
    \label{fig:folding_Lsymbols}
\end{figure}

Given consistent $F$ and $L$ symbols -- a fusion category $\cC$ and a $\cC$-module category~$_\cC \cM$ -- we have defined a topological fixed-point model for a topological phase with boundary.
In fact, this data can also describe interfaces between topological models each of which is described by fusion categories $\cC$ and $\cC'$. Such interfaces are often called \textit{domain walls} and are defined by a \textit{$\cC$-$\cC'$-bimodule category} ~\cite{etingof2016tensor, kitaev2012models, Bridgeman2020computingdatalevin}.
A $\cC$-$\cC'$-bimodule category is defined by a left $\cC$-, and a right $\cC'$-action, each of which is equipped with $L$-, respectively $R$-symbols that are both compatible with fusion in $\cC$, respectively in $\cC'$.
Graphically, the associated moves can be depicted by
\begin{align}
    \raisebox{-0.4\height}{\includegraphics[width=0.35\textwidth]{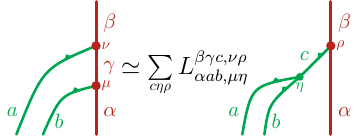}}
\qq{and}
    \raisebox{-0.4\height}{\includegraphics[width=0.35\textwidth]{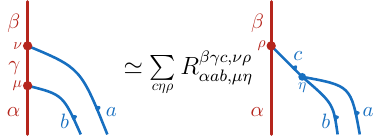}},
\end{align}
where $\cC$-labeled strings are depicted in green and $\cC'$-labeled strings in blue.
Moreover, there are moves including strings on both sides of the domain wall. The associated linear maps are given by so-called \textit{C-symbols} and can be graphically depicted as
\begin{align}
    \raisebox{-0.5\height}{\includegraphics[width=0.35\textwidth]{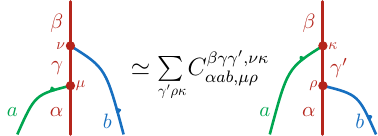}}.
\end{align}
Again, the $L$, $R$ and $C$ symbols have to fulfill consistency conditions coming from different sequences of moves with the same overall effect.
Another way of studying domain walls is by ``folding" one side of the domain wall onto the other side. With this, the $\cC$-$\cC'$ domain wall becomes a boundary of a $\cC\otimes\overline{\cC'}$ model, where the bulk strings are labeled by tuples $(a,a')\in\Obj(\cC)\times\Obj(\cC')$ and the $F$-symbols are inherited from $\cC$ and $\cC'$ but with the $\cC'$ $F$-symbols complex conjugated.
Omitting the multiplicities in $\cC$ and $\cC'$ individually, the folding trick can be depicted graphically as
\begin{align}
    \raisebox{-0.5\height}{\includegraphics[width=0.35\textwidth]{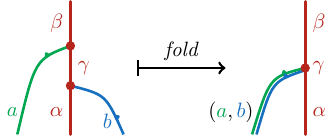}},
\end{align}
Note that the boundary vertex can get a multiplicity even if the two input fusion categories are multiplicity-free. This is less of a feature of the folding trick than a consequence of combining two connected tri-valent vertices into a single four-valent one. We will see that for the model class we are interested in this paper, there is no additional multiplicity coming in. The $L$-symbols of the folded model (see Fig.\,\ref{fig:folding_Lsymbols}) are straight forwardly obtained by a combination of the $(L,R,C)$ symbols using the correspondence above to resolve the $L$ move of the folded model to a sequence of $L,R$ and $C$ moves in the unfolded model,
\begin{align}
     \Tilde{L}^{\beta (\gamma,\beta',\alpha')(c,c')\gamma'}_{\alpha(a,a'),(b,b')} = C^{\beta'\gamma\gamma'}_{\alpha',b,a'} L^{\beta\beta' c}_{\gamma' a b}\overline{R^{\gamma'\alpha'c'}_{\alpha a' b'}}.\label{eq:folding_Lsymbol}
\end{align}

\subsection{Topological gauge theory models}\label{sec:gauge_models}
Let $G$ be a finite group. For the rest of the section, we focus on $\cC=\vV^\omega(G)$, the category of $G$-graded vector spaces. In $\vV^\omega(G)$, the simple objects are group elements and the fusion multiplicities are given by the group multiplication,
\begin{align}
    N_{gh}^k = \delta_{gh,k}.
\end{align}
With that, the vertices take the following form
\begin{align} \label{fig:bulk_vertex}
    \raisebox{-0.5\height}{\includegraphics[width=0.15\textwidth]{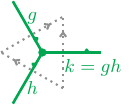}}.
\end{align}
Since there are just two free labels at any vertex, the $F$-moves are defined by a single function $\omega:G\times G\times G\to U(1)$,
\begin{align} \label{fig:Fmove_Abelian}
    \raisebox{-0.5\height}{\includegraphics[width=0.4\textwidth]{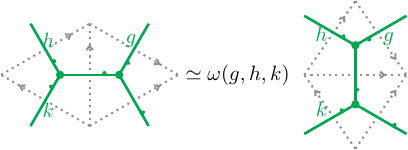}}\qq{represented by} \raisebox{-.5\height}{\includegraphics[width=0.125\textwidth]{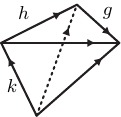}}.
\end{align}
In this case, the pentagon equation reduces to a 3-cocycle condition on $\omega$,
\begin{align}
    \omega(h,k,l)\omega(g,hk,l)\omega(g,h,k) = \omega(gh,k,l)\omega(g,h,kl) \quad\forall g,h,k,l\in G
    .
\end{align}
Moreover, we have a \textit{gauge freedom}. Namely, we can multiply the $F$-symbols by product of local unitaries. In our model these unitaries are simple phases for every vertex, respectively face in the dual picture. In general this phase can depend on the local configuration around the vertex, given by a pair of group elements. Hence, any function $\eta: G\times G\to U(1)$ defines a gauge transformation. Applying the corresponding gauge in \eqref{fig:Fmove_Abelian} maps the associated  3-cocycle to
\begin{align}
    \omega(g,h,k) \mapsto \omega(g,h,k)(\delta \eta)(g,h,k) = \omega(g,h,k)\frac{\eta(h,k)\eta(g,hk)}{\eta(gh,k)\eta(g,h)}.
\end{align}
In fact, the product of $\eta$s with which the 3-cocycle gets multiplied is exactly the coboundary of the 2-cochain $\eta$. This shows that topological lattice models based on $\vV^\omega(G)$ are classified by the third cohomology group $H^3(G,U(1))$, see App.\,\ref{app:cohomology}.

Boundaries in topological gauge theory models correspond to module categories of $\vV^\omega(G)$ and have been studied in detail in the mathematical literature and classified by a (possibly twisted) group algebra of a subgroup $H\subset G$ on which the input 3-cocycle $\omega$ is cohomologically trivial~\cite{davydov2017lagrangian, Bridgeman2020computingdatalevin}. In this section, we see how topological boundaries are classified in our model and connect it to microscopically different, but equivalent, classifications in the literature.
In particular, we explain how to model a boundary associated to a twisted group algebra $\bC^{\psi}[H]$ -- a subgroup $H$ and a 2-cocycle $\psi$ on it -- in our calculations.

Given a subgroup $H$, the labels at the boundary are cosets in $\flatfrac{G}{H}=\{aH\,|\,a\in G\}$.
The bulk $G$-action is then defined by left action on the coset,
\begin{align}
	g\triangleright\alpha = g\triangleright(a H) := (ga)H,
\end{align}
where $a\in\alpha$ is a representative of the coset $\alpha$.

Since every vertex is multiplicity-free and only two of the labels of its incident edges are independent, the $L$-symbols only have three open indices. They are defined by $\psi:G/H\times G\times G\to U(1)$ with
\begin{align} \label{eq:Lmove_gaugemodel}
    \raisebox{-.5\height}{\includegraphics[width=0.35\textwidth]{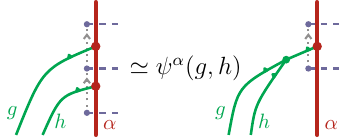}}\qq{represented by}\raisebox{-.5\height}{\includegraphics[width=0.1\textwidth]{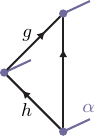}}.
\end{align}
The phase function has to satisfy the mixed pentagon equation in Eq.~\eqref{eq:boundary_pentagon}. In this case, it simplifies to
\begin{align}\label{eq:VecG_Lsymbols_associativity}
    	\omega(g,h,k) =& \frac{ \psi^{\alpha}(h,k)\psi^{\alpha}(g,hk)}{\psi^{\alpha}(gh,k)\psi^{k\triangleright \alpha}(g,h)}=  (\Tilde{\delta}\psi^{\alpha})(g,h,k),
\end{align}
where $\Tilde{\delta}$ is the twisted coboundary operator (for more details, see App.\,\ref{app:cohomology}).
This means $\omega$ has to be the twisted coboundary of $\psi^{\alpha}$ if we interpret $\omega$ as a twisted cochain which is constant in $\alpha$. By setting $\alpha$ to the trivial coset $H$ and restricting all arguments to the subgroup $H$, we see that $\omega$ has to be a (untwisted) coboundary when restricted to $H$.
Given any solution $\psi$ to the equation above and a twisted cocycle $\alpha$ (that is, $\Tilde{\delta}\alpha=0$), $\psi'=\psi\cdot\alpha$ is also a solution.
Furthermore, two different boundaries $\psi$ can be considered equivalent if they differ by on-site unitary gauge on the boundary fusion vertices.
In our case, those fusion spaces are 1-dimensional (or 0-dimensional if the fusion rules are not obeyed), so such an isomorphism is defined by a phase $\xi:\flatfrac{G}{H}\times G\to U(1)$ depending on the labels of the strings adjacent to a boundary fusion vertex.
Since $\psi$ contains three boundary fusion vertices (two at its input and one at its output), the isomorphism acts as
\begin{align}
    \psi^\alpha(g,h) \mapsto \psi^\alpha(g,h)\frac{\xi^\alpha(h)\xi^{h\triangleright\alpha}(g)}{\xi^\alpha(gh)}=\psi^\alpha(g,h)(\Tilde{\delta}\xi^\alpha)(g,h).
\end{align}
We see that the gauge corresponds to a multiplication of a coboundary with the same coboundary operator as in Eq.\,\eqref{eq:VecG_Lsymbols_associativity} but acting on the 1-cochains $\{\xi^\alpha\}$. In total, we see that although the different $\psi$ are not twisted 2-cocycles, their differences are, whereas $\psi$s differing by 2-coboundaries are considered equivalent.
Hence, the gauge equivalence classes of boundary models cannot be directly identified with the twisted cohomology group $H^2(G, (U(1)^{\flatfrac{G}{H}})_G)$, but are equipped with a regular action of the latter. Here, the subscript $_G$ indicates the non-trivial right action of $G$. That is, the set of equivalence classes forms a \emph{torsor} over the second twisted cohomology group.

In App.\,\ref{app:sec:isomorphism} and~\ref{app:thin_boundary}, we show that this classification indeed coincides with known results, i.e.,
\begin{align}\label{eq:isomorphic_classifications}
    H^2(G,(U(1)^{\abs{G/H}})_G)\simeq H^2(H,U(1)).
\end{align}
In particular, this isomorphism holds for every cohomology group $H^n$ for $n\geq 1$ and is induced by the map $m^{(n)}:\flatfrac{G}{H}\times G^{\times n}\to H^{\times n}$ that -- to the best of our knowledge -- has first been mentioned by T. Lawson in Ref.\,\cite{overflow}.

\subsection{Models for Abelian phases}\label{sec:ModelsAbelian}
A topologically ordered phase is called \textit{Abelian} if the fusion outcome of any pair of topological point defects -- anyons -- is unique, i.e. they form an Abelian fusion category. Importantly, not every Abelian fusion category $\cC$ gives rise to an Abelian phase $D(\cC)$ (see Section~\ref{sec:tube_abelian}). However, as we will see later, the Abelianess of the anyons can be traced back to properties of the input category if it is of the form $\vV^{\omega}(G)$. In this case, the group has to be Abelian and the twisting 3-cocycle of a certain form.
Any finite Abelian group is isomorphic to a product of cyclic groups,
\begin{align}\label{eq:AbelianGdecomposition}
    G\simeq \bZ_{p_1^{m_1}}\times\bZ_{p_2^{m_2}}\times \cdots \times \bZ_{p_N^{m_N}},
\end{align}
where $p_i$ is prime and $m_i$ positive integers for any $i=1, \dots, N$. For the rest of this section it suffices to consider $G$ having $N$ independent cyclic factors. Inequivalent $F$-symbols are classified by 3-cocycle classes on the group above. In fact, any 3-cocycle class on such a product group can be represented as a product of so-called type-I, type-II and type-III cocycles (see App.\,\ref{app:cohomology}). The anyons in our model are only Abelian for a cocycle that is cohomologous to a product of type-I and type-II cocycles only, see Section~\ref{sec:tube_abelian}.

From the above decomposition into cyclic factors any subgroup $H$ and the associated cosets $\flatfrac{G}{H}$ can be easily determined. Moreover, its second cohomology group decomposes similarly, see App.\ref{app:cohomology}.
Given a subgroup $H$ (with $k$ factors), we can construct a non-trivial 2-cocycle by taking products of 2-cocycles on each pair $H_{ij} = \bZ_{h_i}\times\bZ_{h_j}$ in the factor decomposition of $H$. In particular, there are $\gcd(h_i,h_j)$ inequivalent 2-cocycle classes on $H_{ij}$ represented by normalized 2-cocycles of the form
\begin{align}\label{eq:AbelianG2cocycle}
    \Omega(a,b)^{q_{ij}} = e^{\frac{2\pi i}{\gcd(h_i,h_j)} q_{ij} a_i b_j}\qcomma\text{with }q_{ij}\in\{0,1, \dots, \gcd(h_i,h_j)\},
\end{align}
where $a_i$($a_j$) is the component of $a$ in $\bZ_{h_i}(\bZ_{h_j})\subset H_{ij}$.
Following the previous subscection, the associated $L$ symbol is obtained by precomposition with $m^{(2)}$,
\begin{align}\label{eq:AbelianGLsymbol}
    \psi^\alpha(a,b)^{q_{ij}} =& (\Omega\circ m^{(2)})(\alpha,a,b)^{q_{ij}}.
\end{align}

\subsection{Examples} \label{sec:model_examples}
In this section, we give some examples of the input data for  $\vV^\omega(G)$ models with boundary.

\subsubsection{\texorpdfstring{$\vV^{\omega}(\bZ_N)$}{Vecω(ZN)}}
The cyclic group of order $N$, $\bZ_N=\{0,1, \dots, N-1\}$ with group operation $a\oplus_p b := a+b \mod N$ has $N$ 3-cocycle classes each of which are generated from a single 3-cocycle class. The normalized 3-cocycle
\begin{align}\label{eq:3cocycletype1}
    \omega_I^n(a,b,c) = e^{\frac{2\pi i}{N} n a(b+c-b\oplus c)/N}\qcomma n\in\bZ_N,
\end{align}
is a canonical representative for the $n$th cocycle class of $H^3(\bZ_N,U(1)) = \bZ_N$~\cite{propitius1995topological}. We say that a 3-cocycle of that form, only supported on a single cyclic factor, is of type I. Using $\bZ_N$ and $\omega_I^n$ as an input gives an Abelian bulk theory.

Let us look at possible boundaries of such a bulk model. For simplicity, we take $N=p$ prime. In this case, $\bZ_p$ only has two subgroups, $H_0=\{0\}$ and $H_1=\bZ_p$. Both $H_1$ and $H_2$ only have trivial 2-cocycles, i.e., only one potential boundary associated to either of them. In the untwisted case both give rise to a boundary and correspond to rough and smooth boundaries of Kitaev's toric code on qu$p$its~\cite{Kitaev2003fault, barter2019domain}. In the twisted case, for $n\neq 0$, $\omega_I^n$ becomes cohomologically trivial only on $H_0$. Hence, these models only have one ``standard" boundary.

\subsubsection{\texorpdfstring{$\vV^{\omega}(\bZ_N\times\bZ_M)$}{Vecω(ZpxZq)}}
An Abelian group with two cyclic factors $\bZ_N\times\bZ_M = \{(a,b)\,|\,a\in\bZ_p,b\in\bZ_q\}$ has $NM$ 3-cocycle classes of type I. Their normalized representatives decompose into a product of cocycles as in Eq.\,\eqref{eq:3cocycletype1}, each supported on one factor only. Additionally, there are non-trivial type-II cocycle classes represented by~\cite{propitius1995topological}
\begin{align}\label{eq:3cocycletype2}
    \omega_{II}^{n_{12}}(a,b,c) = e^{\frac{2\pi i}{\gcd(N,M)} n_{12} a_1(b_2+c_2-b_2\oplus c_2)/M}\qcomma n_{12}\in\bZ_{\gcd(N,M)}.
\end{align}
3-cocycles of type II are supported on two cyclic factors and are gauge inequivalent to any type I cocycle. Together with the two subgroups generated by type I cocycles we have $H^3(\bZ_N\times\bZ_M,U(1)) = \bZ_N\times\bZ_M\times\bZ_{\gcd(N,M)}$. Using $\bZ_N\times\bZ_M$ and a cocycle of type I or II as in input gives an Abelian bulk model.

For simplicity, consider $N=M=p$ prime. The possible boundary models are given by subgroups of $\bZ_p\times\bZ_p$. They are $H_0=\{0\},H_1=\langle (0,1)\rangle\simeq\bZ_p, H_2=\langle (1,0)\rangle\simeq \bZ_p, H_{3,l} = \langle (1,l)\rangle \simeq \bZ_{p}$ and $H_4=\bZ_p\times\bZ_p$.
In the untwisted case any of these subgroups defines a boundary model. Additionally, $H_4$ has non-trivial 2-cocycles of the form
\begin{align}\label{eq:2cocycleZpZq}
	\Omega((a_1,a_2),(b_1,b_2))^m  = e^{\frac{2\pi i}{p}m a_1b_2}\qcomma m\in\bZ_{p}.
\end{align}
Since $H_4$ is the whole group, there is only one coset such that $\psi$ only depends on group labels.
Note that if $H$ has two cyclic factors and is a proper subgroup of $G$ the form of $\psi$ will be different because the coset label $\alpha$ can be non-trivial, see Eq.\,(\ref{eq:AbelianGLsymbol}).

The 3-cocycle in Eq.\,\eqref{eq:3cocycletype2} becomes cohomologically trivial on $H_0$ and the isomorphic subgroups $H_1$, $H_2$ and $H_{3,l}\;\forall l$. All of them have no non-trivial 2-cocycle class which gives $3+l$ inequivalent topological boundaries.

\subsubsection{{\texorpdfstring{$\vV^{\omega}(S_3)$}{Vecω(S3)}}}\label{sec:example_model_s3}
The smallest non-Abelian group is the permutation group of three elements $S_3=\langle t,r\,|\, t^2=r^3=e, tr=r^2t\rangle \simeq \bZ_3\rtimes\bZ_2$ ($e$ being the identiy element). Its third cohomology group is given by~\cite{propitius1995topological}
\begin{align}
    H^3(S_3,U(1)) = \bZ_3\times\bZ_2\simeq \bZ_6.
\end{align}
Interestingly, the third cohomology group is the product of the cohomology groups of the two non-trivial subgroups of $S_3$. However, since $S_3$ is a a semidirect product of the two, the 3-cocycles are not simple products of 3-cocycles of the subgroups.
The $6$ inequivalent classes can be represented by the normalized cocycles~\cite{propitius1995topological}
\begin{align}\label{eq:3cocycle_S3}
    \omega^p(t^Ar^a,t^Br^b,t^Cr^c) = e^{\frac{2\pi i}{3} p (-1)^{B+C}a\left((-1)^Cb+c -[(-1)^Cb\oplus c]\right)/3} (-1)^{p ABC}\qcomma p\in\bZ_6,
\end{align}
where the elements in $S_3$ are represented by $t^Ar^a$ with $A\in\bZ_2$, $a\in\bZ_3$ and $\oplus$ denotes addition mod 3.
When comparing the individual factors with a 3-cocycle of a cyclic group (see Eq.\,\eqref{eq:3cocycletype1}), we note that the second factor, $(-1)^{ABC}$, corresponds to the non trivial 3-cocycle of the (Abelian) subgroup $\bZ_2\subset S_3$ and the first factor corresponds to the non-trivial 3-cocycle of the subgroup $\bZ_3\subset S_3$ precomposed with the automorphism $\rho_{A,B,C}\in\Aut(\bZ_3^{\times 3})$ defined by $\rho_{A,B,C}(a,b,c) = ((-1)^{B+C}a,(-1)^C b, c)$
for any $(A,B,C)\in\bZ_2^{\times 3}$.

$S_3$ has four subgroups up to conjugation, $H_0 =\{e\}$, $H_t = \langle t\rangle\simeq \bZ_2$, $H_r=\langle r\rangle\simeq\bZ_3$ and $H_G = S_3$. All of them have a trivial second cohomology group~\cite{Bridgeman2020computingdatalevin, Cong2017Defects}. Hence, there is one boundary type associated to each of these three subgroups in the untwisted case. For a twisted bulk model, only the subgroups on which the 3-cocycle in Eq.\,\eqref{eq:3cocycle_S3} is trivial, define a consistent boundary. Specifically, $\omega^p$ is cohomologically trivial on $H_r$ for $p=0,3$, and is trivial on $H_t$ for $p=0,2,4$.

\section{Bulk anyons: Tube algebra}\label{sec:tube}
In this section, we will revisit the question of how to add anyons to the fixed-point models from Section~\ref{sec:prelim_bulk}.
Anyons are (irreducible) point-defects in the bulk of a topological phase. Their world-lines live in a three-dimensional space-time and are equipped with compatible fusion and braiding structure. In mathematical terms, anyons form the simple objects in a \textit{(unitary) modular tensor category} (U)MTC. In fact, all their defining data can be calculated with fixed-point models.
In this manuscript, we mainly focus on finding the anyons themselves, i.e., the set of irreducible sub-spaces of the point-defects and comment shortly on how to derive the fusion and braiding data with similar methods in Section~\ref{sec:fusionbraiding}.

\subsection{General fixed-point models}
Anyons are point-like topological defects in the bulk, and are known to be characterized by string-nets on an annulus, respectively a ``tube", $(0,1)\times S_1$. In fact, the associated vector space can be equipped with a multiplicative action on itself, rendering it an algebra. The irreducible representations of this \textit{tube algebra}~\cite{bullivant2019tube, bullivant2018phd, evans1995ocneaunu} can be associated to the simple objects in the UMTC describing the bulk anyons. In this section, we will illustrate this concept by first calculating the irreducible sub-spaces of the tube algebra in topological gauge theory models and sketch how to obtain a consistent fusion and braiding on them.

\begin{figure}
    \centering
    \includegraphics[width=0.9\textwidth]{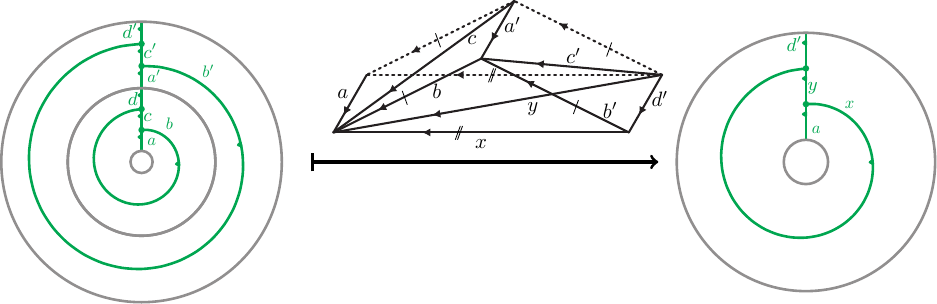}
    \caption{The multiplication of two tube algebra basis elements $(a,b,c,d)_T$ and $(a',b',c',d')_T$ is defined via gluing the two associated string diagrams together (left) and using $F$-moves to reduce it to the cellulation on the right. The phase acquired by the sequence of moves can be derived by evaluating the space-time complex that maps the two dual triangulations to each other, which is composed of three tetrahedra (middle). Note that the front- and the back-side edges of the space-time complex above are identified.}
    \label{fig:tube_mult}
\end{figure}

In principle, one can choose any tesselation of the annulus for the upcoming analysis. It is beneficial to choose a simple representative. For the rest of this section, we define a basis element of the tube algebra $T$, labeled by $(a,b,c,d)\in \Obj(\cC)^4$ (again, omitting the multiplicity labels at the vertices), via the following cellulation:
\begin{align}\label{eq:tube_basis}
    (a,b,c,d)_T := \raisebox{-0.5\height}{\includegraphics[width=0.2\textwidth]{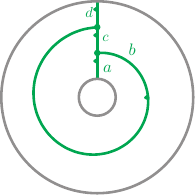}}\;.
\end{align}
We can define an (associative) multiplication on the vector space spanned by the diagrams of the above form by gluing two tessellated tubes together and using topological moves to map back to the initial cellulation, see Fig.\,\ref{fig:tube_mult}.
Evaluating the space-time complex corresponding to the recellulation, we obtain
\begin{align}
    (a,b,c,d)_T\ast(a',b',c',d')_T = \delta_{d,a'}\sum_{x,y} F_{yac}^{bb'x} F_{ybx}^{b'd'c'} \overline{F_{ybc}^{a'b'c'}} (a,x,y,d')_T.
\end{align}

\subsection{Fusion and braiding in the bulk}\label{sec:fusionbraiding}
The anyons in topological fixed-point models form simple objects of a \textit{unitary modular tensor category} (UMTC). This means they are equipped with additional data/topological quantum numbers related to their fusion and braiding. In fact, this data together is known as the Drinfeld center of the fusion category defining the lattice model. Although it is outside of the scope of this paper to compute the full center, i.e., the fusion and braiding data, we want to comment on how they can be calculated with similar techniques that we have already introduced. For related discussions we refer to App.\,\ref{app:state_sum} and Refs.\,\cite{Hu2013Twisted, bullivant2019tube}.

Before deducing new quantities of the topological phase of the lattice model, let us take a step back and put the tube algebra into a larger context. The fact that the topological vector spaces associated to a tube, $S_1\times(0,1)$, forms an algebra, comes from the fact that gluing a tube onto a tube again gives a tube. Hence, we can define an action of $T$ onto itself, i.e., there exist a map
\begin{align}
    T\to\End(T),
\end{align}
defining an (multiplicative) action of $T$ onto itself.
Similarly, when considering cellulations (modulo local moves) of other manifolds, one can define endomorphisms on the topological vector space from gluing operations that leave the manifold invariant.
We will now turn our attention to how this allows us to calculate the fusion multiplicities in the UMTC formed by the anyons.
Consider a fusion process in 2+1-dimensional space-time. It can represented as a vertex where three anyon world-lines meet.
The boundary of the regular neighborhood of this vertex is a pair of pants -- or three-punctured sphere.
Just as before we can define a vector space (in terms of its basis) by tessellating this manifold with trivalent vertices from the input fusion category (see \eqref{fig:bulk_vertex_general}).
Let's call this space $F$.
In fact, gluing a tube onto any of the three holes of the pair of pants does not change its topology. Hence, we can define a ``tri-representation"
\begin{align}\label{eq:pantsTrirep}
    T\times T\times T \to \End(F).
\end{align}
As a representation, $F$ decomposes into a direct sum of triples of irreducible representations of $T$,
\begin{align}\label{eq:anyonfusionspace_multiplicities}
    F=\bigoplus_{a,b,c} \Tilde{N}_{ab}^c a\otimes b\otimes c,
\end{align}
where $a,b,c$ label the irreducible representations of $T$, i.e., the anyons of the bulk.
The multiplicities in this decomposition $\Tilde{N}_{ab}^c$ coincide with the fusion multiplicities of the bulk anyons.
In order to calculate them, we consider the tri-representation in Eq.\,\eqref{eq:pantsTrirep} for a triple of central idempotents $(P^T_a,P^T_b,P^T_c)\in T^{\times 3}$ yielding a projector $P_F$, and then take the trace of $P_F$. We will see that we use a similar approach to get the fusion multiplicities into the boundary, see Section~\ref{sec:bulkbdryfusion}.
In App.\,\ref{app:state_sum} we give a formula for $\Tilde{N}_{ab}^c$ for an input category of the form $\vV^\omega(G)$. We will use the same method to calculate the dimension of the ``fusion" vector spaces for another type of space-time event, namely the partial condensation of an anyon in the bulk to an anyon in the boundary, see Section~\ref{sec:bulkbdryfusion}.

To complete the UMTC, we have to define a braiding and an associator on the fusion spaces of the irreducible representations of $T$. One way to fully define a braiding is via so-called $R$-symbols, defining a half-braiding on anyon world-lines,
\begin{align}\label{eq:Rsymbol}
     \raisebox{-0.5\height}{\includegraphics[width=0.3\textwidth]{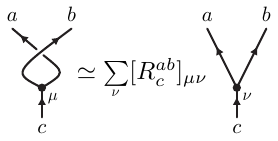}}.
\end{align}
The associator on the other hand is defined by so-called $F$-symbols changing the order of fusion.
Both $R$- and $F$-symbols have to obey consistency conditions known as \textit{pentagon} and \textit{hexagon conditions}~\cite{etingof2016tensor}. Given this set of data, one can derive the topological quantum numbers like self-exchange statistics (topological spin $\theta_a$) and the mutual-exchange statistics ($S$ matrix) for any anyon with a simple calculation. For more details, see App.\,E of Ref.\,\cite{kitaev2006anyons}.

Topological fixed-point models allow for a direct calculation of both $R$ and $F$-symbols.\footnote{In fact, one can directly calculate the modular data of the anyons without deriving the $R$ tensor first. For this, one considers the vector space defined by a cellulation of a torus and analyzes the endomorphism induced by the mapping class group of the torus, generated by $S$ and $T$ matrices. For a detailed derivation,
see Ref.\,\cite{Hu2013Twisted}.} In App.\,\ref{app:state_sum}, we give an explicit prescription of how to obtain the $F$-symbols and give a formula for the case of a $\vV^\omega(G)$ model. Let us here shortly lay out how to obtain the $R$-symbols with microscopic models. For this we again consider the vector space $F$ defined via a tessellated pair of pants. This space represents the fusion space of three bulk anyons. The half-braiding acts on this vertex by interchanging two of its legs as in Eq.\,\eqref{eq:Rsymbol}. This induces a non-trivial action on the cellulation. In the same spirit as before we can use local moves to map it back to the original cellulation and thereby define an action on $F$. To express this action as a tensor in the basis of anyons, i.e., irreducible representation of $T$, we have to project the $R$-action onto this associated sub-spaces by precomposing it
with a triple of central projectors $(P^T_a,P^T_b,P^T_c)\in T^{\times 3}$. Similarly, one can calculate the $F$-symbols from microscopic models by mapping between the two ways of decomposing a sphere with four holes into two pairs of pants.

\subsection{Topological gauge theory models}\label{sec:tube_gaugetheory}
For gauge theory models the input fusion category is $\vV^{\omega}(G)$, each vertex is multiplicity-free.
The basis state in Eq.\,\eqref{eq:tube_basis} are parametrized by two group elements.
We label them by $(g,h)_T = (hgh^{-1}, h, hg, g)_T$.
Plugging in the defining data from Section~\ref{sec:gauge_models}, we obtain the \emph{tube algebra} $T$ for the group case as,
\begin{align}\label{eq:tube_multiplication}
    (g',h')_T \ast(g,h)_T = \delta_{g',h gh^{-1}}\beta_{g}(h',h)(g,h' h)_T,
\end{align}
defined over the vector space $\bC^{G\times G}$, where $\beta_g(h,h')$ is the phase assigned to the sequence of moves from the left to the right hand side of Fig.\,\ref{fig:tube_mult}. Evaluating the space-time complex representing the moves in Fig.\,\ref{fig:tube_mult}, we get
\begin{align}\label{eq:betag_general}
    \beta_g(h',h) = \omega(h'hg(h'h)^{-1}, h',h)\,\omega(h',h,g) \,\overline{\omega(h',hgh^{-1},h)},
\end{align}
which can be seen as a slant product $i_g\omega$, see App.\,\ref{app:cohomology}.
The tube algebra represents $D^{\omega}(G)$, the \emph{twisted quantum double} of $G$, introduced by Dijkgraaf et al in Ref.\,\cite{dijkgraaf1991quasi} and plays an important role in the study of (finite) gauge theory models for topological phases and its applications to topological stabilizer codes~\cite{Hu2013Twisted, bullivant2019tube,Magdalena2021nonpauli, Ellison2022Pauli}.

In fact, we find that Eq.\,\eqref{eq:tube_multiplication} is of the form of Eq.\,\eqref{eq:mult_algebrawithaction} with $X=G$ acting on itself via conjugation, i.e., $h\triangleright g = hgh^{-1}$.
Indeed, $\beta_g$ plays the role of $\Psi^x$ and fulfills
\begin{align}\label{eq:twisted2cocyclebeta}
    \beta_{x}(h,k)\beta_x(g,hk)= \beta_x(gh,k)\beta_{kxk^{-1}}(g,h) 
\end{align}
in analogy to Eq.\,\eqref{eq:twisted2cocyclegeneral}. Choosing a normalized 3-cocycle $\omega$ makes $\beta_g$ normalized and with that we can use the algorithm from Section~\ref{sec:twistedwithaction}.

To construct the irreducible representations of $T$ and the associated central idempotents we proceed as described in Section~\ref{sec:twistedwithaction}.
First, the transitive subsets $\{X_i\}$ are the conjugacy classes $\{c\}$.
For each conjugacy class $c$, we pick a representative $\hat{c}$.
It is stabilized by its centralizer $Z(\hat{c}) = \{g\in G\;|\; gc=cg\}$.
Following the general considerations from Section~\ref{sec:twistedwithaction} the irreducible sub-spaces of the tube algebra are additionally labeled by irreducible $\beta_{\hat{c}}$-projective representation of $Z(\hat{c})$.
Combined, we identify the irreducible sub-spaces of $T$ with a pair $(c,\rho_c)$, a conjugacy class and an irreducible $\beta_g$-projective representation of $Z(\hat{c})$.
The central idempotents are given by
\begin{align}
    P^T_{(c,\rho_c)} = \frac{\dim(\rho_c)}{\abs{Z(c)}} \sum_{g\in c}\sum_{h\in Z(g)} \overline{\Tilde{\chi}_{\rho_c}^g(h)} (g,h)_T,
\end{align}
where $\Tilde{\chi}_{\rho_c}^g(h) := \Tr(\rho_c^g(h))$ denotes the projective character of the IPR $\rho_c^g$ of $Z(g)$.
Note that the stabilizer group of any element in $c$ is given by conjugating the centralizer of the chosen representative, $Z(g\hat{c}g^{-1}) = gZ(\hat{c})g^{-1}$.


In practice, finding the irreducible projective character functions is not a straight forward task.
In the special case, where $\beta_g$ is a coboundary, i.e.,
\begin{align}\label{eq:trivial_2cocycle}
    \exists \epsilon_g:G\to U(1):\quad \beta_g(a,b) = \frac{\epsilon_{aga^{-1}}(b)\epsilon_g(a)}{\epsilon_g(ab)} = (\Tilde{\delta}\epsilon_g)(a,b),
\end{align}
the IPRs of $Z(g)$ are in bijection with the irreducible linear representations. In particular, one can ``gauge" away the twist $\beta_g$ with the cochain $\epsilon_g$ and the projective characters are given by
\begin{align}\label{eq:projectiveChar_trivial}
    \Tilde{\chi}_{\rho}^g(h) = \epsilon_g(h)\chi^g_\rho(h),
\end{align}
where $\chi^g_\rho(h)$ is the linear (not projective) character of the irrep $\rho$ of $Z(g)$. With that, the central idempotents simplify to
\begin{align}\label{eq:PT_trivialbeta}
    P^T_{(c,\rho_c)} = \frac{\dim(\rho_c)}{\abs{Z(c)}}\sum_{g\in c}\sum_{h\in Z(g)} \overline{\epsilon_g(h)\chi_{\rho_c}^g(h)}(g,h)_T.
\end{align}
We will see in the following sections that the examples considered in this paper are all of the above form.
A well-studied example where $\beta_g$ defines a non-trivial projective irrep is obtained from $G=\bZ_N^{\times 3}$ and $\omega$ being cohomologous to a type-III 3-cocycle.
In this case, there are higher-dimensional irreducible sub-spaces in the tube algebra even though the input group is Abelian, see, for example,
Refs.\,\cite{propitius1995topological, Hu2013Twisted}.

\subsection{Tube algebra for Abelian anyons}\label{sec:tube_abelian}
In an Abelian anyon model the fusion of two anyons $a$ and $b$ yields a unique anyon $c$, i.e., $\Tilde{N}_{ab}^c=1$ for exactly one $c$ and $0$ otherwise. In other words, the fusion space of any pair of anyons is one-dimensional.
If we want our microscopic model to yield an Abelian anyon theory, we thus require $\abs{c}=\dim(\rho_c)=1$ for all conjugacy classes and IPR $\rho_c$. The fact that all conjugacy classes consist of a single element implies that $G$ is Abelian. All linear irreducible representations of an Abelian group are one-dimensional and can be labeled by group elements. In order for the IPRs to be one-dimensional as well, $\beta_g$ has to be trivial in the sense of Eq.\,\eqref{eq:trivial_2cocycle}. In this case, the central idempotents simplify to
\begin{align}
    P^T_{(g,k)} = \frac{1}{\abs{Z(c)}}\sum_{h\in G}\overline{\epsilon_g(h)\chi_k(h)} (g,h)_T \qcomma g,k\in G.
\end{align}
With these $\abs{G}^2$ independent central idempotents in the $\abs{G}^2$-dimensional algebra $T$ we have found all irreducible sub-spaces. Each is labeled by a pair of group elements $(g,k)$ and one-dimensional. The latter shows that the associated anyons are indeed Abelian and is directly related to the fact that $\beta_g$ is a coboundary. If it is a non-trivial 2-cocycle, the resulting anyon theory is non-Abelian~\cite{Hu2013Twisted, propitius1995topological}.

\subsection{Examples}\label{sec:tube_examples}
In this section, we will first give two examples for Abelian models that are representative for any Abelian lattice model. To illustrate the generic procedure of explicitly finding the central idempotents also for non-Abelian models we further discuss twisted versions of a $S_3$ lattice model. For any $\vV^\omega(G)$ model, the tube algebra is diagonalized by the irreducible $\beta$-projective representations of the centralizers of all the conjugacy classes of $G$.

\subsubsection{\texorpdfstring{$\vV^{\omega}(\bZ_p)$}{Vecω(Zp)}}
Any 3-cocycle class of $\bZ_N=\{0,1, \dots, N-1\}$ is represented by a type-I 3-cocycle as in Eq.\,\eqref{eq:3cocycletype1}.
Note that this cocycle is symmetric in the latter two arguments, $\omega_I^n(a,b,c) = \omega_I^n(a,c,b)\;\forall a,b,c$.
Using this and the fact that $\bZ_p$ is an Abelian group, $\beta_a(b,c)$ reduces to $\omega_I^n(a,b,c)$ (see Eq.\,\eqref{eq:betag_general}) and the multiplication in $T$ is given by
\begin{align}
    (a,b)_T\ast(a',b')_T = \delta_{a,a'}\omega_I^n(a,b,b')(a,b\oplus b')_T = \delta_{a,a'}e^{\frac{2\pi i}{p^2}na(b+b'-b\oplus b')}(a,b\oplus b')_T.
\end{align}
This tube algebra is Abelian in the sense described above since for any every $a\in\bZ_p$, we can define the 1-cochain
\begin{align}
    \epsilon_a^n(b) := e^{\frac{2\pi i}{p^2}nab},
\end{align}
such that $\beta_a(b,c) = \omega_I^n(a,b,c) = (\delta\epsilon_a^n)(b,c)$. Together with the group character function $\chi_a(b) = e^{\frac{2\pi i}{p}ab}$, the central idempotents of the tube algebra -- labeled by $(a,k)\in\bZ_p^{\times 2}$ -- are given by
\begin{align}\label{eq:Zp_tube_centralidempotents}
    P^T_{(a,k)} = \frac{1}{p}\sum_{b\in\bZ_p} e^{-\frac{2\pi i}{p^2}( n ab + p kb)}(a,b)_T.
\end{align}

\subsubsection{\texorpdfstring{$\vV^{\omega}(\bZ_p\times\bZ_p)$}{Vecω(ZpxZp)}}
For a type-I cocycle on either of the two factors of $\bZ_p\times\bZ_p$, the central idempotents will have the same form as the ones in Eq.\,\eqref{eq:Zp_tube_centralidempotents}. In this example, we consider a type-II cocycle defined in Eq.\,\eqref{eq:3cocycletype2}. The multiplication in $T$ is given by
\begin{subequations}
\begin{align}
    \begin{split}
        ((a_1,a_2),(b_1,b_2))_T\ast((a'_1,a'_2),(b'_1,b'_2))_T =& \delta_{a,a'}\omega_{II}^{n_{12}}((a_1,a_2),(b_1,b_2),(b'_1,b'_2))\\
    &\qq\ ((a_1,a_2),(b_1\oplus b'_1,b_2\oplus b'_2))_T
    \end{split}\\
    \begin{split}
        =& \delta_{a,a'}e^{\frac{2\pi i}{p}a_1(b_2+b'_2-b_2\oplus b'_2)/q}\\
        &\qq\ ((a_1,a_2),(b_1\oplus b'_1,b_2\oplus b'_2))_T,
    \end{split}
\end{align}
\end{subequations}
where we abbreaviated $\delta_{a_1,a'_1}\delta_{a_2,a'_2}$ with $\delta_{a,a'}$.
Again, we can find a 1-cochain
\begin{align}
    \epsilon^{n_{12}}_a(b) := e^{\frac{2\pi i}{p^2}n_{12}a_1b_2}
\end{align}
with the coboundary $(\delta\epsilon^{n_{12}}_a)(b,c)=\omega_{II}^{n_{12}}(a,b,c)$. Together with the character function $\chi_a(b)=e^{\frac{2\pi i}{p} (a_1b_1 + a_2b_2)}$ the central idempotents -- labeled by $(a,k)\in(\bZ_p\times\bZ_p)^{\times 2}$ -- are given by
\begin{align}
    P^T_{(a,k)} = \frac{1}{p^2}\sum_{b_1,b_2\in\bZ_p} e^{-\frac{2\pi i}{p^2}(n_{12}a_1b_2 + p k_1b_1 + p k_2b_2)} ((a_1,a_2),(b_1,b_2))_T.
\end{align}

\subsubsection{\texorpdfstring{$\vV^{\omega}(S_3)$}{Vecω(S3)}}\label{sec:tube_examples_s3}
In this section, we will first consider the tube algebra of an untwisted $\vV(S_3)$ model in detail and then sketch how the twisting by a 3-cocycle affects the tube algebra. Note that the untwisted model has been studied in various contexts in the past~\cite{laubscher2019universal, Bridgeman2020computingdatalevin, Cong2017Defects}.

$S_3$ has two independent generators, $r$ and $t$. They satisfy $t^2=r^3=e$, where $e$ is the identity element, and $tr=r^2t=r^{-1}t$. The subgroup $\bZ_3$ generated by $r$ is normal. $S_3$ has three conjugacy classes
\begin{align}
    \overline{e}=& \{e\}\qcomma
    \overline{r}= \{r, r^2\}\qq{and}
    \overline{t}= \{t, tr, tr^2\}.
\end{align}
with the respective centralizers
\begin{align}
    Z(\overline{e}) \simeq& Z(e) = S_3\qcomma
    Z(\overline{r}) \simeq Z(r) = \langle r\rangle \simeq \bZ_3\qq{and}
    Z(\overline{t}) \simeq Z(t) = \langle t\rangle \simeq \bZ_2.
\end{align}
To obtain the central idempotents of the tube algebra, we need to find the irreducible representations of the centralizers. Let us first consider the trivial conjugacy class $A$. Its centralizer is all of $S_3$ and has three irreducible representations, a trivial one $\Gamma^0$, a one-dimensional one $\Gamma^1$ and a two-dimensional one $\Gamma^2$. The characters are given by
\begin{center}
    \begin{tabular}{c|c c c}
       $S_3$ & $\overline{e}$ & $\overline{r}$ & $\overline{t}$ \\ \hline
    $\Gamma^0$ & 1 & 1 & 1\\
         $\Gamma^1$ & 1 & 1 & -1\\
         $\Gamma^2$ & 2 & -1 & 0
    \end{tabular}
\end{center}
We obtain three anyons associated to the trivial conjugacy class. The corresponding idempotents read
\begin{subequations}
\begin{align}
    P_{(\overline{e},\Gamma^0)}^T =& \frac{1}{6}\left( (e,e)_T + (e,r)_T + (e,r^2)_T + (e,t)_T + (e,tr)_T + (e,tr^2)_T\right),\\
    P_{(\overline{e},\Gamma^1)}^T =& \frac{1}{6} \left( (e,e)_T + (e,r)_T + (e,r^2)_T - (e,t)_T - (e,tr)_T - (e,tr^2)_T\right),\\
    P_{(\overline{e},\Gamma^2)}^T =& \frac{1}{3} \left( 2(e,e)_T - (e,r)_T - (e,r^2)_T \right).
\end{align}
\end{subequations}
Note that $(\overline{e},\Gamma^0)$ is the trivial anyon -- or vacuum -- subspace. The dimensions of the respective sub-spaces coincide with the dimensions of the irreducible representations $\Gamma^0$, $\Gamma^1$ and $\Gamma^2$ respectively, i.e., $d_{(\overline{e},\Gamma^0)} = d_{(\overline{e},\Gamma^1)} = 1$ and $d_{(\overline{e},\Gamma^2)} = \dim(\Gamma^2) =  2$.

Now consider the two non-trivial conjugacy classes. Their centralizers are cyclic groups, so their irreducible representations are one-dimensional and labeled by group elements. The characters read
\begin{center}
    \begin{tabular}{c|c c c}
         $\bZ_3$ & $\{0\}$ & $\{1\}$ & $\{2\}$ \\ \hline
         0 & $1$ & $1$ & $1$\\
         1 & $1$ & $e^{2\pi i/3}$ & $e^{-2\pi i/3}$\\
         2 & $1$ & $e^{-2\pi i/3}$ & $e^{2\pi i/3}$
    \end{tabular} \hspace{2cm} %
    \begin{tabular}{c|c c }
         $\bZ_2$ & $\{0\}$ & $\{1\}$  \\ \hline
         0 & 1 & 1\\
         1 & 1 & -1
    \end{tabular}.
\end{center}
Evaluating Eq.\,\eqref{eq:PT_trivialbeta} for the two conjugacy classes and the character functions shown above gives the projectors
\begin{subequations}
\begin{align}
    P^T_{(\overline{r},0)} =& \frac{1}{6}\left((r,e)_T + (r,r)_T + (r,r^2)_T +(r^2,e)_T + (r,r)_T + (r,r^2)_T\right)\;,\\
    P^T_{(\overline{r},1)} =& \frac{1}{6}\left((r,e)_T + (r^2,e)_T + e^{2\pi i/3}((r,r)_T + (r^2,r)_T) + e^{-2\pi i/3}((r,r^2)_T + (r^2,r^2)_T)\right)\;,\\
    P^T_{(\overline{r},2)} =& \frac{1}{6}\left((r,e)_T + (r^2,e)_T + e^{-2\pi i/3}((r,r)_T + (r^2,r)_T) + e^{2\pi i/3}((r,r^2)_T + (r^2,r^2)_T)\right)\;,\\[12pt]
    P^T_{(\overline{t},0)} =& \frac{1}{6}\left( (t,e)_T + (t,t)_T + (tr,e)_T + (tr,tr)_T + (tr^2,e)_T + (tr^2,tr^2)_T\right)\;,\\
    P^T_{(\overline{t},1)} =& \frac{1}{6}\left( (t,e)_T + (tr,e)_T + (tr^2,e)_T - (t,t)_T  - (tr,tr)_T - (tr^2,tr^2)_T\right)\;.
\end{align}
\end{subequations}
The dimensions of the associated irreducible representations are
\begin{align}
    d_{(\overline{r},0)} = d_{(\overline{r},1)} = d_{(\overline{r},2)} = \abs{\overline{r}} = 2 \qq{and} d_{(\overline{t},0)} = d_{(\overline{t},1)} = \abs{\overline{t}} = 3\;.
\end{align}
So in total we have found $8$ different anyons and the corresponding central idempotents of the tube algebra.

Next, let us consider a non-trivial 3-cocycle from Eq.\,\eqref{eq:3cocycle_S3}. This yields the phase (see Eq.\,\eqref{eq:tube_multiplication}),
\begin{align}
    \beta_{t^A r^a}^p(t^Br^b,t^{B'}r^{b'}) = \frac{\omega^p(t^A r^{(-1)^{B+B'}[a+((-1)^{B'}b+b')((-1)^A)-1]},t^Br^b,t^{B'}r^{b'})\omega^p(t^Br^b,t^{B'}r^{b'},t^{A}r^{a})}{\omega^p(t^Br^b,t^{A}r^{(-1)^{B'}(a+b'((-1)^A-1)},t^{B'}r^{b'})},
\end{align}
in the multiplication in $T$, which turns out to be a (twisted) coboundary~\cite{propitius1995topological}
\begin{align}
    \beta_{t^A r^a}^p(t^Br^b,t^{B'}r^{b'}) = (\Tilde{\delta}\epsilon_{t^A r^a}^p)(t^Br^b,t^{B'}r^{b'}) \qq{with} \epsilon^p_{t^A r^a}(t^Br^b) = e^{\frac{2\pi i}{9}p(b((-1)^Ba\oplus_3 2Ab)-Ab^2)} i^{pAB}.
\end{align}
The cochain $\epsilon_{t^Ar^a}$ defines the projective characters in terms of the linear ones, see Eq.\,\eqref{eq:projectiveChar_trivial}. With that, the central idempotents of the $\omega^p$-twisted $S_3$ theory are given by Eq.\,\eqref{eq:PT_trivialbeta}. The number and dimensions of the irreducible sub-spaces are the same as in the untwisted case but the phases in the central idempotents are modified by $\epsilon_g^p$. This in turn changes their fusion and braiding (see Refs.\,\cite{Hu2013Twisted, propitius1995topological, coste2000finite}).

\section{Boundary anyons: Semi-tube algebra}\label{sec:RectangleAlgebra}
After having modeled point-defects (anyons) in the bulk, we continue with point-defects at the boundary. Even though these defects are not equipped with a braiding structure (they can only move along a boundary), we call them boundary anyons.
However, they form a fusion category. A framework to calculate the fusion multiplicities thereof was introduced in Refs.\,\cite{Bridgeman2020computingdatalevin, bridgeman2019fusing} as ``vertical fusion". In fact, the boundary anyons have to form a Morita equivalent category to the one that defines the bulk strings~\cite{Cong2017Defects}.

\subsection{General fixed-point models}

\begin{figure}
    \centering
    \includegraphics[width=0.5\textwidth]{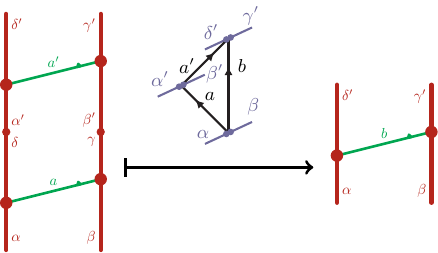}
    \caption{The multiplication of two basis elements in $S$ is defined via gluing the two associated semi-tube string diagrams together (left) and using $L$ (and $F$) moves to reduce it to the cellulation on the right. The associated phase can be derived by considering the space-time complex that maps from the initial to the final cellulation.
    }
    \label{fig:semitube_basis_multiplication}
\end{figure}

Boundary anyons are determined by string-nets on a ``semi-tube", a tube that is cut in half by the physical boundary. One might also think of one half being replaced by vacuum.
Note that for a domain wall between two phases, modeled by fusion categories $\cC$ and $\cC'$, the other half would not be vacuum but another bulk for the fusion category $\cC'$.
However, as illustrated in Section~\ref{sec:folding}, these domain wall diagrams can be related to an equivalent boundary diagram of the form considered in this section.
This semi-tube can be tessellated by a string diagram as follows,
\begin{align}\label{eq:semitube_basis}
    \raisebox{-0.5\height}{\includegraphics[width=0.2\textwidth]{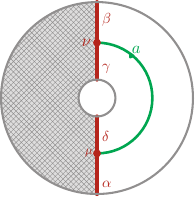}} =: (\alpha, \beta, \gamma,\delta, a; \mu,\nu)_S,
\end{align}
which defines an orthonormal basis of the defect space $S$.
In the following, we will omit the multiplicity labels at the vertices.
Again, we can use gluing operations and topological moves to define a multiplication on $S$.
In Eq.\,\ref{fig:semitube_basis_multiplication} we show the sequence of local moves that defines the multiplication in terms of its associated space-time complex.
Evaluating the complex gives rise to the multiplication
\begin{align}\label{eq:semi-tube_multiplication_general}
    (\alpha',\beta',\gamma',\delta',a')_S \ast (\alpha,\beta,\gamma,\delta,a)_S = \delta_{\alpha',\delta}\delta_{\beta',\gamma} \sum_{b} L_{\beta a a'}^{\gamma' \beta' b} \overline{L_{\alpha a a'}^{\delta' \alpha' b}} (\alpha,\beta, \delta', \gamma', b)_S.
\end{align}
By analogy to the tube algebra, we call the resulting algebra
\textit{semi-tube algebra}.
Similar structures have been already introduced in various places in the literature as ``module tube algebra", ``module annular algebra" or ``ladder category"~\cite{kitaev2012models, bridgeman2022invertible, barter2021computingassociators, Bridgeman2020computingdatalevin}.

\subsection{Topological gauge theory models}\label{sec:semitube_gaugetheory}
In the case where the bulk is defined via $\vV^\omega(G)$, the boundary model is given by a subgroup $H\subseteq G$ (and a 2-cocycle in $Z^2(H,U(1))$), the boundary labels are cosets in $\flatfrac{G}{H}$ and every vertex is multiplicity-free, see Section~\ref{sec:gauge_models}.
Hence, a semi-tube basis element is parameterized by $(\alpha,\beta,g)_S\in \flatfrac{G}{H}\times\flatfrac{G}{H}\times G$.
The associated basis state from Eq.\,\eqref{eq:semitube_basis} are given by $(\alpha,\beta,g)_S := (\alpha, \beta, g\triangleright \beta,g\triangleright\alpha, g)_S$.
Plugging in the defining data form Section~\ref{sec:gauge_models} into Eq.\,\eqref{eq:semi-tube_multiplication_general}, we get the multiplication
\begin{align}\label{eq:rectangle_multiplication}
    (\alpha',\beta',g')_S \ast (\alpha,\beta,g)_S = \delta_{\alpha',g\triangleright\alpha}\delta_{\beta',g\triangleright\beta} \overline{\psi^\alpha(g',g)} \psi^\beta(g',g)(\alpha,\beta,g'g)_S
\end{align}
over the vector space $\bC^{G/H\times G/H\times G}$.

The anyons on the boundary correspond to the irreducible sub-spaces of $S$ and we again want to find the associated indecomposable central idempotents in $S$.

Again, $S$ is an algebra of the form discussed in Section~\ref{sec:twistedwithaction} where $X = \flatfrac{G}{H}\times \flatfrac{G}{H}$ with $G$ acting via simultaneous left-translation: $g\triangleright (aH,bH) = (gaH,gbH)$.
Using the (twisted) 2-cocycle condition of $\psi$ we can easily see that
\begin{equation}
\Psi^{\alpha,\beta}(g',g):=\overline{\psi^\alpha(g',g)}\psi^\beta(g',g)
\end{equation}
is a twisted 2-cocycle fulfilling
\begin{align}
\Psi^{\alpha,\beta}(g,h)\Psi^{\alpha,\beta}(g,hk) = \Psi^{\alpha,\beta}(gh,k)\Psi^{k\triangleright\alpha,k\triangleright\beta}(g,h)\quad \forall g,h,k\in G;\; \alpha,\beta\in \flatfrac{G}{H},
\end{align}
in analogy to Eq.\,\eqref{eq:twisted2cocyclegeneral}.
Following Section~\ref{sec:twistedwithaction}, we first have to find the $G$-orbits of $\flatfrac{G}{H}^{\times2}$.
Let $\alpha^{-1}\in H\backslash G$ be the left inverse of $\alpha$, i.e., $\alpha^{-1}\alpha = H$.
Then we find that the double coset $x:=\alpha^{-1}\beta\in H\backslash\flatfrac{G}{H}$, is invariant under the $G$ action.
Moreover, all the subsets $S_x := \{(\alpha,\beta)\in \flatfrac{G}{H}\;|\; \alpha^{-1}\beta = x\}$ are transitive as we will show now.

Consider two pairs of cosets $(\alpha_1,\beta_1),(\alpha_2,\beta_2)\in S_x$ represented by $(a_1,b_1),(a_2,b_2)\in G$.
Since both pairs define the same double coset, there exist $h,h'\in H$ such that $h^{-1}a_1^{-1}b_1 h' = a_2^{-1}b_2$.
Hence, there exists a group element, namely $g=b_2h'^{-1}b_1^{-1} = a_2h^{-1}a_1^{-1}$, such that $g\triangleright\alpha_1 = \alpha_2$ and $g\triangleright\beta_1=\beta_2$.
So any two coset pairs $(\alpha_1,\beta_1),(\alpha_2,\beta_2)\in S_x$ are related via the action of some $g\in G$.

Next, we find the stabilizer subgroup for any element in $S_x$.
In particular, we can pick a representative $\hat{x} = (\hat{\alpha},\hat{\beta})$, explicitly calculate its stabilizer subgroup and then use Eq.\,\eqref{eq:stabilizerrelation} to derive the stabilizer groups of the other elements in $S_x$.
Let $(\hat{\alpha},\hat{\beta}) = (aH,bH)\in S_x$. By construction, $(H,a^{-1}bH)$ will also be in $S_x$. The stabilizer of this element,
\begin{align}
    \Stab_G((H,a^{-1}bH)) = \{g\in G\;|\;(gH,ga^{-1}bH) = (H,a^{-1}bH)\},
\end{align}
is particularly easy to compute and given by
\begin{align}\label{eq:Kx_intersection}
    K_x := H\cap a^{-1}bHb^{-1}a.
\end{align}
This formula holds for any element in $S_x$, replacing  $a^{-1}b$ with the respective double-coset representative of $x$.
From these general considerations above, we can see that $K_x$ simplifies in some special cases:
\begin{itemize}
    \item For the trivial double coset $x=H$, $K_x=H$.
    \item For $H$ normal in $G$, $K_x=H\;\forall x$.
\end{itemize}

This result already appeared in Ref.\,\cite{cong2016topological} in the derivation of boundary defects in untwisted Quantum Double Models. In the twisted case, boundary anyons are labeled by irreducible $\Psi^{\alpha,\beta}$-projective representations of $K_x$ instead of linear ones. In the following, if not stated otherwise, we label these representations with $\kappa_x$.

Combined, the irreducible sub-spaces of $S$ are labeled by a doble-coset $x\in H\backslash\flatfrac{G}{H}$ and an irreducible $\Psi^{\hat{\alpha},\hat{\beta}}$-projective representation of $K_x$.
The associated central idempotents are given by
\begin{align}\label{eq:PS_general}
    P^S_{(x,\kappa_x)} = \frac{\dim(\kappa_x)}{\abs{K_x}}\sum_{\stackrel{\alpha, \beta\in\flatfrac{G}{H}}{\alpha^{-1}\beta=x}}\sum_{g\in \Stab_G((\alpha,\beta))} \overline{\Tilde{\chi}_{\kappa_x}^{(\alpha,\beta)}(g)} (\alpha,\beta, g)_S,
\end{align}
where $\Tilde{\chi}_{\kappa_x}^{(\alpha,\beta)}(g):= \Tr(\kappa_x^{(\alpha,\beta)}(g))$ is the projective character of the IPR $\kappa_x$ of $\Stab_G((\alpha,\beta))$.

\subsection{Semi-tube algebra with Abelian anyons}\label{sec:AbelianBoundaries}
In this section, we will consider models for which both the fusion of the bulk and the boundary anyons is Abelian.
As we have seen, the bulk is then defined by a finite Abelian group $G$ with $N$ cyclic factors and a 3-cocycle cohomologous to a product of type-I and type-II cocycles.
The boundary is defined by a subgroup $H$ on which the bulk 3-cocycle is cohomologically trivial.
Since $G$ is an Abelian group any subgroup $H$ is normal and there is a one-to-one correspondence of double cosets and cosets.
In particular, $\alpha^{-1}\beta = x$ implies that we can rewrite $\beta = \alpha x$ and the sum over the two coset labels in Eq.\,\eqref{eq:PS_general} reduces to one.
Moreover, $K_x=H$ for all $x$ such that we can omit the $x$-label from the irrep label $\kappa_x$.

The fusion of the boundary anyons is described by how the irreps of the semi-tube algebra multiply, respectively \textit{fuse}, to a direct sum of other irreps.
If every fusion product only consist of a single irrep, we say the boundary anyons are Abelian.
As discussed in Sec.\,\ref{sec:tube_gaugetheory} for the tube algebra there is a direct correspondence to the dimensions of every irrep being one-dimensional, meaning that the multiplcation in the tube algebra is commutative.%
In general, however, this correspondence does not hold.
One can have a non-commutative algebra, captured by the existence of higher-dimensional irreps, that describe Abelian anyons.
At the same time, every irrep being one-dimensional gives a sufficient condition on the associated anyons being Abelian.
For models with Abelian bulk anyons (Abelian $G$ and no type-III 3-cocycles) every irrep is one-dimensional iff $\Psi^{\alpha,\alpha x}$ is cohomologous to a trivial 2-cocycle.
In this case, we can express the irreducible projective characters in terms of linear ones,
\begin{align}
    \Tilde{\chi}_{\kappa}^{(\alpha,\alpha x)}(g) = \eta^{(\alpha)}_x(g)\chi_\kappa(g).
\end{align}
Here, $\eta^{(\alpha)}_x: G\to U(1)$ is the cochain that trivializes the 2-cocycle $\Psi^{\alpha,\alpha x}$.
Linear characters $\chi_\kappa$ of (finite) Abelian groups are well known, see Section~\ref{sec:tube_examples}.
Combining the observations above, the central idempotents read
\begin{align}
    P^S_{(x,\kappa)} = \frac{1}{\abs{H}}\sum_{\alpha\in\flatfrac{G}{H}}\sum_{g\in H}\overline{\eta^{(\alpha)}_x(g)\chi_{\kappa}(g)}(\alpha,\alpha x, g)_S\qcomma \begin{array}{l}
    x\in\flatfrac{G}{H},\\
    \kappa\in\text{Irreps}(H).
    \end{array}
\end{align}
We believe that any boundary models to a bulk that leads to Abelian anyons only result in an Abelian fusion of the boundary anyons, also when the semi-tube algebra has higher-dimensional irreps.

Note the different notions of ``Abelian''.
In particular, the semi-tube algebra might not be Abelian (commutative) as an algebra but its irreducible representations still be described by an Abelian fusion category.
For a recipe to calculate that fusion data explicitly, we refer to Refs.\,\cite{bridgeman2019fusing,Bridgeman2020computingdatalevin}.

\subsection{Examples}\label{sec:boundaryExamples}
In this subsection we will sketch the construction for some exemplary cases.
First, we note how the form reduces in the two extreme cases where the subgroup is one of the two trivial subgroups.
Then, we will focus on small Abelian groups and as a simple example for an non-Abelian model we will give a full description of how to obtain the central idempotents in the case of $G=S_3$.
In particular, we will see how Eq.\,\eqref{eq:Kx_intersection} helps significantly in constructing the stabilizer group whose irreducible representations labels parts of the boundary anyons.

\subsubsection{Trivial Subgroups}
\label{sec:trivial_subgroups}
Before we consider specific groups (and their subgroups), let us take a closer look on how Eq.\,\eqref{eq:PS_general} simplifies in the case of the two trivial subgroups $\{1_G\}$ and $G$.
In the former case, we denote the resulting semi-tube algebra $S$.
The central idempotents are labeled by $x\in \{1_G\}\backslash\flatfrac{G}{\{1_G\}}\simeq G$ alone.
All the prefactors in Eq.\,\eqref{eq:PS_general} evaluate to 1 resulting in
\begin{align}\label{eq:PR_trivialsubgroup1}
    P^{S}_{(x)} = \sum_{\alpha\in G}(\alpha, \alpha x, 1_G)_S\qcomma\text{with }x\in G.
\end{align}

In the latter case, where $H=G$, we denote the resulting semi-tube algebra by $S$.
In this case the central idempotents are labeled by irreducible (linear) representations of $G$ since there is only one double coset, namely the trivial one.
Moreover, since $H=G$ only defines a boundary in the case of a trivial 3-cocycle in the bulk, $S$ is isomorphic to an untwisted group algebra.
With that, only the $G$-character functions $\chi_\kappa$ enter into the prefactor in Eq.\,\eqref{eq:PS_general} and the central idempotents take the form
\begin{align}\label{eq:PR_trivialsubgroup2}
    P^{S}_{(\kappa)} = \frac{\dim(\kappa)}{\abs{G}} \sum_{g\in G} \overline{\chi_\kappa(g)} (\overline{1_G},\overline{1_G},g)_S\qcomma\text{with }\kappa\in\text{Irreps}(G).
\end{align}
Note that with $H=\{1_G\}$, the fusion category of boundary anyons is the same as the input fusion category $\vV(G)$, whereas for $H=G$, they form the Morita equivalent category $\Rep(G)$.

\subsubsection{\texorpdfstring{$\vV^{\omega}(\bZ_p)$}{Vecω(Zp)}}
Consider $\bZ_p=\{0,1, \dots, p-1\}$ for $p$ prime with the type-I 3-cocycle $\omega_I^n$ in Eq.\,\eqref{eq:3cocycletype1}. $\bZ_p$ has only the trivial and full subgroup which were already discussed in the previous section.
For the trivial subgroup $H_0=\{1_G\}$, the $p$ central idempotents are given as in Eq.~\eqref{eq:PR_trivialsubgroup1} by
\begin{align}
    P^S_{(x)} = \sum_{\alpha=0}^{p-1} (\alpha, \alpha\oplus_p x,0)_S\qcomma\text{with }x\in\{0,1, \dots, p-1\},
\end{align}
where $\oplus_p$ denotes addition modulo $p$. The full subgroup only defines a boundary for the trivial 3-cocycle $n=0$. The central idempotents are then given as in Eq.\,\eqref{eq:PR_trivialsubgroup2}.

\subsubsection{\texorpdfstring{$\vV^{\omega}(\bZ_p\times\bZ_p)$}{Vecω(ZpxZq)}}
Consider the type-II cocycle $\omega_{II}^{n_{12}}$ on $\bZ_p\times\bZ_p$ from Eq.\,\eqref{eq:3cocycletype2}.
The cases of the boundary given by a trivial and full subgroups are discussed in Section~\ref{sec:trivial_subgroups}. Note that the latter only defines a boundary for $n_{12}=0$.
The three types of $\bZ_p$ subgroups $H_1$, $H_2$, $H_3\simeq\bZ_p$ from Section~\ref{sec:example_model_s3} define valid boundaries for any $n_{12}$.

$\bZ_p$ only has trivial 2-cocycles so it defines a unique boundary.
Let us consider the particularly simple case of $H_2=\langle (0,1)\rangle\simeq\bZ_p\subset\bZ_p\times\bZ_p$, on which $\omega_{II}$ directly evaluates to $1$.
In the microscopic model, the corresponding boundary labels are double cosets $H_2\backslash\flatfrac{\bZ_p\times\bZ_p}{H_2}$, which can be identified with the integers $\{0,1, \dots, p-1\}$.
With that, the $p^2$ central idempotents are labeled by $(x, \kappa)\in\bZ_p\times\bZ_p$ and are given by
\begin{align}
    P^S_{(x,\kappa)} = \frac{1}{p}\sum_{\alpha=0}^{p-1}\sum_{g=0}^{p-1} e^{-\frac{2\pi i}{p} g_2\kappa} (\alpha,\alpha\oplus_p x, (0,g))_S.
\end{align}

\subsubsection{\texorpdfstring{$\vV^\omega(S_3)$}{Vecω(S3)}}
As an examplary case for non-Abelian models, consider $G=S_3$.
It has four conjugacy classes of subgroups, see Section~\ref{sec:example_model_s3}.
The trivial subgroup $H_0=\{e\}$ defines a boundary for any choice of bulk 3-cocycle.
In this case, the central idempotents of $S$ the take the form of Eq.\,\eqref{eq:PR_trivialsubgroup1}.

In the case of a trivial  bulk 3-cocycle, any other subgroup defines a boundary as well.
Consider the non-trivial normal subgroup $H_r=\langle r\rangle\simeq\bZ_3$.
It defines a boundary of a bulk model twisted by $\omega^p$ for $p=0,3$.
Since $H_r$ is normal in $G$ the double cosets $H_2\backslash\flatfrac{G}{H_2}$ are in one-to-one correspondence with cosets $\flatfrac{G}{H_2}\simeq\bZ_2$.
The central idempotents are labeled by $x\in\flatfrac{G}{H_r}\simeq\bZ_2$ and an irreducible representation of $H_r$, $\kappa\in\bZ_3$.
Moreover, $H_r$ only has trivial 2-cocycles, so we use linear characters as in Eq.\,\eqref{eq:PS_general}.
Taken together, the central idempotents for an $H_r$-boundary read
\begin{align}\label{eq:S3Pr2}
    P^{S_r}_{(x,\kappa)} = \frac{1}{3}\sum_{\alpha=0}^1\sum_{g=0}^2e^{-\frac{2\pi i}{3}\kappa g} (\alpha,\alpha \oplus_2 x, r^g)_S\qcomma\text{with }\kappa\in\{0,1,2\},x\in\{0,1\},
\end{align}
for both $p=0$ and $3$.
This shows that there are $6$ inequivalent boundary anyons. In fact, when considering their fusion on the boundary, they form the fusion category $\vV(S_3)$ which is the same as the input category for the bulk model and with that -- as expected -- Morita equivalent to it, compare Ref.\,\cite{Cong2017Defects}.

Let us now consider the more subtle case of the non-normal subgroup $H_t\simeq\bZ_2$.
It defines a boundary of a bulk model defined by the 3-cocycle $\omega^p$ for $p=0,2,4$ (see Eq.\,\eqref{eq:3cocycle_S3}).
There are two double cosets $x\in H_t\backslash\flatfrac{G}{H_t} = \{\{e,t\}=H_t, \{r,r^2,tr,tr^2\} =  H_t rH_t\}$.
We obtain the stabilizer subgroup of both double cosets with Eq.\,\eqref{eq:Kx_intersection}.
For the trivial double coset it is given by $H_t$ itself whose irreducible representations are labeled by $\bZ_2 = \{0,1\}$.
For the non-trivial double coset $x=H_trH_t$ we pick the representative $r\in S_3$.
Noting that $rH_tr^{-1} = \langle tr \rangle$ only shares the identity element with $H_t$, the stabilizer group for this double coset is the trivial subgroup $\{e\}$.
Hence, we find only one boundary anyon for $x=H_trH_t$.

Combined, the $S_t$ boundary can host three different anyons, associated to the indecomposable central idempotents
\begin{subequations}
    \begin{align}
P^{S_t}_{(H_t,\kappa)} =& \frac{1}{2}\sum_{\alpha\in\flatfrac{S_3}{H_t}}\sum_{g=0}^1 (-1)^{\kappa g} (\alpha,\alpha, t^g)_S\qcomma \kappa\in\{0,1\} \qq{and}\\
    P^{S_t}_{(H_trH_t)} =& \sum_{\alpha\neq\beta} (\alpha,\beta,e)_S.
\end{align}
\end{subequations}
In contrast to the $H_r$-boundary the $H_t$-boundary has only 3 anyons.
Including fusion along the boundary, they form the fusion category $\Rep(S_3)$ which is indeed Morita-equivalent to $\vV(S_3)$ (compare Ref.\,\cite{Cong2017Defects}).

\section{Bulk-to-boundary fusion events}\label{sec:bulkbdryfusion}
After defining our model in the bulk of a spacial 2-manifold and on its boundary, we modeled point-like defects, namely anyons, both in the bulk and on the boundary.
In this section, we will look at fusion events between bulk and boundary anyons, which in space-time can be interpreted as point defects on the boundary where a boundary anyon world-line meets a (bulk) anyon world-line.
In space such a fusion event can be interpreted as the process of moving a bulk anyon to the boundary and turning it into a boundary anyon.

\subsection{Bimodule construction in general fixed-point models}
\begin{figure}
    \centering
    \includegraphics[width=0.9\textwidth]{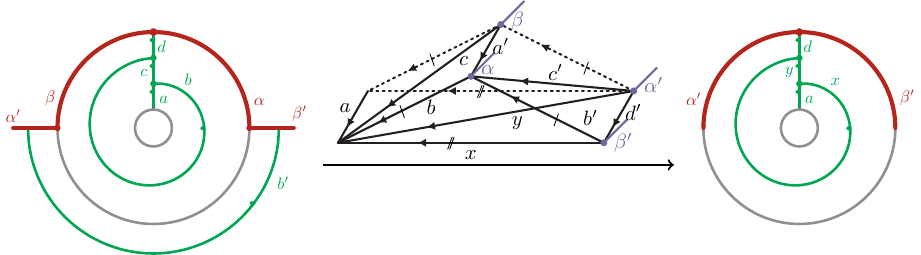}
    \caption{The $S$-action on $C$ is defined by gluing a semi-tube element onto a $C$ element from the top and using local moves to reduce the cellulation back to the reference one \eqref{eq:bimoodule_basis_general}. These moves can be represented by the space-time complex shown above. It decomposes into three bulk tetrahedra and two boundary triangles.}
    \label{fig:bimodule_Raction}
\end{figure}

As we have seen in Section~\ref{sec:tube} and Section~\ref{sec:RectangleAlgebra}, the (boundary) anyons can be computed by considering string-net diagrams on a (semi-)tube. Stacking a (semi-)tube to another one defines an action of the vector space of (semi-)tubes on that same vector space, which can be interpreted as an algebra, namely the (semi-)tube algebra. Since a bulk-to-boundary fusion event maps a bulk to a boundary anyon, it should be related to a 1-manifold which connects between tube segments and semi-tube segments. This can be achieved by an annulus where one boundary circle connects to tubes, and the other boundary circle half consists of the physical boundary and half connects to semi-tubes,
\begin{align}\label{eq:bimoodule_basis_general}
    \raisebox{-0.5\height}{\includegraphics[width=0.25\textwidth]{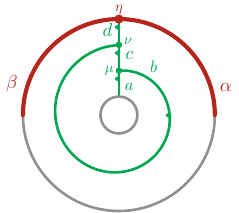}
} =: (a,b,c,d,\alpha, \beta;\mu,\nu,\eta)_C.
\end{align}
A tube can be attached to the small gray circle in the center, and a semi-tube to the large gray half circle at the bottom.
As shown, we choose a minimal string diagram to cover the 2-manifold.
This 2-manifold defines a bimodule $C$ with an action of the tube algebra $T$ and an action of the semi-tube algebra $S$, such that the representations commute with each other.
The vector space of $C$ is spanned by labeled string diagrams of the above form.
The $T$-action is defined by gluing a tube into the inner hole and using $F$ moves to rearrange the string diagrams to the above form. Similarly, the $S$-action is defined by gluing a semi-tube segment to the exterior gray half circle and using $F$ and $L$ moves.
The general actions are given by
\begin{subequations}
   \begin{align}
    (a',b',c',d')_T \triangleright (a,b,c,d,\alpha,\beta)_C =& \delta_{d',a}\sum_{x,y} F_{ya'c'}^{b'bx} F_{yb'x}^{bdc} \overline{F_{yb'c'}^{abc}} (a',x,y,d,\alpha,\beta)_C\\
    (a,b,c,d,\alpha,\beta)_C\triangleleft (\alpha',\beta',\gamma',\delta',a')_S =& \delta_{\gamma',\alpha}\delta_{\delta',\beta} \sum_{x,y,z, e} L_{\beta'a'z}^{\beta\alpha' e} \overline{L_{\beta'da'}^{\beta\alpha e}} F_{yac}^{ba'x} F_{ybx}^{a'ze} \overline{F_{ybc}^{da'e}} (a,x,y,z,\beta',\alpha')_C,
\end{align}
\end{subequations}
where we again omitted the multiplicity labels. In Fig.\,\ref{fig:bimodule_Raction} we show the $S$-action as a space-time complex. The $T$-action is essentially given by the complex in Fig.\,\ref{fig:tube_mult} with two additional boundary labels at two of the vertices.

As a bimodule, $C$ is a equivalent to a representation of $T\otimes S$, and thus decomposes into pairs $(i,j)$ of irreducible representations $i$ of $T$ and $j$ of $S$ with multiplicities $m_{ij}$,
\begin{align}
    C \simeq \bigoplus_{i,j} m_{ij} T_i\otimes S_j\qcomma m_{ij}\in\bZ_{\geq 0},
\end{align}
where $T_i$ and $S_j$ are the irreducible representations of the tube, respectively rectangle, algebra.
Physically, $m_{ij}$ the dimension of the vector space of the bulk-boundary fusion event ``$i$ fuses to $j$".
Acting with the associated central idempotents of these irreducible representations from the left and right constructs a projector onto the respective sub-spaces of dimension $m_{ij}$ which we will denote with $P^C_{ij}$.
The multiplicities are obtained by taking the trace of this projector,
\begin{equation}
\label{eq:general_multiplicities}
m_{ij} = \frac{1}{\dim(i)\dim(j)} \Tr(P^C_{ij})\;,\text{with} \quad P^C_{ij}(\bullet) \coloneqq P^T_i\triangleright \bullet \triangleleft P^S_j\;.
\end{equation}

\subsection{Bulk-boundary fusion in gauge theory models}
In gauge theory models the bulk strings are given by $\vV^\omega(G)$ and the boundary strings by a $\vV^\omega(G)$-module category, classified by a subgroup $H$ and a 2-cocyle class in $H^2(H,U(1))$. The basis of $C$ can by fully parameterized by two group and one coset element. We choose the following assignment:
\begin{align}\label{eq:biomdule_basis}
    (g,h,\alpha)_C := (hgh^{-1}, h, hg, g,\alpha, g\triangleright \alpha)
\end{align}
with $g,h\in G$ and $\alpha\in\flatfrac{G}{H}$.
The $T$ action reduces to
\begin{align}
    (g',h')_T\triangleright (g,h,\alpha)_C =& \delta_{g',hgh^{-1}} \beta_{g}(h',h) (g, h'h,\alpha)_C.
\end{align}
Similarly, the $S$-action is given by
\begin{align}
    (g,h,\alpha)_C\triangleleft (\alpha',\beta',h')_S =& \delta_{h'\triangleright \beta',\alpha}\delta_{g\triangleright \alpha, h'\triangleright\alpha'} \psi^{\beta'}(h',h'^{-1}gh') \overline{\psi^{\beta'}(g,h')} \beta_g(h,h') (g,h h',\beta')_C.
\end{align}

Given these two actions we can use the central idempotents derived in Sections ~\ref{sec:tube_gaugetheory} and~\ref{sec:semitube_gaugetheory} respectively their regular representation to project onto the associated irreducible representations of $T$ and $R$ within $C$.
In combination, Eq.\,\eqref{eq:general_multiplicities} defines a projector $P^C_{(c,\rho_c),(x,\kappa_x)}\in \End(C)$, projecting onto the pair of Irreps $(c,\rho_c)$ and $(x,\kappa_x)$ in $C$.
With the $T$ and $S$ action above, the consecutive action of $P^T_{(c,\rho_c)}$ and $P^S_{(x,\kappa_x)}$ onto a basis element $(g,h,\alpha)_C$ is given by
\begin{subequations}
\begin{align}
\begin{split}
    P^T_{(c,\rho_c)}\triangleright (g,h,\alpha)_C =& \frac{\dim(\rho_c)}{\abs{Z(c)}}\sum_{g'\in c}\sum_{h'\in Z(g')} \overline{\Tilde{\chi}^{g'}_{\rho_c}(h')} (g',h')_T\triangleright(g,h,\alpha)_C\\
    =& \delta_{g\in c} \frac{\dim(\rho_c)}{\abs{Z(c)}}\sum_{h'\in Z(g)}\beta_{g}(h,h') \overline{\Tilde{\chi}^g_{\rho_c}(h')} (g, hh',\alpha)_C
\end{split}\\
\begin{split}
    (g,hh',\alpha)_C\triangleleft P^S_{(x,\kappa_x)} =& \frac{\dim(\kappa_x)}{\abs{K_x}}\sum_{\stackrel{\alpha',\beta'\in\flatfrac{G}{H}}{\alpha'^{-1}\beta'=x}}\sum_{g'\in K_x} \overline{\Tilde{\chi}^{(\alpha',\beta')}_{\kappa_x}(g')} (g,hh',\alpha)_C\triangleleft(\alpha',\beta',g')_S\\
    =& \frac{\dim(\kappa_x)}{\abs{K_x}}   \delta_{\alpha^{-1} g^{-1} \alpha,x} \sum_{g'\in \Stab_G((g\triangleright\alpha,\alpha))} \psi^{\alpha}(g',g'^{-1}gg')\overline{\psi^{\alpha}(g,g')}\\
    &\times \beta_g(hh',g') \overline{\Tilde{\chi}^{(g\triangleright\alpha, \alpha)}_{\kappa_x}(g')}\; (g,hh'g',\alpha)_C,
\end{split}
\end{align}
\end{subequations}
where we have used Eq.\,\eqref{eq:relation_projective_irreps} to relate the projective characters of different -- but isomorphic --- stabilizer groups.
Taken together,
\begin{align}
\begin{split}
    P^C_{(c,\rho_c),(x,\kappa_x)} \left((g,h,\alpha)_C \right) 
    =& \delta_{g\in c\cap x} \frac{\dim(\rho_c)\dim(\kappa_x)}{\abs{Z(c)}\abs{K_x}}\sum_{h'\in Z(g)} \sum_{g'\in\Stab_G((g\triangleright\alpha,\alpha))} \beta_{g}(h,h')\beta_g(hh',g') \\
    &\times\psi^{\alpha}(g',g'^{-1}gg')\overline{\psi^{\alpha}(g,g')}\overline{\Tilde{\chi}^g_{\rho_c}(h')}\overline{\Tilde{\chi}^{(g\triangleright\alpha,\alpha)}_{\kappa_x}(g')} (g, hh'g',\alpha)_C\;,
\end{split}
\end{align}
where we have used that $h'\in Z(g)$ in every term of the sum above.
The multiplicities $m_{(c,\rho_c),(x,\kappa_x)}$ can now be obtained using Eq.\,\eqref{eq:general_multiplicities}.
Using the dimensions of the irreps,
\begin{equation}
\dim((c,\rho_c)) = |c|\dim(\rho_c)\;,\quad
\dim((x,\kappa_x)) = \frac{G}{|K_x|} \dim(\kappa_x)\;,
\end{equation}
we obtain
\begin{align}\label{eq:fusion_mult_general}
    \begin{split}
    m_{(c,\rho_c),(x,\kappa_x)} =&  \frac{\abs{K_x}}{\abs{c}\dim(\rho_c)\dim(\kappa_x)\abs{G}} \Tr(P^C_{(c,\rho_c),(x,\kappa_x)}) \\
        =& \frac{1}{\abs{G}} \sum_{g\in c} \sum_{\stackrel{\alpha\in\flatfrac{G}{H}}{\alpha^{-1}g^{-1}\alpha = x} }\sum_{h\in Z(g)\cap \Stab_G((g\triangleright\alpha,\alpha))}  \psi^{\alpha}(h,g)\overline{\psi^{\alpha}(g ,h)}
        \Tilde{\chi}^{g}_{\rho_c}(h) \overline{\Tilde{\chi}^{(g\triangleright\alpha,\alpha)}_{\kappa_x}(h)}
    \end{split}
\end{align}
where we have used the twisted 2-cocycle condition for $\beta_g$, Eq.\,\eqref{eq:twisted2cocyclebeta}, and Eq.\,\eqref{eq:relation_projective_irreps} to simplify the trace to the expression in the last line.

Eq.\,\eqref{eq:fusion_mult_general} describes the fusion events at the boundary of any topological twisted gauge theory model. The condensation formula in Ref.\,\cite{Beigi11} can be seen as a special case where the bulk 3-cocycle is trivial and the boundary defect is set to the trivial one. Especially for non-Abelian models this formula gives insight into how non-Abelian bulk anyons split into boundary anyons when approaching the boundary.

\subsubsection{Abelian Models}
For Abelian models, the above expression simplifies significantly. First note, that each group element is its own conjugacy class and the double coset $x$ corresponds to a unique left coset. Moreover, the projective representations are in one-to-one correspondence to the linear representations (see Section~\ref{sec:tube_abelian}) which can be labeled by group elements. Similarly, the boundary anyon labels simplify (see Section~\ref{sec:AbelianBoundaries}). We get
\begin{align}
    m_{(g,k),(x,\kappa)} =& \delta_{g\triangleright x, H} \frac{1}{\abs{G}} \sum_{\alpha\in\flatfrac{G}{H}}\sum_{h\in H} \psi^{\alpha}(h,g)\overline{\psi^{\alpha}(h,g)} \epsilon_g(h) \chi_k^G(h) \overline{\eta^{(\alpha)}_x(h)\chi_\kappa^H(h)},
\end{align}
where $\chi^G_k$ denotes an irreducible $G$-character and $\chi^H_\kappa$ an irreducible $H$-character.

\subsection{Calculating Lagrangian algebra objects in gauge theory models}
Macroscopically a (topological) boundary is defined via the set of anyons that can condense at the boundary, i.e., fuse to the trivial boundary charge.
This set has to form consistency conditions involving their fusion and exchanges statistics which render the set of condensable anyons a \textit{Lagrangian algebra object}~\cite{davydov2017lagrangian, Kapustin2011Topological, levin2013protected} in the UMTC describing the anyon theory.
Our framework connects the microscopic description of the boundary as a lattice model to the macroscopic picture of anyon condensation.
In particular, Eq.\,\eqref{eq:fusion_mult_general} allows us to explicitly calculate the Lagrangian algebra object\footnote{Note that the full Lagrangian algebra object is not only described by an object in the UMTC, i.e. the set of condensable anyons, but also by an algebra morphism. This morphism can also be computed explicitly combining structures and techniques described in this manuscript. We plan to address this in future work.} corresponding to a $(H,\psi)$-boundary by setting $(x,\kappa_x)=(H,\Gamma^0)$, the trivial boundary defect. We get
\begin{align}\label{eq:purecondensation}
    m_{(c,\rho_c),(H,0)} =& \frac{1}{\abs{G}} \sum_{\alpha\in G/H}\sum_{g\in c\cap \Stab_G(\alpha)}\sum_{h\in Z(g)\cap \Stab_G(\alpha)} \psi^{\alpha}(h,g)\overline{\psi^{\alpha}(g,h)} \Tilde{\chi}_{\rho_c}^g(h),
\end{align}
noting that $\alpha^{-1} H\alpha   = \Stab_G(\alpha) = \{g\in G \,|\, g\triangleright\alpha = \alpha\}$.
Note that for a trivial 3-cocycle, this formula can be derived from Ref.\,\cite{Beigi11}. In particular, we have shown that in twisted models the linear character gets replaced by a projective character.

In the Abelian case, it simplifies further to
\begin{align}\label{eq:purecondensation_Abelian}
    m_{(g,k),(H,0)} =& \delta_{g\in H} \frac{1}{\abs{H}}\sum_{h\in H}  \psi(h,g)\overline{\psi(g,h)} \epsilon_g(h)\chi_k^G(h),
\end{align}
where $\chi^G_k$ denotes an irreducible $G$-character.

\subsubsection{\texorpdfstring{$\vV^{\omega}(\bZ_N)$}{Vecω(ZN)}}
As first example we consider the simplest class of Abelian models, when $G=\bZ_N$, the cyclic group of order $N$. Its character function reads $\chi_k(g)=e^{\frac{2\pi i}{N}k g}$. The bulk can be twisted with a type-I 3-cocycle of the form of Eq.\,\eqref{eq:3cocycletype1} giving rise to $\epsilon_g(h)^q=e^{\frac{2\pi i}{N^2}q gh}, q\in \bZ_N$.

For any $q$, the trivial subgroup $H_0=\{0\}$ defines a boundary. Plugging the defining data into Eq.\,\eqref{eq:purecondensation_Abelian} yields the associated Lagrangian subgroup
\begin{align}
    \cL_0 = \{(0,k)\,|\,k=0,1, \dots, N-1\},
\end{align}
i.e., the boundary condenses all the pure charges.

For an untwisted bulk, $q=0$, the other trivial subgroup $H_1=\bZ_N$ defines a boundary as well. Plugging in its defining data into Eq.\,\eqref{eq:purecondensation_Abelian} gives rise to the associated Lagrangian subgroup
\begin{align}
    \cL_1 = \{(g,0)\,|\,g=0,1, \dots, N-1\}
\end{align}
consisting of all the pure fluxes.

If $N$ is not prime $\bZ_N$ has more subgroups. Depending on $q$, these might also define a boundary. For example, for $N=4n$ for $n\in\mathbb{N}$ and $q=N/2 = 2n$, the subgroup $H_2 = \langle N/2 \rangle \simeq\bZ_2$ also defines a boundary, since $\omega_I^{N/2}\big|_{H_2} \equiv 1$.
Using Eq.\,\eqref{eq:purecondensation_Abelian} yields the associated Lagrangian subgroup
\begin{align}
    \cL_2 = \{(g N/2,2k+g)\,|\,g=0,1,\,k=0,1, \dots, N/2-1\}.
\end{align}

\subsubsection{\texorpdfstring{$\vV^{\omega}(\bZ_p\times\bZ_p)$}{Vecω(ZpxZp)}}
Consider the group $G=\bZ_p^{\times 2}$ with the character function $\chi_{(k_1,k_2)}((g_1,g_2)) = e^{\frac{2\pi i}{p}(k_1g_1+k_2g_2)}$ and a bulk twist of type-II
$\omega^q$ from Eq.\,\eqref{eq:3cocycletype2}, giving rise to $\epsilon^q_{(g_1,g_2)}((h_1,h_2))=e^{\frac{2\pi i}{p^2}g_1h_2}$.

A non-trivial subgroup defining a boundary is $H_3 = \langle (0,1)\rangle$ on which $\omega^q$ evaluates to 1. It only has a trivial 2-cocycle class, giving rise to the lagrangian subgroup
\begin{align}
    \cL_3 = \{((0,g),(k,0))\,|\,g,k=0,1, \dots, N-1\},
\end{align}
independent of $q$.

For $q=0$, there are $p$ boundaries associated to the subgroup $H_4=G=\bZ_p^{\times 2}$ (see Eq.\,\eqref{eq:2cocycleZpZq}), parametrized by $m\in\bZ_p$.
Plugging a non-trivial 2-cocycle $\Omega^m$ into Eq.\,\eqref{eq:purecondensation_Abelian} yields the associated Lagrangian subgroups
\begin{align}
    \cL^m_4 = \{((g_1,g_2),(mg_2,-mg_1))\,|\,g_1,g_2=0,1, \dots, N-1\}.
\end{align}
We see that the non-trivial 2-cocycle on $H_4$ ($m\neq 0$) ``couples" the flux with the charge that is condensed at the associated boundary.

\subsubsection{\texorpdfstring{$\vV^\omega(S_3)$}{Vecω(S3)}}
For the untwisted $\vV(S_3)$ model, there are 4 inequivalent boundaries, one for each subgroup of $S_3$: $H_0$, $H_1$, $H_2$ and $H_3$ (as defined in Section~\ref{sec:example_model_s3}).
Let us first consider the normal subgroup $H_1\simeq\bZ_3$.
It defines a valid boundary for the trivial bulk cocycle as well as the cocycle $\omega^3$ (see Eq.\,\eqref{eq:3cocycle_S3}).
The associated Lagrangian algebra objects in these two cases are
\begin{subequations}
\begin{align}
    \cL_1^{p=0} =& (\overline{e},\Gamma^0)\oplus (\overline{e},\Gamma^1)\oplus 2 (\overline{r},0) \\
    \cL_1^{p=3}=& (\overline{e},\Gamma^0)\oplus (\overline{e},\Gamma^1)\oplus (\overline{r},1)\oplus (\overline{r},2)
\end{align}
\end{subequations}
Note that we have included the fusion multiplicities in the above expressions.
The boundary anyons are labeled by a double coset $x\in H_1\backslash\flatfrac{S_3}{H_1} = \{H_1,H_1tH_1\}$ and irreducible representations of $H_1$, $\kappa\in\{0,1,2\}$ all of which are one-dimensional.
The multiplicities for the bulk-boundary fusion events can be found in Tab\,\ref{tab:fusions3H1trivial}.

As a second example, we consider the other non-trivial subgroup $H_2$.
It defines a boundary in the untwisted case as well as for the non-trivial 3-cocycles $\omega^2$ and $\omega^4$.
(see Eq.\,\eqref{eq:3cocycle_S3}).
The boundary anyons are labeled by a double coset $x\in H_2\backslash\flatfrac{S_3}{H_2} = \{H_2,H_2rH_2\}$ and an irrep of $H_2$, $\kappa\in\{0,1\}$.
Interestingly, the fusion multiplicities are the same for both twisted models and the untwisted model. The associated Lagrangian algebra object is
\begin{align}
    \cL_2^{p=0} = \cL_2^{p=2} = \cL_2^{p=4} = (\overline{e},\Gamma^0)\oplus (\overline{e},\Gamma^2) \oplus (\overline{t},0) 
\end{align}
The explicit fusion multiplicities, also for non-trivial boundary defects, can be found in Tab.\,\ref{tab:fusions3H2}.

\subsection{Condensation domain walls for Abelian phases}
It is known that any Abelian twisted quantum double of a finite group can be obtained from boson condensation in an untwisted quantum double of a group of larger cardinality~\cite{Ellison2022Pauli,Duivenvoorden2017Entanglement}. The condensation process can be viewed as a non-invertible domain wall from the parent phase (an untwisted quantum double of an Abelian group) to the condensate (the twisted quantum double in question). In this section, we illustrate how such a domain wall is described on the microscopic level in two exemplary cases.

\subsubsection{Condensation domain walls between type-I twisted and untwisted quantum doubles}
The simplest examples of the above condensation is
the double semion phase. On the one hand, it is realized by the twisted quantum double of $\bZ_2$. On the other hand it can be described as the phase obtained when condensing $e^2m^2$ in the $\bZ_4$ toric code, the (untwisted) quantum double of $\bZ_4$. More generally, a $\bZ_N$ quantum double twisted by a type-I cocycle $\omega_I^n$ (see Eq.\,\eqref{eq:3cocycletype1}) is equivalent to a condensate of a quantum double of $\bZ_{N^2}$ (where $e^Nm^N$ is condensed). Such a condensation process can be viewed as a non-invertible domain wall between the parent phase $D(\vV(\bZ_{N^2}))$ and the condensate $D(\vV^{\omega_I}(\bZ_N))$. In this section, we will see how these condensation domain walls can be obtained from our microscopic models. Interestingly, we see how the non-trivial type-I cocycle of $\bZ_N$ becomes trivial when considered as a cocycle of a larger subgroup of $\bZ_N\times\bZ_{N^2}$.

Via the folding trick (see Section~\ref{sec:folding}), a domain wall between an untwisted $\bZ_{N^2}$ and a $\bZ_{N}$ quantum double twisted by a type-I cocycle $\omega_I^n$ (see Eq.\,\eqref{eq:3cocycletype1}) corresponds to a boundary of a $\vV^{\omega_I^n}(\bZ_{N^2}\times\bZ_N)$ model, where $\omega_I^n$ is trivial on $\bZ_{N^2}$ and of the form of Eq.\,\eqref{eq:3cocycletype1} on $\bZ_N$. In the following we show that the subgroup $H_{\text{cond}}^I = \langle (1,1)\rangle\simeq\bZ_{N^2}\subset \bZ_N\times\bZ_{N^2}$ defines a valid boundary and that it corresponds to the condensation domain wall discussed in the previous paragraph when unfolded.

On $H_{\text{cond}}^I$, the 3-cocycle reads
\begin{align}
    \omega_I^n\Big|_{H_{\text{cond}}^I}(a,b,c) = e^{\frac{2\pi i}{N} n A(B+C-B\oplus_N C)/N}\qcomma a,b,c\in \bZ_{N^2}
\end{align}
and $A,B,C$ are $a,b,c\mod N$. In fact, it is a coboundary of the cochain
\begin{align}
    \psi^n(a,b) = e^{\frac{2\pi i}{N} n A(b-B)/N},
\end{align}
which we can check explicitly
\begin{subequations}
\begin{align}
    (\delta\psi^n)(a,b,c) =& e^{\frac{2\pi i}{N^2}n \left[B(c-C) + A(b\oplus_{N^2}c - B\oplus_N C) - (A\oplus_N B)(c-C) - A(b-B) \right]}\\
    =& e^{\frac{2\pi i}{N} n A(B+C- B\oplus_N C)/N} = \omega_I^n\Big|_{H_{\text{cond}}^I}(a,b,c),
\end{align}
\end{subequations}
where we have used that addition of the integers in the exponent is modulo $N^2$. This shows that $\omega_I^n$ is a coboundary on $H_{\text{cond}}^I$ and with that, that this subgroup defines a valid boundary with $L$-symbols given by $\psi^n$ (see Eq.\,\eqref{eq:VecG_Lsymbols_associativity}). In Tab\,\ref{tab:condensigDWtype1_tunneling} we list the fusion multiplicities for $N=2$ and $n=1$ which align with what one would expect from the anyon condensation picture, a transition from the $\bZ_4$ toric code to the double-semion phase.
In general, if $H$ has non-trivial 2-cocycles, we can multiply the trivializing cochain with such a 2-cocycle to obtain a boundary of a different type.

\begin{table}[ht]
    \centering
    \caption{Fusion multiplicities at the the domain wall implementing the condensation from a $\bZ_4$ toric code to the double-semion phase. We see that $e^2m^2$ is condensed since it fuses with the vacuum charge in the double-semion phase. Moreover, for a fixed double-semion anyon, the two toric-code anyons with which it has non-zero fusion multiplicity differ by $e^2m^2$. An all-zero row shows that the associated toric-code anyon is confined to one side of the domain wall, i.e., there is no valid fusion event with any of the double-semion anyons.}
    \label{tab:condensigDWtype1_tunneling}
    \begin{tabular}{r|c c c c }
& $1$ & $b$& $\overline{s}$ & $s$\\ \hline
$1 = (0,0)$ & 1 & 0 & 0 & 0\\
$e = (0,1)$ & 0 & 0 & 0 & 0\\
$e^2 = (0,2)$ & 0 & 1 & 0 & 0\\
$e^3=(0,3)$ & 0 & 0 & 0 & 0\\
$m=(1,0)$ & 0 & 0 & 0 & 0\\
$me=(1,1)$ & 0 & 0 & 0 & 1\\
$me^2=(1,2)$ & 0 & 0 & 0 & 0\\
$me^3=(1,3)$ & 0 & 0 & 1 & 0\\
$m^2=(2,0)$ & 0 & 1 & 0 & 0\\
$m^2e=(2,1)$& 0 & 0 & 0 & 0\\
$m^2e^2=(2, 2)$ & 1 & 0 & 0 & 0\\
$m^2e^3=(2, 3)$ & 0 & 0 & 0 & 0\\
$m^3=(3, 0)$ & 0 & 0 & 0 & 0\\
$m^3e=(3, 1)$ & 0 & 0 & 1 & 0\\
$m^3e^2=(3, 2)$ & 0 & 0 & 0 & 0\\
$m^3e^3=(3, 3)$ & 0 & 0 & 0 & 1
    \end{tabular}
\end{table}

\subsubsection{Condensation domain walls between fully twisted and untwisted quantum doubles}
As a second example we consider the ``fully twisted" quantum double of $\bZ_N\times\bZ_N$, i.e., we consider a 3-cocycle cohomologous to a product of type-I cocycles on both $\bZ_N$ factors, see Eq.\,\eqref{eq:3cocycletype1}, and the type-II cocycle on the pair, see Eq.\,\eqref{eq:3cocycletype2}, with $n_1=n_2=n_{12}\eqqcolon n$.
For $N=2$, it describes the six-semion phase where all elementary fluxes have semionic self-exchange statistics, see Ref.\,\cite{Ellison2022Pauli}.
This twisted quantum double can also be reppresented as a non-trivial condensate of the untwisted quantum double of $\bZ_{N^2}\times \bZ_{N^2}$.

Via the folding trick, the domain wall in question is equivalent to a boundary of a $\vV^\omega(\bZ_{N^2}^{\times 2}\times\bZ_N^{\times 2})$ model with $\omega$ being a product of type-I cocycles (see Eq.\,\eqref{eq:3cocycletype1}) and a type-II cocycle (see Eq.\,\eqref{eq:3cocycletype2}) on both $\bZ_N$ factors,
\begin{align}
    \omega^n((a_1,a_2,A_1,A_2), (b_1,b_2,B_1,B_2), (c_1,c_2,C_1,C_2)) = e^{\frac{2\pi i}{N^2} n (\sum_i A_i(B_i+C_i - [B_i\oplus_N C_i]) + A_1(B_2+C_2 - [B_2\oplus_N C_2]))}
\end{align}
where $a_i,b_i,c_i\in\bZ_{N^2}$ and $A_i,B_i,C_i\in \bZ_N$.

The domain wall that models the condensation transition is associated to the subgroup $H_{\text{cond}}^{I,II} = \langle (1,0,1,0),(0,1,0,1)\rangle\subset \bZ_{N^2}^{\times 2}\times\bZ_N^{\times 2}$. We parametrize it by two $\bZ_{N^2}$ variables, $H_{\text{cond}}^{I,II} = \{(a_1,a_2,a_1 \mod N,a_2\mod N))\;|\; a_i\in\bZ_{N^2}\}$. Similar to the previous example we can find a coboundary $\beta^n((a_1,a_2),(b_1,b_2)) = e^{\frac{2\pi i}{N^2} n \sum_i A_i(b_i-B_i) + A_1(b_2-B_2)}$ whose coboundary is $\omega^n\big|_{H_{\text{cond}}^{I,II}}$,
\begin{subequations}
\begin{align}
    \omega^n\Big|_{H_{\text{cond}}^{I,II}}((a_1,a_2),(b_1,b_2),(c_1,c_2)) =& e^{\frac{2\pi i}{N^2} n \left( \sum_i A_i(B_i+C_i - [B_i\oplus_N C_i]) + A_1(B_2+C_2 - [B_2\oplus_N C_2]) \right)}\\
    =& (\delta\beta^n)((a_1,a_2),(b_1,b_2),(c_1,c_2)),
\end{align}
\end{subequations}
where $A_i,B_i,C_i$ stand for $a_i,b_i,c_i\mod N$. Hence, this subgroup indeed defines a valid boundary.
In contrast to $H_{\text{cond}}^I$, $H_{\text{cond}}^{I,II}$ has $N^2-1$ non-trivial 2-cocycle classes, represented by 2-cocycles of the form of Eq.\,\eqref{eq:2cocycleZpZq} with $p=q=N^2$.
We can obtain the 2-cochains $\psi$ defining all different domain walls by multiplying the cochain $\beta^n$ defined above with one of those 2-cocycles.
We find that a condensing domain wall from $\vV(\bZ_{N^2}\times\bZ_{N^2})$ to $\vV^{\omega}(\bZ_N\times\bZ_N)$ corresponds to non-trivial 2-cocycle of order $N$. Plugging in the defining data into Eq.\,\eqref{eq:purecondensation_Abelian} and reverting the fold, we recover that the associated domain wall indeed implements the condensation from one phase to the other, compare Ref.\,\cite{Ellison2022Pauli}.
In Tab.\,\ref{tab:condensationZ4Z4sixsemion} we printed a snippet of the condensation table for the domain wall implementing the transition from two copies of the $\bZ_4$ toric code to the six-semion phase. In particular, we find that the correct toric code anyons either 1) condense, i.e., fuse to $1$ in the double-semion phase, 2) confine, i.e., do not fuse to any anyon in the double-semion phase and 3) tunnel to the expected anyon in the double-semion phase.

\begin{table}[ht]
    \centering
    \caption{Snippet of tunneling table from two $\bZ_4$ Toric Codes to 6-semion model for $H_{\rm cond}^{I,II}$ and trivial 2-cocycle on it (top) and $m=1,2,3$ cocycle on it (from top to bottom) and $m=3$, see Eq.\,\eqref{eq:2cocycleZpZq} with $p=N^2$. $m=2$ seems to have the right condensation pattern of $e_2^2m_2^2, m_1^2e_1^2e_2^2\mapsto 1$ and two inequivalent semions $m_2e_1e_2$ and $m_1e_1$ get mapped to valid generators of the 6-semion model.}
    \label{tab:condensationZ4Z4sixsemion}
    \begin{tabular}{r|c c c c c c c c c c c c c c c c}
%
& $1$& $b_1$& $b_2$& $b_3$& $\overline{s_1}$& $s_1$& $\overline{s_2}$& $s_2$& $s_3$& $\overline{s_4}$& $s_3$& $s_4$& $s_5$& $\overline{s_5}$& $\overline{s_6}$& $s_6$\\ \hline
$1=(0, 0, 0, 0)$ & 1 & 0 & 0 & 0 & 0 & 0 & 0 & 0 & 0 & 0 & 0 & 0 & 0 & 0 & 0 & 0\\
$m_2e_1e_2 = (0, 1, 1, 1)$ & 0 & 0 & 0 & 0 & 0 & 0 & 0 & 1 & 0 & 0 & 0 & 0 & 0 & 0 & 0 & 0\\
$m_2^2e_2^2 = (0, 2, 0, 2)$ & 1 & 0 & 0 & 0 & 0 & 0 & 0 & 0 & 0 & 0 & 0 & 0 & 0 & 0 & 0 & 0\\
$e_1^2 = (0, 0, 2, 0)$ & 0 & 0 & 1 & 0 & 0 & 0 & 0 & 0 & 0 & 0 & 0 & 0 & 0 & 0 & 0 & 0\\
$m_2^3e_1^3e_2^3 = (0, 3, 3, 3)$ & 0 & 0 & 0 & 0 & 0 & 1 & 0 & 0 & 0 & 0 & 0 & 0 & 0 & 0 & 0 & 0\\
$m_1e_1 = (1, 0, 1, 0)$ & 0 & 0 & 0 & 0 & 0 & 0 & 0 & 0 & 0 & 0 & 1 & 0 & 0 & 0 & 0 & 0\\
$m_1^2e_1^2e_2^2 = (2, 0, 2, 2)$ & 1 & 0 & 0 & 0 & 0 & 0 & 0 & 0 & 0 & 0 & 0 & 0 & 0 & 0 & 0 & 0\\
$m_1^3e_1^3 = (3, 0, 3, 0)$ & 0 & 0 & 0 & 0 & 0 & 0 & 0 & 0 & 0 & 0 & 0 & 1 & 0 & 0 & 0 & 0 %
%
\\ \vdots
    \end{tabular}
\end{table}

\subsection{Non-Abelian islands in Abelian phases}

\emph{Topological stabilizer-based quantum error correction} (QEC) is believed to be a promising framework to protect a logical qubit against (local) noise. Any Abelian phase can be used to construct a topological stabilizer code~\cite{Magdalena2021nonpauli, Ellison2022Pauli}. In such codes, the logical Pauli operators can be associated with certain (non-trivial) loops of anyon ribbon operators. However, one cannot topologically protect a universal gate set with an Abelian phase alone~\cite{Beverland2016Protected, Webster2022Universal}. One approach to achieve a topologically protected universal gate set is to go beyond stabilizer-based topological QEC and use non-Abelian phases. However, even though first analysis show that -- in principle -- non-Abelian QEC is possible~\cite{feng2015non, Schotte2022Quantum}, it does not appear to be the best approach to simply store a qubit and perform simple operations, like Clifford gates.
In Ref.\,\cite{laubscher2019universal} Laubscher et al. have introduced the concept of \textit{non-Abelian islands} within an Abelian phase.
The authors construct a protocol that allows to teleport a qubit encoded within punctures of an untwisted $\bZ_2$ model into a puncture-encoded qubit within an untwisted $S_3$ model (and vice versa).
This allows to perform a non-Clifford gate within the $S_3$ phase that can be teleported back into the $\bZ_2$ phase.
In this section, we show that the same method can be used to interface twisted models on a microscopic level by constructing the associated tunneling tables from Eq.\,\eqref{eq:purecondensation}.

In the following, consider puncture-encoded qubits within a topological phase, as in Ref.\,\cite{laubscher2019universal}. To be able to teleport logical information from one phase to the other, one has to interface the two codes with a suitable domain wall.
It has to tunnel exactly the right anyons whose ribbon operators span the logical Pauli group.
To connect the macroscopic physics described by the tunneling of anyons with a microscopic model, Eq.\,\eqref{eq:fusion_mult_general} is of essence.
Starting from a microscopic description in terms of a subgroup and a 2-cocycle on it, it allows to check which domain walls can be used to teleport logical information from side of the domain wall to the other.
In this section, we show how twisted non-Abelian models can be interfaced with twisted Abelian models to teleport a qubit, respectively qudit.
In particular, we use Eq.\,\eqref{eq:purecondensation} to construct tunneling tables that show which anyons can be transported though a given domain wall with local operations.
We illustrate the concept via two examples, namely how different twisted $S_3$ models can be interfaced with a twisted $\bZ_2$ (double-semion) model and a twisted $\bZ_3$ model.
After that we comment on how to find suitable microscopic models for more general phases that allow for teleportation of logical information from an Abelian to a non-Abelian phase.
Moreover, we comment on the origin of universal quantum computation with non-Abelian islands in an Abelian codes.

Imagine we want to interface a non-Abelian model, for example $\vV^\omega(S_3)$, with an Abelian one, for example $\vV^{\omega'}(\bZ_N)$.
Any domain wall between such two models is equivalent to a boundary of a stacked model, for example $\vV^{\omega\overline{\omega'}}(S_3\times \bZ_N)$, see Section~\ref{sec:folding}.
In the following, we will illustrate how this folding trick is used to construct the tunneling tables in two examples, where $\vV^\omega(S_3)$ is interfaced with $\vV^{\omega'}(\bZ_N$) for $N=2,3$.
In particular, the 3-cocycles $\omega$ on $S_3$ have to be in different cohomology classes to tunnel anyons non-trivially through the domain wall.

\subsubsection{\texorpdfstring{$\vV^\omega(S_3)\leftrightarrow \vV^\omega(\bZ_2)$}{Vecω(S3)<->Vecω(Z2)}}
As a first example, we consider a domain wall between a $\omega^3$-twisted $S_3$ model (see Eq.\,\eqref{eq:3cocycle_S3}) and the double-semion phase, the twisted $\bZ_2$ model.
After folding, the domain wall corresponds to a boundary of a $S_3\times\bZ_2$ model twisted by the  3-cocycle
\begin{align}
    \omega((A,a,x),(B,b,y),(C,c,z)) = (-1)^{ABC+xyz}.
\end{align}
The non-trivial subgroup $H=\{(x,0,x)\,|\,x\in\bZ_2\}\simeq \bZ_2$ defines a boundary since $\omega$ evaluates to 1 on it. It has only trivial 2-cocycles. Plugging in the defining data ($G=S_3\times\bZ_2$, $\omega$, $H$, $\psi\equiv 1$) into Eq.\,\eqref{eq:purecondensation} and unfolding yields the tunneling table Tab.\,\ref{tab:tunnelingS3Z2}.

\begin{table}[ht]
    \centering
    \caption{Tunneling table for a $\vV^{\omega^3}(S_3)-\vV^{\omega}(\bZ_2)$ domain wall defined by $H=\langle (1,0,1)\rangle\simeq\bZ_2$.}
    \label{tab:tunnelingS3Z2}
    \begin{tabular}{c|c c c c}
         & 1 & $b$ & $s$ & $\overline{s}$  \\ \hline
        $(\overline{e}, \Gamma^0)$ & 1 & 0 & 0 & 0 \\
        $(\overline{e}, \Gamma^1)$ &0 & 1 & 0 & 0 \\
        $(\overline{e}, \Gamma^2)$ & 1 & 1 & 0 & 0 \\
        $(\overline{r}, 0)$ & 0 & 0 & 0 & 0 \\
        $(\overline{r}, 1)$ & 0 & 0 & 0 & 0 \\
        $(\overline{r}, 2)$ & 0 & 0 & 0 & 0 \\
        $(\overline{t}, 0)$ & 0 & 0 & 1 & 0 \\
        $(\overline{t}, 1)$ & 0 & 0 & 0 & 1
    \end{tabular}
\end{table}

Note that the generator of $H$ is supported on $S_3$ as well as on $\bZ_2$ which in turn makes the domain wall (partly) transparent.

\subsubsection{\texorpdfstring{$\vV^{\omega}(S_3)\leftrightarrow \vV^\omega(\bZ_3)$}{Vecω(S3)<->Vecω(Z3)}}
Similarly, one can interface an $S_3$ and a $\bZ_3$ model non-trivially.
In this example, we consider a $\omega^4$-twisted $S_3$ model (see Eq.\,\eqref{eq:3cocycle_S3}) and a type-I twisted $\bZ_3$ model.
After folding, the boundary in question is a boundary of a $S_3\times\bZ_3$ model twisted by a 3-cocycle of the form
\begin{align}
    \omega((A,a,x),(B,b,y),(C,c,z))^ =e^{\frac{2\pi i}{9}( (-1)^{B+C}a((-1)^Cb +c - [(-1)^Cb\oplus_3 c]) - x(y+z-[y\oplus_3 z]))}.
\end{align}
It evaluates to 1 on the subgroup $H=\langle (0,1,1)\rangle\simeq \bZ_3$.
There are only trivial 2-cocycles on $H$ so it defines a unique boundary. Plugging in the defining data $(G=S_3\times \bZ_2,\omega,H,\Psi\equiv 1)$ into Eq.\,\eqref{eq:purecondensation} and reinterpreting the results via unfolding, we obtain the tunneling table in Tab.\,\ref{tab:tunnelingS3Z3}.
\begin{table}[ht]
    \centering
    \caption{Tunneling table for a $\vV^{\omega^4}(S_3)-\vV^{\omega}(\bZ_3)$ domain wall defined by $H=\langle (0,1,1)\rangle\simeq\bZ_3$.}
    \label{tab:tunnelingS3Z3}
    \begin{tabular}{c|c c c c c c c c c}
         & $(0,0)$ & $(0,1)$ & $(0,2)$ & $(1,0)$ & $(1,1)$ & $(1,2)$ & $(2,0)$ & $(2,1)$ & $(2,2)$  \\ \hline
        $(\overline{e}, \Gamma^0)$ & 1 & 0 & 0 & 0 & 0 & 0 & 0 & 0 & 0 \\
        $(\overline{e}, \Gamma^1)$ & 1 & 0 & 0 & 0 & 0 & 0 & 0 & 0 & 0 \\
        $(\overline{e}, \Gamma^2)$ & 0 & 1 & 1 & 0 & 0 & 0 & 0 & 0 & 0 \\
        $(\overline{r}, 0)$ & 0 & 0 & 0 & 0 & 1 & 0 & 0 & 0 & 1 \\
        $(\overline{r}, 1)$ & 0 & 0 & 0 & 0 & 0 & 1 & 1 & 0 & 0  \\
        $(\overline{r}, 2)$ & 0 & 0 & 0 & 1 & 0 & 0 & 0 & 1 & 0 \\
        $(\overline{t}, 0)$ & 0 & 0 & 0 & 0 & 0 & 0 & 0 & 0 & 0 \\
        $(\overline{t}, 1)$ & 0 & 0 & 0 & 0 & 0 & 0 & 0 & 0 & 0
    \end{tabular}
\end{table}
Comparing the two tunneling tables Tab.\,\ref{tab:tunnelingS3Z2} and~\ref{tab:tunnelingS3Z3}, we see that -- in a way -- $\overline{r}$ and $\overline{t}$ have changed roles.

\subsubsection{Remarks on Non-Abelian islands for quantum computation}
In order to use islands of Non-Abelian codes within an Abelian code, one needs a way to transfer logical information from one code to the other fault-tolerantly.
As layed out in Ref.\,\cite{laubscher2019universal} this can be achieved by a certain code-deformation protocol that effectively moves holes (that encode some logical information) through an interfacing region from the Non-Abelian code to the Abelian code.
In fact, this interface can be understood as a domain wall between the two codes and as such, it is described by tunneling tables as the ones calculated above.
To be able to transfer logical information, the domain wall has to be transparent for a subset of the anyons of both models.
Consider interfacing two group theoretical models, each defined by a finite group $G (G')$ and a 3-cocycle.
In this case, the interface is defined by a subgroup $H$ of $G\times G'$ (on which the 3-cocycle is trivial) and a 2-cocycle $\psi$ on $H$.
From Eq.\,\eqref{eq:purecondensation} we can derive conditions on $(H,\psi)$ for the associated domain wall to be (semi-)transparent.
We observe that there are two ways in which one can ``couple'' the two models non-trivially:
\begin{enumerate}
    \item \textbf{Coupling via the subgroup:} The subgroup $H$ cannot be generated a set of generators $\{(g_i,g_i')\}$ each of which is of the form $(g,1_{G'})$ or $(1_G,g')$.
    We say, $H$ does not \textit{factorize} over the two sides of the domain wall.

    \item \textbf{Coupling via a 2-cocycle:} Let $H$ have two subfactors $H_1\subseteq G\times\{1_{G'}\}$ and $H_2\subseteq \{1_G\}\times G'$.
    A 2-cocycle that is non-trivial on $H_1\times H_2$ but not cohologous to a product of 2-cocycles on $H_1$ and $H_2$ individually.
\end{enumerate}
Note that a (semi-)transparent domain wall can also fulfill both of the above conditions.
The examples considered above are all transparent due to condition 1.
For example, $S_3$ has both a $\bZ_2$ and a $\bZ_3$ subgroup which allow it to be interfaced with $\bZ_2$ and $\bZ_3$ models.
The 2-cocycles fulfilling condition 2 are in most cases 2-coycles on an (Abelian) subgroup isomorphic to $\bZ_N\times \bZ_M$ for some pair of integers $N,M$ that are not coprime\footnote{$N,M$ have to share a divisor in order for there to exist a non-trivial 2-coycle class, see App.\,\ref{app:cohomology}}.
The simplest example of a domain wall that is transparent for that reason is the $e-m$ duality domain wall in a $\vV(\bZ_2)$ model.
Microscopically, it is defined by subgroup of the folded model $H\simeq\bZ_2\times\bZ_2$ and the non-trivial 2-cocycle on it.

Ref.\,\cite{laubscher2019universal} nicely describes a computational scheme based on Non-Abelian island for the simplest case of $\vV(S_3)$ islands in a $\vV(\bZ_2)$ model and can be straight forwardly generalized.
It relies on preparing an auxiliary non-stabilizer state within the Non-Abelian patches, encoded into certain punctures.
Since the preparation involves topological charge measurements, universality can be achieved even in group-theoretical MTCs that wouldn't be universal by braiding alone \cite{naidu2011finiteness}.

\section{Conclusion and outlook}\label{sec:conslusion}
In this manuscript, we have explored bulk-to-boundary anyon fusion events in
non-chiral topologically ordered quantum models, aimed at understanding
how anyonic defects interact with external defects such as
boundaries or domain walls.
It has been motivated by considerations both in the study of quantum phases of matter and quantum information theory.
Specifically, we have calculated bulk-to-boundary fusion multiplicities in topological fixed-point models in 2+1 space-time dimensions.
Our framework allows for a step-by-step calculation of the fusion multiplicities in all such models.
Apart from the calculation of projective irreducible representations of the $G$-subgroups which define the action in the tube and semi-tube algebras, the calculation only involves the evaluation of linear expressions.
The fusion multiplicities allow to characterize the behavior of anyonic bulk excitations when approaching a boundary and -- via the folding trick -- to calculate the effect an anyon has on a domain wall when moved through it.

At the core of our construction lies a bimodule that is a representation of the tube algebra defining the bulk anyons as well as the semi-tube algebra defining the boundary anyons.
We have defined this bimodule for any fixed-point model and explicitly derived a closed formula for the fusion multiplicities in the subclass of topological lattice gauge theories, where the fusion category defining the bulk is given by $\vV^\omega(G)$.
We have used this formula to calculate Lagrangian algebra objects in various gauge theory models without the need to explicitly solve the consistency conditions defining a Lagrangian algebra object.
This is particularly useful for more involved non-Abelian topological phases where finding Lagrangian algebra objects is more intricate.
We showcase this in the derivation of the Lagrangian algebra objects for $\vV^\omega(S_3)$ models with non-trivial 3-cocycles.
Moreover, using the folding trick we can use bulk-to-boundary fusion events to study the tunneling of anyons through domain walls.
In particular, our formula allows to keep track of which anyon is left behind at a domain wall when moving a bulk anyon from one side to the other.
As a proof of principle, we have calculated the fusion multiplicities of a special class of non-invertible domain walls which implement anyon condensation from Abelian untwisted to twisted quantum doubles.
This is a well-known transition on the level of the anyon models~\cite{Ellison2022Pauli, Duivenvoorden2017Entanglement} but as far as we know we are the first to give a microscopic description of the corresponding domain walls in space-time.
Moreover, we have shown how to interface Abelian twisted quantum double phases with non-Abelian ones on a microscopic level.

In our construction, we have observed that both bulk and boundary anyons are characterized by a special type of algebra which we call \textit{twisted group algebra with action}.
We give a step-by-step recipe to derive the irreducible representations of such algebras from simpler projective representations of the subgroups stabilizing the action.
In fact, any sort of line-like defects in space-time in a $\vV^\omega(G)$ state-sum model will be
classified by irreducible representations of a certain group algebra with action.
Hence, our techniques can be used to study these defects and their interaction with membrane-like defects in 3+1-dimensional models.
Generalizing this recipe to groupoid-like algebras (see Ref.\,\cite{bullivant2018phd}) and thereby extending the methodology beyond twisted quantum double models might be an interesting avenue for further research.

Together with previous work~\cite{bridgeman2019fusing, Bridgeman2020computingdatalevin, barter2019domain, bridgeman2020computing, kitaev2012models} the fusion multiplicities calculated in this paper contributes to the algebraic description of the anyons in topological models with boundaries and domain walls.
An understanding of similar depth of defects in higher-dimensional models is lacking.
Further research directions can include the application of the techniques used in this paper to line-defects in 3+1-dimensional models and extending the techniques to further understand the interaction of defects of different (co-)dimensions in higher-dimensional models.

Again, our work is not only interesting from the
perspective of the mathematically minded study of
topological phases of matter but it has also important
applications in different areas of physics.
Above all, any practical \emph{topological quantum error correction} (QEC) scheme involves boundaries or domain walls in one way of the other.
On the one hand, stabilizer-based topological QEC can be understood as storing a qudit in the ground space of an Abelian topological phase modeled by a $\vV^\omega(G)$ fixed-point model~\cite{Magdalena2021nonpauli, Ellison2022Pauli}.

On the other hand, going beyond stabilizer-based approaches allows to natively perform universal topological quantum computation~\cite{feng2015non}. Given the overhead in resources of protocols that uplift non-universal stabilizer-based approaches of quantum computing to universal ones by means of magic-state distillation~\cite{RevModPhys.87.307,LitinskiMagic}, such an avenue may well have its benefits.
In both cases a thorough understanding of the interaction of anyons with other types of defects is important.
For example, finding the logical operators in a given planar code including boundaries and domain walls reduces to characterizing which anyons can condense at which boundary.
Our work shows how to calculate these quantities in the most general case, in particular we extend the results of Ref.~\cite{Beigi11} to twisted quantum doubles.
The bulk-to-boundary fusion multiplicities can also be used to study computational protocols including boundaries and domain walls~\cite{Kesselring2018boundariestwist, kesselring2022condensation, laubscher2019universal}.
This includes lattice surgery~\cite{HorsmanSurgery,Litinski2019gameofsurfacecodes} schemes in Abelian topological codes and allows for systematic study of the computational possibilities of a given code via anyon condensation~\cite{kesselring2022condensation}.
Moreover, domain walls between non-Abelian phases und Abelian ones can be used to generalize the scheme presented in Ref.\,\cite{laubscher2019universal} where a partly-transparent domain wall between a $\vV(S_3)$ and a $\vV(\bZ_2)$ phase is used to teleport a topologically encoded qubit from one model to the other.
Starting there, it will be interesting to investigate universal computing schemes based on twisted quantum doubles, particularly when combined with a Pauli-based
description of the Abelian phase from Ref.\,\cite{Ellison2022Pauli}.

Lastly, to fully understand the computational capabilities of \emph{non-Abelian quantum error correction}~\cite{feng2015non, Schotte2022Quantum} we want to investigate domain walls between non-Abelian phases to see how external defects can extend non-Abelian codes.
This is again partially motivated by the quest to find schemes for quantum computing without the need for magic state distillation.
These few examples should illustrate the wide applicability of bulk-to-boundary fusion events, especially in topological QEC.
We hope our work sparks inspiration to develop new QEC and computing protocols based on more exotic topological phases. On a higher level, this
work is aimed at contributing to building new interfaces between quantum information theory and mathematical condensed matter physics which seems a mutually inspiring intersection.

\begin{acknowledgments}
    The authors would like to thank Markus Kesselring for fruitful discussions on the application of our work to topological QEC schemes. JCMdlF also wants to thank Tyler Ellison for discussions on condensation domain walls. This work is supported by the DFG (CRC 183) and the BMBF (RealistiQ, QSolid), and the BMWK (PlanQK).
\end{acknowledgments}

\appendix\section{Examples of fusion multiplicities}
In this appendix, we summarize the fusion multiplicities at boundaries of an untwisted $\vV(\bZ_2\times\bZ_2)$ and $\vV^\omega(S_3)$ models with different 3-cocycles.

\subsection{\texorpdfstring{$\vV(\bZ_2\times\bZ_2)$}{Vec(Z2xZ2)}}
For the $\bZ_2\times\bZ_2$ model, we first show the table of fusion multiplicities at the trivial boundary defined by the order 1 subgroup $H = \{(0,0)\}$ (Tab.\,\ref{tab:Z2Z2_H00}). In Tabs.\,\ref{tab:Z2Z2_H01}  and~\ref{tab:Z2Z2_H11} we give the fusion multiplicities at the boundaries defined by the subgroups $\langle (0,1)\rangle$ and $\langle (1,1)\rangle$.
The only subgroup with non-trivial 2-cocycles is $H=G=\bZ_2\times\bZ_2$. It has two 2-cocycle classes and hence two boundaries associated to it. The corresponding fusion multiplicities can be found in Tab.\,\ref{tab:Z2Z2_HG}.
All these fusion events can either be seen as fusions into the boundary of a $\bZ_2\times\bZ_2$ model (which is in the phase as the topological color code~\cite{kubica2015unfolding}) or -- equivalently -- as domain wall tunneling events between two $\vV(\bZ_2)$ models (toric code phase).

\begin{table}[ht]
    \centering
    \caption{The fusion multiplicities for all pairs of bulk and boundary anyons for $G=\bZ_2\times\bZ_2$ and the standard boundary modeled by the subgroup $H=\{(0,0)\}$. The bulk anyons (rows) are labeled by $G^{\times 2}$, the boundary anyons by cosets $G/H\simeq G$, represented by $G$ elements. The fact that they are actually coset labels is marked with an overline.}
    \label{tab:Z2Z2_H00}
    \begin{tabular}{l| c c c c}
    & $\overline{(0, 1)}$& $\overline{(1, 0)}$& $\overline{(0, 0)}$& $\overline{(1, 1)}$\\ \hline
    ((0, 1), (0, 1)) & 1 & 0 & 0 & 0\\
    ((0, 1), (1, 0)) & 1 & 0 & 0 & 0\\
    ((0, 1), (1, 1)) & 1 & 0 & 0 & 0\\
    ((0, 1), (0, 0)) & 1 & 0 & 0 & 0\\
    ((1, 0), (0, 1)) & 0 & 1 & 0 & 0\\
    ((1, 0), (1, 0)) & 0 & 1 & 0 & 0\\
    ((1, 0), (1, 1)) & 0 & 1 & 0 & 0\\
    ((1, 0), (0, 0)) & 0 & 1 & 0 & 0\\
    ((1, 1), (0, 1)) & 0 & 0 & 0 & 1\\
    ((1, 1), (1, 0)) & 0 & 0 & 0 & 1\\
    ((1, 1), (1, 1)) & 0 & 0 & 0 & 1\\
    ((1, 1), (0, 0)) & 0 & 0 & 0 & 1\\
    ((0, 0), (0, 1)) & 0 & 0 & 1 & 0\\
    ((0, 0), (1, 0)) & 0 & 0 & 1 & 0\\
    ((0, 0), (1, 1)) & 0 & 0 & 1 & 0\\
    ((0, 0), (0, 0)) & 0 & 0 & 1 & 0
    \end{tabular}
\end{table}

\begin{table}[ht]
    \centering
        \caption{Fusion multiplicities at the boundary of a model where the bulk is defined by $G=\bZ_2\times\bZ_2$ and a trivial 3-cocycle and the boundary by the non-trivial subgroup $H=\langle(0,1)\rangle$. The bulk anyons (rows) are labeled by $G^{\times 2}$ and the boundary anyons (columns) by $G/H \times G$. The coset label marked with an overline.}
    \label{tab:Z2Z2_H01}
    \begin{tabular}{l| c c c c}
& ($\overline{(1, 0)}$, (0, 1))& ($\overline{(1, 0)}$, (0, 0))& ($\overline{(0, 0)}$, (0, 1))& ($\overline{(0, 0)}$, (0, 0))\\ \hline
((0, 1), (0, 1)) & 0 & 0 & 1 & 0\\
((0, 1), (1, 0)) & 0 & 0 & 0 & 1\\
((0, 1), (1, 1)) & 0 & 0 & 1 & 0\\
((0, 1), (0, 0)) & 0 & 0 & 0 & 1\\
((1, 0), (0, 1)) & 1 & 0 & 0 & 0\\
((1, 0), (1, 0)) & 0 & 1 & 0 & 0\\
((1, 0), (1, 1)) & 1 & 0 & 0 & 0\\
((1, 0), (0, 0)) & 0 & 1 & 0 & 0\\
((1, 1), (0, 1)) & 1 & 0 & 0 & 0\\
((1, 1), (1, 0)) & 0 & 1 & 0 & 0\\
((1, 1), (1, 1)) & 1 & 0 & 0 & 0\\
((1, 1), (0, 0)) & 0 & 1 & 0 & 0\\
((0, 0), (0, 1)) & 0 & 0 & 1 & 0\\
((0, 0), (1, 0)) & 0 & 0 & 0 & 1\\
((0, 0), (1, 1)) & 0 & 0 & 1 & 0\\
((0, 0), (0, 0)) & 0 & 0 & 0 & 1\\
    \end{tabular}
\end{table}

\begin{table}[ht]
    \centering
    \caption{The bulk-to-boundary fusion multiplicities in a model where the bulk is defined by $G=\bZ_2\times\bZ_2$ and a trivial 3-cocycle and  the boundary by the non-trivial subgroup $H=\langle(1,1)\rangle$. The bulk anyons (rows) are labeled by $G^{\times 2}$ and the boundary anyons (columns) by $G/H \times G$. The coset label marked with an overline.}
    \label{tab:Z2Z2_H11}
    \begin{tabular}{l| c c c c}
& ($\overline{(1, 0)}$, (1, 1))& ($\overline{(1, 0)}$, (0, 0))& ($\overline{(0, 0)}$, (1, 1))& ($\overline{(0, 0)}$, (0, 0))\\ \hline
((0, 1), (0, 1)) & 1 & 0 & 0 & 0\\
((0, 1), (1, 0)) & 1 & 0 & 0 & 0\\
((0, 1), (1, 1)) & 0 & 1 & 0 & 0\\
((0, 1), (0, 0)) & 0 & 1 & 0 & 0\\
((1, 0), (0, 1)) & 1 & 0 & 0 & 0\\
((1, 0), (1, 0)) & 1 & 0 & 0 & 0\\
((1, 0), (1, 1)) & 0 & 1 & 0 & 0\\
((1, 0), (0, 0)) & 0 & 1 & 0 & 0\\
((1, 1), (0, 1)) & 0 & 0 & 1 & 0\\
((1, 1), (1, 0)) & 0 & 0 & 1 & 0\\
((1, 1), (1, 1)) & 0 & 0 & 0 & 1\\
((1, 1), (0, 0)) & 0 & 0 & 0 & 1\\
((0, 0), (0, 1)) & 0 & 0 & 1 & 0\\
((0, 0), (1, 0)) & 0 & 0 & 1 & 0\\
((0, 0), (1, 1)) & 0 & 0 & 0 & 1\\
((0, 0), (0, 0)) & 0 & 0 & 0 & 1
    \end{tabular}
\end{table}

\begin{table}[ht]
    \centering
    \caption{$G=\bZ_2\times\bZ_2$ $H=G$ with trivial (left) and non-trivial 2-cocycle (right).}
    \label{tab:Z2Z2_HG}
    \begin{tabular}{l| c c c c}
& (0, 1)& (1, 0)& (1, 1)& (0, 0)\\ \hline
((0, 1), (0, 1)) & 0 & 1 & 0 & 0\\
((0, 1), (1, 0)) & 1 & 0 & 0 & 0\\
((0, 1), (1, 1)) & 0 & 0 & 1 & 0\\
((0, 1), (0, 0)) & 0 & 0 & 0 & 1\\
((1, 0), (0, 1)) & 0 & 1 & 0 & 0\\
((1, 0), (1, 0)) & 1 & 0 & 0 & 0\\
((1, 0), (1, 1)) & 0 & 0 & 1 & 0\\
((1, 0), (0, 0)) & 0 & 0 & 0 & 1\\
((1, 1), (0, 1)) & 0 & 1 & 0 & 0\\
((1, 1), (1, 0)) & 1 & 0 & 0 & 0\\
((1, 1), (1, 1)) & 0 & 0 & 1 & 0\\
((1, 1), (0, 0)) & 0 & 0 & 0 & 1\\
((0, 0), (0, 1)) & 0 & 1 & 0 & 0\\
((0, 0), (1, 0)) & 1 & 0 & 0 & 0\\
((0, 0), (1, 1)) & 0 & 0 & 1 & 0\\
((0, 0), (0, 0)) & 0 & 0 & 0 & 1
    \end{tabular}
    ~
    \begin{tabular}{l|cccc}
         & (0, 1)& (1, 0)& (1, 1)& (0, 0)\\ \hline
((0, 1), (0, 1)) & 0 & 0 & 1 & 0\\
((0, 1), (1, 0)) & 0 & 0 & 0 & 1\\
((0, 1), (1, 1)) & 1 & 0 & 0 & 0\\
((0, 1), (0, 0)) & 0 & 1 & 0 & 0\\
((1, 0), (0, 1)) & 0 & 0 & 0 & 1\\
((1, 0), (1, 0)) & 0 & 0 & 1 & 0\\
((1, 0), (1, 1)) & 0 & 1 & 0 & 0\\
((1, 0), (0, 0)) & 1 & 0 & 0 & 0\\
((1, 1), (0, 1)) & 0 & 1 & 0 & 0\\
((1, 1), (1, 0)) & 1 & 0 & 0 & 0\\
((1, 1), (1, 1)) & 0 & 0 & 0 & 1\\
((1, 1), (0, 0)) & 0 & 0 & 1 & 0\\
((0, 0), (0, 1)) & 1 & 0 & 0 & 0\\
((0, 0), (1, 0)) & 0 & 1 & 0 & 0\\
((0, 0), (1, 1)) & 0 & 0 & 1 & 0\\
((0, 0), (0, 0)) & 0 & 0 & 0 & 1
    \end{tabular}
\end{table}

\subsection{\texorpdfstring{$\vV^\omega(S_3)$}{Vecω(S3)}}
Depending on which 3-cocycles we use in $\vV^\omega(S_3)$ different subgroups define topological boundaries.
In Tab.\,\ref{tab:fusions3H1trivial} we give the fusion multiplicities for the cases when $H_r$ defines a boundary.
In Tab.\,\ref{tab:fusions3H2} we contrast it with the fusion multiplicities for the cases where $H_t$ defines a boundary.

\begin{table}[ht]
    \centering
    \caption{All the fusion multiplicities for a model where the bulk is defined by $G=S_3$ and the boundary by the non-trivial subgroup $H_r=\langle r\rangle$ (see Section~\ref{sec:example_model_s3}). The bulk anyons (rows) are labeled as in Section~\ref{sec:tube_examples_s3} and the boundary anyons by $H_r\backslash G/H_r \times \bZ_3$. We show the fusion multiplicites for the 3-cocycles $\omega^p$ (see Eq.\,\eqref{eq:3cocycle_S3}) with $p=0,3$. Where they differ we give the value for $p=3$ in brackets.
    These are all models where $H_r$ defines a valid boundary.}
    \label{tab:fusions3H1trivial}
    \begin{tabular}{l|c c c c c c}
         & $(H_r,0)$ & $(H_r,1)$ & $(H_r,2)$ & $(H_rtH_r,0)$ & $(H_rtH_r,1)$ & $(H_rtH_r,2)$ \\ \hline
         $(\overline{e},\Gamma^0)$ & 1 & 0 & 0 & 0 & 0 & 0 \\
         $(\overline{e},\Gamma^1)$ & 1 & 0 & 0 & 0 & 0 & 0 \\
         $(\overline{e},\Gamma^2)$ & 0 & 1 & 1 & 0 & 0 & 0 \\
         $(\overline{r},0)$ & 2(0) & 0(1) & 0(1) & 0 & 0 & 0 \\
         $(\overline{r},1)$ & 0(1) & 2(0) & 0(1) & 0 & 0 & 0 \\
         $(\overline{r},2)$ & 0(1) & 0(1) & 2(0) & 0 & 0 & 0 \\
         $(\overline{t},0)$ & 0 & 0 & 0 & 1 & 1 & 1 \\
         $(\overline{t},1)$ & 0 & 0 & 0 & 1 & 1 & 1
    \end{tabular}
\end{table}


\begin{table}[ht]
    \centering
    \caption{All the fusion multiplicities for a model where the bulk is defined by $G=S_3$ and the boundary by the non-trivial subgroup $H_t=\langle t\rangle$ (see Section~\ref{sec:example_model_s3}). The bulk anyons (rows) are labeled as in Section~\ref{sec:tube_examples_s3} and the boundary anyons by $H_t\backslash G/H_t$ and the irreducible representations of the associated stabilizer group. For $H_t$ as the trivial double coset, this is $\bZ_2$, for $H_t r H_t$, the only non-trivial double coset, the stabilizer group is trivial, i.e., $\bZ_1$. We show the fusion multiplicites for the 3-cocycles $\omega^p$ (see Eq.\,\eqref{eq:3cocycle_S3}) with $p=0,2,4$. Interestingly, they coincide for all of these models. In fact, for other values of $p$ $H_t$ does not define a valid boundary.}
    \label{tab:fusions3H2}
    \begin{tabular}{l|c c c }
         & $(H_t,0)$ & $(H_t,1)$ & $(H_trH_t,0)$ \\ \hline
         $(\overline{e},\Gamma^0)$ & 1 & 0 & 0  \\
         $(\overline{e},\Gamma^1)$ & 0 & 1 & 0  \\
         $(\overline{e},\Gamma^2)$ & 1 & 1 & 0  \\
         $(\overline{r},0)$ & 0 & 0 & 1  \\
         $(\overline{r},1)$ & 0 & 0 & 1  \\
         $(\overline{r},2)$ & 0 & 0 & 1  \\
         $(\overline{t},0)$ & 1 & 0 & 1  \\
         $(\overline{t},1)$ & 0 & 1 & 1
    \end{tabular}
\end{table}

\newpage

\section{Cohomology of finite groups}\label{app:cohomology}

In this section, we provide an algebraic definition of cohomology theory for finite groups. For a more detailed background, see, for example, Refs.\,\cite{chen2013symmetry, brown2012cohomology}.

\begin{definition}[(left) $G$-module]\label{app:module_def}
  Consider a finite group $(G,\cdot)$. An Abelian group $(M,\ast)$
  together with a (left) group action $\triangleright : G\times M\to M, (g,a)\mapsto g\triangleright a$ that $\forall \;a,b\in M$, $g,h\in G$ fulfills
  \begin{itemize}[nosep]
    \item $g\triangleright (a\ast b) = (g\triangleright a)\ast (g \triangleright b)$ and
    \item $(g\cdot h)\triangleright a = g\triangleright (h\triangleright a)$,
  \end{itemize}
  is called a \textit{left $G$-module}.
 Analogously, a \textit{right $G$-module} is defined with a $G$-action from the right which we denote by $\triangleleft$. An Abelian group $M$ equippped with both a $G$-action from the left and a $G'$-action from the right is called \textit{$G$-$G'$-bimodule}.
\end{definition}

\begin{remark}
  Any Abelian group can be made into a $G$-module with a trivial group action by defining $g\triangleright a = a\;\forall a\in M, g\in G$. 
\end{remark}

\begin{definition}[$n$-cochain]
Let $(M,\ast)$ be a $G$-module. A map $\eta_n: G^{\times n} \to M$ is called \textit{$n$-cochain} (of $G$ over $M$). We denote the space of all such $n$-cochains by $\cC^n(G,M)$.
In fact, $\cC^n(G,M)$ is a group with the group multiplication inherited from $M$.
\end{definition}

\begin{definition}[(twisted) $n$-coboundary]
Let $(M,\ast)$ be a $G$-bimodule, potentially with trivial (left/right) $G$-action. The map $\delta_n:\cC^{n}(G,M)\to \cC^{n+1}(G,M)$ defined by
\begin{align}
\begin{split}
    (\delta_n\eta_n)(g_0,g_1, \dots, g_n) :=& (g_0\triangleright \eta_n(g_1, \dots, g_n))\ast(\eta_n(g_0, \dots, g_{n-1})\triangleleft g_n)^{(-1)^{n+1}}\\
    &\ast \prod_{i=0}^{n-1}\eta_n(g_0, \dots, g_{i-1},g_ig_{i+1}, \dots, g_n)^{(-1)^{i+1}},
\end{split}
\end{align}
is called \textit{$n$-coboundary operator}.

Most importantly, for any $n$ and $\eta\in\cC^n(G,M)$, $(\delta_{n+1}\circ\delta_{n})(\eta) = 1_{\cC^n(G, M)}$, where $1_{\cC^n(G, M)}$ denotes the trivial (constant) map from $G^{\times n}$ to $M$.
We say that the coboundary of a coboundary is trivial.
Often, we will shortly write $\delta=\bigoplus_n\delta_n$ which can act on any (combination of) cochain(s).
\end{definition}

\begin{remark}
In this work we encounter two sorts of $G$-modules. First, $U(1)$ with trivial action, and second, $U(1)^A$ for a finite set $A$ whose action is determined by a permutation action of $G$ on $A$. We will distinguish the latter \textit{twisted coboundary operator} by explicitly writing $\Tilde{\delta}$.
\end{remark}

\begin{definition}[(twisted) cohomology groups]
Let $(M,\ast)$ be a $G$-module. The cochains in the kernel of $\delta_n$,
\begin{align}
    \cZ^n(G,M) := \ker(\delta_n) =  \{\eta\in\cC^n(G,M)\,|\, \delta_n\eta_n \equiv 1_{\cC^{n+1}(G, M)}\},
\end{align}
are called \textit{$n$-cocycles}. They form a subgroup of the group of cochains $\cC^n(G,U(1))$. The $n+1$ cochains in the image of $\delta_n$,
\begin{align}
    \cB^n(G,M) := \Im(\delta_{n-1}) =  \{\eta\in\cC^{n+1}(G,M)\,|\,\eta = \delta_{n-1}\beta,\,\beta\in\cC^{n-1}(G,M)\},
\end{align}
are called \textit{$n$-coboundaries}.

The quotient group
\begin{align}
    H^n(G,M) := \faktor{\cZ^n(G,M)}{\cB^n(G,M)}
\end{align}
is called \textit{$n$th cohomology group of $G$ over $M$}.
\end{definition}

\begin{remark}
If two $n$-cocycles are in the same equivalence class in $H^n(G,M)$ we call them \textit{cohomologous}.\\
In any $n$-cocycle class we can find a representative $\omega\in \cZ^n(G,M)$ that fulfills
\begin{align}
    \omega(1_G,g, h, ...) =  \omega(g,1_G,h,...) = ... = \omega(g,h, \dots, 1_G) = 1_M.
\end{align}
We call such a cocycle \textit{normalized}.
\end{remark}

\begin{definition}[slant product]
Let $(M,\ast)$ be a $G$-module. For any $x\in G$ and integer $n>1$, the map $i_x^n:\cC^{n}(G,M)\to\cC^{n-1}(G,M)$,
\begin{align}
\begin{split}
    (i_x^n\eta_n)(g_1, \dots, g_{n-1}) :=& \eta_n(x,g_1, \dots, g_{n-1})^{(-1)^{n-1}}\\
    &\ast\prod_{k=1}^{n-1} \eta_n(g_1, \dots, g_k,g_k^{-1}\cdots g_1^{-1}x g_1\cdots g_k,g_{k+1}, \dots, g_{n-1})^{(-1)^{n-1+k}},
\end{split}
\end{align}
is called \textit{slant product}.
Again, we write $i_x=\bigoplus_n i_x^n$ to abbreviate the collection of all slant products.
\end{definition}

\begin{remark}
$i_g^n$ induces a homomorphism on the cohomology groups $H^n(G,M)\to H^{n-1}(G,M^G)$ where $M^G$ is a $G$-module
whose action is a combination of the action on $M$ plus conjugation of the $G$-label.
Indeed, one can check that $(\Tilde{\delta}_{n-1}\circ i^n) = (i^{n+1}\circ\delta_{n})$.
This explains why $\beta_g$ in Section~\ref{sec:tube} is a twisted 2-cocycle.
\end{remark}

In the following, we will give some examples for cohomology groups of Abelian and non-Abelian groups over $U(1)$ as a $G$-module with trivial action.
\begin{exmpl}[$\bZ_N$]
The cyclic group $\bZ_N$ is one of the few finite groups where one can simply derive all its cohomology groups. Following Ref.\,\cite{chen2013symmetry}, we get
\begin{align}
    H^n(\bZ_N,U(1)) = \begin{cases}
        U(1) & n=0,\\
        \bZ_1 & n=0\mod 2, n>0,\\
        \bZ_N & n=1\mod 2.
    \end{cases}
\end{align}
\end{exmpl}

\begin{exmpl}[$\bZ_N\times\bZ_M$]
Using the Künneth formula for group cohomology~\cite{chen2013symmetry}, we can relate the cohomology groups of $\bZ_N\times\bZ_M$ to the ones of $\bZ_N$ and $\bZ_M$. For
$n=0,1,2,3$, we obtain
\begin{subequations}
\begin{align}
    H^0(\bZ_N\times\bZ_M, U(1)) =& U(1),\\
    H^1(\bZ_N\times\bZ_M, U(1)) =& \bZ_N\times\bZ_M,\\
    H^2(\bZ_N\times\bZ_M, U(1)) =& \bZ_{\gcd(N,M)},\\
    H^3(\bZ_N\times\bZ_M, U(1)) =& \bZ_N\times\bZ_M\times\bZ_{\gcd(N,M)}.
\end{align}
\end{subequations}
\end{exmpl}

\begin{exmpl}[$\bZ_{n_1}\times\bZ_{n_2}\times\cdots\times\bZ_{n_k}$]
To calculate the cohomology groups of product groups, like $\bZ_{n_1}\times\bZ_{n_2}\times\cdots\times\bZ_{n_k}$, we can use the Künneth formula~\cite{chen2013symmetry}. For $n=0,1,2,3$,
we obtain
\begin{subequations}
\begin{align}
    H^0(\bZ_{n_1}\times\bZ_{n_2}\times\cdots\times\bZ_{n_k}, U(1)) =& U(1),\\
    H^1(\bZ_{n_1}\times\bZ_{n_2}\times\cdots\times\bZ_{n_k}, U(1)) =& \bigotimes_{i=1}^k\bZ_{n_i},\\
    H^2(\bZ_{n_1}\times\bZ_{n_2}\times\cdots\times\bZ_{n_k}, U(1)) =& \bigotimes_{\text{pairs }(i,j)}\bZ_{\gcd(n_i,n_j)},\\
    H^3(\bZ_{n_1}\times\bZ_{n_2}\times\cdots\times\bZ_{n_k}, U(1)) =& \bigotimes_{i=1}^k\bZ_{n_i}\bigotimes_{\text{pairs }(i,j)}\bZ_{\gcd(n_i,n_j)}\bigotimes_{\text{triples }(i,j,k)}\bZ_{\gcd(n_i,n_j,n_k)}.\label{app:eq:3cocyclesGAbelian}
\end{align}
\end{subequations}
\begin{remark}
The 3-cocycles classes in Eq.\,\eqref{app:eq:3cocyclesGAbelian} decompose into products of 3-cocycle classes of three types: The ones only depending on the  single tensor factor (type-I), a pair of factors (type-II) and depending on a triple of factors (type-III). Remarkably, the slant product maps 3-cocycles of type I and II to trivial 2-cocycles and only 3-cocycles of type III to non-trivial ones.
\end{remark}
\end{exmpl}

\begin{exmpl}[$S_3$] Following Refs.~\cite{Bridgeman2020computingdatalevin, propitius1995topological} the first 3 cohomology groups of $S_3$ are given by
\begin{subequations}
\begin{align}
    H^0(S_3,U(1)) =& U(1),\\
    H^1(S_3,U(1)) =& \bZ_2,\\
    H^2(S_3,U(1)) =& \bZ_1 ,\\
    H^3(S_3,U(1)) =& \bZ_2\times\bZ_3\simeq \bZ_6.
\end{align}
\end{subequations}
\end{exmpl}

\section{\texorpdfstring{The isomorphism $m^{(n)}$}{The isomorphism m}}\label{app:sec:isomorphism}
Let $(G,\cdot)$ be a finite group and $X$ a left $G$-set. In particular, there exist a map $\triangleright:G\times X\to X$ such that
\begin{align}
    (g\cdot h)\triangleright x = g\triangleright(h\triangleright x)\; \forall g,h\in G, x\in X.
\end{align}
Given such a $G$-action we consider the $G$-module $U(1)^X$ with the non-trivial (right) action defined by
\begin{align}
    (\phi\triangleleft g)^x = \phi^{g\triangleright x}\qq{}\forall x\in X,\phi\in U(1)^{X}
\end{align}
An $n$-cochain for this $G$-module can be represented by a map $\psi:G^{\times n}\times \flatfrac{G}{H}\to U(1)$. In this representation, the twisted coboundary operator acts as
\begin{align}\label{app:eq:twistedcoboundarycosets}
\begin{split}
    (\Tilde{\delta}\psi)^\alpha(g_n,...,g_1,g_0) =& \psi^{\alpha}(g_{n-1}, \dots, g_0) \psi^{g_0\triangleright  \alpha}(g_n,g_{n-1},\dots ,g_1)^{(-1)^{n+1}}\\
    &\times \prod_{i=1}^{n} \psi^{\alpha}(g_n, \dots, g_ig_{i-1}, \dots, g_0)^{(-1)^{n+i+1}}.
\end{split}
\end{align}
Every $G$-set $X$ is isomorphic to a disjoint union of cosets
\begin{align}
    X  \simeq \bigcup_i \flatfrac{G}{H_i}
\end{align}
for a collection of subgroups $\{H_i\subset G\}$. $G$ acts transitively on each component by left-translation $g\triangleright(xH_i) = (g\cdot x)H_i$.
This makes $U(1)^X$ a cartesian product of decoupled modules, so the total cohomology group is isomorphic to a product
\begin{align}
    H^n(G,U(1)^{X}) \simeq \bigotimes_i H^n(G,U(1)^{\flatfrac{G}{H_i}}).
\end{align}

The goal of this appendix is to show the following isomorphism that holds for each component:
\begin{align}
    H^n(G,U(1)^{\flatfrac{G}{H}}) \simeq H^n(H,U(1)),
\end{align}

This isomorphism is induced by the following two maps
\begin{align}
\includegraphics[valign=c]{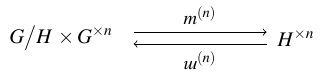}.
\end{align}
The easy direction of the isomorphism is given by $\minv^{(n)}$, the embedding of $H^{\times n}$ into $G^{\times n}\times \flatfrac{G}{H}$, i.e.
\begin{align}
    \minv^{(n)}(a,b,...) = (a,b,...,H).
\end{align}
The map in the converse direction is obtained from first choosing a representative for each of the cosets.
Let $(g_n, \dots, g_1, \alpha)\in G^{\times n}\times\flatfrac{G}{H}$ and consider the cosets $\alpha,g_1\triangleright\alpha, \dots, (g_n\cdots g_1)\triangleright\alpha$.
We denote their respective representatives by $r_0,r_1, \dots, r_n$.
With that, $m^{(n)}$ is defined by
\begin{align}\label{app:eq:cohomisomorph}
m^{(n)}(g_n, \dots, g_1,\alpha) = (r_n^{-1}g_n r_{n-1},r_{n-1}^{-1}g_{n-1}r_{n-2}, \dots, r_1^{-1}g_1r_0).
\end{align}
The image is guaranteed to lie in $H^{\times n}$ because, by construction, $g_k(r_{k-1} H) = r_kH\; \forall k$ and hence $r_i^{-1}g_ir_{i-1}\in H,\;\forall i$.

\begin{remark}
For $g_1, \dots, g_n\in H$, $m^{(n)}(g_n, \dots, g_1,H) = (g_n, \dots, g_1)$. In the case of an Abelian group, any $g_i\in H$ itself is left invariant by $m^{(n)}$, independent of the remaining arguments.
\end{remark}

We can apply $m^{(n)}$, respectively $\minv^{(n)}$, on cochains $\psi\in\cC^n(G,U(1)^{G/H})$ and $\rho\in\cC^n(H,U(1))$, via precomposition
\begin{align}
    \hat{\rho}:= \rho\circ\minv^{(n)} \in \cC^n(H,U(1)) \qq{and} \Tilde{\psi}:=\psi\circ m^{(n)}.
\end{align}
In fact, both are chain maps, i.e. mapping cocycles onto cocycles.
This can be seen with a straight forward calculation
\begin{subequations}
\begin{align}
    \begin{split}
        (\Tilde{\delta}\Tilde{\psi})^\alpha(g_n,\dots, g_1,g_0)
        =& \psi(m^{(n)}(g_{n-1}, \dots, g_0,\alpha)) \psi(m^{(n)}(g_n,\dots,g_1,g_0\triangleright\alpha)^{(-1)^{n+1}}\\
        &\times\prod_{i=1}^{n} \psi(m^{(n)}(g_n, \dots, g_ig_{i-1}, \dots, g_0,\alpha))^{(-1)^{n-i+1}}.
    \end{split}\\
    \begin{split}
        =&\psi(r_{n-1}^{-1}g_{n-1}r_{n-2}, \dots, r_0^{-1}g_0r) \psi(r_n^{-1}g_nr_{n-1}, \dots, r_1^{-1}g_1r_0)^{(-1)^{n+1}}\\
        &\times \prod_{i=1}^{n-1} \psi(r_n^{-1}g_nr_{n-1}, \dots, r_{i}^{-1}g_ig_{i-1}r_{i-2}, \dots, r_0^{-1}g_0r)^{(-1)^{i+1}}
    \end{split}\\
    =&(\delta\psi)(r_{n}^{-1}g_nr_{n-1},\dots, r_0^{-1}g_0r) =  \widetilde{(\delta\psi)}^\alpha(g_n,\dots,g_1,g_0),
\end{align}
\end{subequations}
where $r_i$ is the chosen representative for coset $(g_ig_{i-1}\cdots g_0)\triangleright\alpha$ and $r$ the representative of $\alpha$.
The converse, $\delta\hat{\rho} = \widehat{\Tilde{\delta}\rho}$ follows simply from the fact that the $G$-action on $U(1)^{G/H}$ becomes trivial when every argument of the cochain is restricted to $H$.

To complete the proof, we have to show that $\Tilde{\bullet}$ and $\hat{\bullet}$ induced by $m^{(n)}$ and $\minv^{(n)}$ are inverses of each other up to coboundaries and thereby preserve the cohomology class.
One direction is easy to see from
\begin{align}
    m^{(n)}\circ\minv^{(n)} = \Id_{H^{\times n}}.
\end{align}
To show the converse, we explicitly construct a $n-1$-cochain $\Omega: G^{n-1}\times G/H\to U(1)$ such that
\begin{align}\label{app:eq:proof_cohomologous}
    \psi^\alpha(g_n, \dots, g_1)\overline{{\psi(\minv^{(n)}(m^{(n)}}(g_n, \dots, g_1,\alpha)))} = (\Tilde{\delta}\Omega)^\alpha(g_n, \dots, g_1),
\end{align}
namely
\begin{align}
    \Omega^\alpha(g_{n-1},\dots, g_1) = \prod_{i=0}^{n-1}\psi^H(g_{n-1}, \dots, g_{i+1}, r_{i}, r_{i}^{-1}g_{i} r_{i-1},\dots r_1^{-1}g_1r_0)^{(-1)^{n+i}},
\end{align}
with $r_0$ being the representative for $\alpha$ and $r_i$ for $(g_{n-1}\cdots g_1)\triangleright\alpha$.
This can be seen explicitly by rewriting $\psi^\alpha(g_n, \dots, g_1)$ with $(\Tilde{\delta}\psi)^\alpha(g_n,g_{n-1},\dots, g_1,r_0) = 1$ in terms of $\psi^H$.

This isomorphism is of particular importance for our work. It not only defines the $L$-symbols of boundaries in Section~\ref{sec:gauge_models} but also enters in the algebra diagonalization in Section~\ref{sec:twistedwithaction}.
Moreover, Eq.\,\eqref{app:eq:proof_cohomologous} shows that the twisted 2-cocycle defining the semi-tube algebra $\Psi^{\alpha,\alpha x}$ is cohomologous for all $\alpha$ when restricted to $H$.
This is implicitly used in the central idempotents since we sum over $\alpha$.
In App.\,\ref{app:thin_boundary}, we give a geometric interpretation of this isomoprhism as an invertible domain wall on a boundary state-sum.

\section{State-sum picture}
\label{app:state_sum}
In this appendix we give a concise and systematic rederivation of what is discussed in the main text, using the language of state-sum models in space-time. We will make use of the notions of \emph{extended manifolds/cellulations} as defined in Appendix~B of Ref.~\cite{walkerwang}. Roughly, an extended manifold is a composite of manifolds-with-boundary called \emph{regions} of different dimension attached to each other in different ways. The \emph{link} of a region is the intersection of a small-enough $\epsilon$-sphere around a point within the space normal to that region, and has to be the same for all points of that region. For example, if we have a 1-dimensional region embedded into a 3-dimensional one, the normal space at a point is a plane, so the link of the 1-dimensional region is a circle. To get an extended cellulation, we triangulate the Cartesian product of each region with its link. This triangulation is identified with the boundary triangulation of higher-dimensional regions.

We now associate state-sum variables and weights to different cells of the different regions, and demand their invariance under topology-preserving moves such as Pachner moves. The highest-dimensional region then defines the bulk of a state-sum model, and the lower-dimensional regions define topological boundaries, anyon world-lines, domain walls, and other sorts of defects. In the following, we will present the twisted quantum double model and its boundaries, anyons, etc. in this language.

\subsection{Bulk}
The bulk of the twisted quantum double model is a state-sum path integral on 3-dimensional space-time triangulations with a branching structure consisting of tetrahedra,
\begin{equation}
\label{eq:statesum_tetrahedron}
\includegraphics[valign=c]{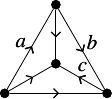}\;.
\end{equation}
At every edge of the triangulation there is a state-sum variable (such as $a$, $b$, and $c$ above) taking values in the group $G$. On every triangle,
\begin{equation}
\label{eq:state_sum_triangle}
\includegraphics[valign=c]{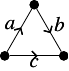}\;,
\end{equation}
the group labels $a$, $b$, and $c$ have to satisfy the constraint
\begin{equation}
ab=c\;.
\end{equation}
At every tetrahedron with labels as in Eq.~\eqref{eq:statesum_tetrahedron}, we have a weight
\begin{equation}
\omega(a,b,c)\;.
\end{equation}
Note that the three labels $a$, $b$, and $c$ determine all other labels of the tetrahedron via the triangle constraint. If the tetrahedron has the opposite orientation from Eq.~\eqref{eq:statesum_tetrahedron}, we instead associate the complex conjugate $\overline{\omega}$. This is how unitarity/Hermiticity is implemented in the state-sum language. Retriangulation invariance is then imposed via Pachner moves as shown in Eq.~\eqref{eq:pentagon_move} in the main text. This makes $\omega$ a group 3-cocycle representing an element of $H^3(G, U(1))$.

\subsection{Boundary}
Next we include boundaries. The normal space to a point in the boundary is a half-line, and an $\epsilon$-sphere restricted to that half line is a single point. So the link of the boundary is a point, and an extended cellulation is essentially just a 3-dimensional triangulation with boundary. However, it better to ``thicken'' the boundary triangles into 3-cells which look like triangle prisms,
\begin{equation}
\label{eq:boundary_triangle}
\includegraphics[valign=c]{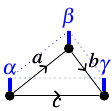}\;.
\end{equation}
The bottom triangle formed by the three edges labeled $a$, $b$, $c$ is attached to a bulk tetrahedron, the three rectangles on the sides are attached to other boundary 3-cells, and the top triangle formed by the three dotted edges corresponds to the actual boundary. By replacing boundary triangles with boundary 3-cells, we can also make sense of two boundary triangles separated by an ``infinitely thin'' bulk as for example later in Eq.~\eqref{eq:boundary_anyon_compactification}. At every boundary vertex (or more precisely, every of the short thick blue lines in Eq.~\eqref{eq:boundary_triangle}), there is a state-sum variable taking values in some finite set $A$, which is equipped with a right action of $G$,
\begin{equation}
\triangleleft: A\times G\rightarrow A\;.
\end{equation}
At each ``thickened'' boundary edge,
\begin{equation}
\label{eq:boundary_edge}
\includegraphics[valign=c]{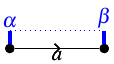}\;,
\end{equation}
such as at the three sides of the boundary triangle in Eq.~\eqref{eq:boundary_triangle}, the labels fulfil the constraint
\begin{equation}
\label{eq:space-time_coset_action}
\beta=\alpha\triangleleft a\;.
\end{equation}
To each triangle as in Eq.~\eqref{eq:boundary_triangle}, we associate a weight
\begin{equation}
\psi^\alpha(a,b)\;,
\end{equation}
noting that the labels $c$, $\beta$, and $\gamma$ are determined from $\alpha$, $a$, and $b$ via the constraints. If the triangle has opposite orientation from Eq.~\eqref{eq:boundary_triangle}, we associate the complex conjugate instead, just as for the bulk tetrahedra. Note that this Hermiticity condition holds generally for all weights/cells in a state-sum, and we will in the following always assume it without explicitly saying so.

Topological invariance can be imposed by a move attaching/removing a tetrahedron to the boundary,
\begin{equation}
\label{eq:boundary_move}
\includegraphics[valign=c]{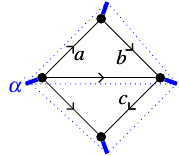}
=
\includegraphics[valign=c]{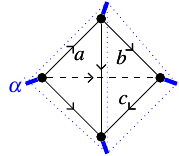}
\;.
\end{equation}
The same move was depicted slightly differently in Eq.~\eqref{eq:boundary_pentagon} in the main text. On the left side, there are two boundary triangles only, drawn with the bulk at the back such that the dashed blue lines are in front of the black lines with arrows. On the right hand side, we have two boundary triangles, and a bulk tetrahedron attached to both at the back. The boundary triangles on both sides alone are related by a 2-2 Pachner move. So $\psi$ has to fulfil
\begin{equation}
\psi^\alpha(a,b)\psi^\alpha(ab,c)=\omega(a,b,c)\psi^\alpha(a,bc)\psi^{\alpha\triangleleft a}(b,c)\;,
\end{equation}
which is the same (apart from slightly different conventions) as in Eq.~\eqref{eq:VecG_Lsymbols_associativity} in the main text.


\subsection{Anyons}
Anyons in the state-sum can be represented by considering triangulations of 3-manifolds with an embedded 1-manifold representing the anyon world-line. The link of an embedded 1-manifold is a circle, which can be triangulated with a single ``looping'' edge whose end vertices coincide,
\begin{equation}
\label{eq:looping_edge}
\includegraphics[valign=c]{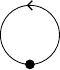}\;.
\end{equation}
Since the 1-manifold itself is triangulated with edges, we need to trianglate an edge times the looping edge. We pick a triangulation of this ``tube segment'' with two triangles,
\begin{equation}
\label{eq:anyon_segment}
\includegraphics[valign=c]{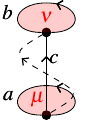}\;.
\end{equation}
Even though such a tube segment is formally something two-dimensional, it is good to think of it as a cylinder-like 3-cell, with a red shaded 1-gon at the top and bottom. An extended cellulation is a 3-manifold triangulation into which we embed sequences of tube segments as above. The two red shaded 1-gons attach to other tube segments, whereas the two triangles wrapping round the side attach to bulk tetrahedra. Those triangles may also be directly attached to the triangles of other tube segments with an infinitely thin bulk in between.

The state-sum associates one additional variable to each red shaded 1-gon where two tube segments meet,
\begin{equation}
\includegraphics[valign=c]{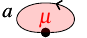}\;.
\end{equation}
The dimension of this variable $\mu$ may to depend on the group label $a$ at the corresponding loop edge. To each tube segment with labels as in Eq.~\eqref{eq:anyon_segment}, we associate a weight
\begin{equation}
(\rho^a(c))_\mu^\nu\;.
\end{equation}
The topological invariance is again imposed by combinatorial moves. This time, we have two tube segments stacked on one side, and one tube segment on the other side. The one tube segment needs to be surrounded by bulk tetrahedra such that the two sides become compatible.

In order to concisely state the axioms giving rise to topological invariance, and to classify the different anyons, it is useful to apply a \emph{dimensional reduction}, more specifically a \emph{compactification}. To this end, consider a mapping from two-dimensional triangulations to 3-dimensional ones by taking the cartesian product with the circle. Pulling back the 3-dimensional state-sum along this mapping we obtain a two-dimensional state-sum. On a combinatorial level, we take the cartesian product of a two-dimensional triangulation with the triangulation of the circle link, i.e., the looping edge in Eq.~\eqref{eq:looping_edge}. Every triangle then becomes a ``triangle prism'' with top and bottom identified,
\begin{equation}
\label{eq:anyon_compactification}
\includegraphics[valign=c]{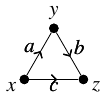}
\coloneqq
\includegraphics[valign=c]{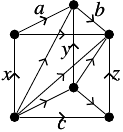}\;.
\end{equation}
We need to choose a simple 3-dimensional triangulation for this triangle prism which can be done with 3 tetrahedra as shown. We then redistribute the (independent) 3-dimensional state-sum variables onto the two-dimensional triangulation as shown. This way, we get a two-dimensional state-sum with one $G$-variable at each edge and one at each vertex. At each edge,
\begin{equation}
\includegraphics[valign=c]{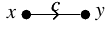}\;,
\end{equation}
we have a constraint $y=c^{-1}xc$. The state-sum weight at each triangle is
\begin{equation}
\beta^x(a,b)\coloneqq \omega(a,b,(ab)^{-1}xab)\overline{\omega(a,a^{-1}xa,b)}\omega(x,a,b)\;.
\end{equation}
One of the tetrahedra above has a different orientation as in Eq.~\eqref{eq:statesum_tetrahedron}, thus the according weight $\omega$ is complex conjugated. Up to differences in conventions, this is Eq.~\eqref{eq:tube_multiplication} from the main text. By construction, the topological invariance of the two-dimensional state-sum follows from the topological invariance of the 3-dimensional state-sum. The resulting state-sum is a thick state-sum as discussed in Appendix~\ref{app:thin_boundary}, whose vertex label set is $G$, with $G$-action
\begin{equation}
x \triangleleft g = g^{-1}xg\;.
\end{equation}
Accordingly, $\beta$ is a $G$ 2-cocycle with module $U(1)^G$, such that $g\in G$ acts on $\phi\in U(1)^G$ by $g\triangleright \phi(x)=\phi(x\triangleleft g^{-1})$.

Next, we extend the above mapping to a mapping from 2-manifolds with boundary to 3-manifolds with anyon world-lines. We do this by mapping the 1-dimensional boundary to a 1-dimensional anyon world-line. Combinatorially, we map a ``thickened'' boundary edge to an anyon tube segment. Pulling back the anyon state-sum along this mapping yields a two-dimensional boundary state-sum,
\begin{equation}
\includegraphics[valign=c]{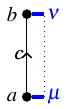}
\coloneqq
\includegraphics[valign=c]{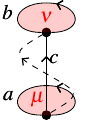}\;.
\end{equation}
Now, the boundary retriangulation invariance of the compactified boundary,
\begin{equation}
\label{eq:2d_boundary_invariance}
\includegraphics[valign=c]{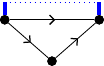}
=
\includegraphics[valign=c]{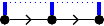}\;,
\end{equation}
is equivalent to the retriangulation invariance of the anyon world-lines. Thus, we see that line-like defects are in one-to-one correspondence with the boundaries of the compactified two-dimensional state-sum. So the weights $(\rho^a(c))_\mu^\nu$ are determined by the equation
\begin{equation}
\beta^a(c,d) (\rho^{a}(cd))_\lambda^\nu = \sum_{\mu} (\rho^a(c))_\lambda^\mu (\rho^{c^{-1} ac}(d))_\mu^\nu\;.
\end{equation}
Note that there is a finite number of \emph{irreducible} $\rho$, which are what is commonly understood by anyons. To see this we notice that a two-dimensional state-sum with variables only on the edges is an associative (in fact, $\dagger$-Frobenius-) algebra, and is a direct sum of full matrix algebras. We can copy the vertex variables of our state-sum onto each of the adjacent edges, such that at every edge we have a triple $(x,a,y)$. The triples have to satisfy $y=x\triangleleft a$ and are thus fully specified by $(x,a)$. The resulting algebra equals the tube algebra discussed in the literature and in the main text in Section~\ref{sec:tube}. Its irreducible matrix blocks, or the according irreducible representations, are the anyons.

\subsection{Boundary anyons}
To describe anyons within the boundary we need to define the state-sum on triangulations of manifolds with a 1-manifold embedded into the boundary. The normal space to the 1-manifold is a half-plane, and an $\epsilon$-sphere within this half-plane is an interval. Thus the link is an interval, and it can be triangulated with only two (thickened) boundary vertices separated by an infinitely thin bulk,
\begin{equation}
\label{eq:interval_link}
\includegraphics[valign=c]{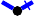}\;.
\end{equation}
The tick on the right short blue line is in order to remove the symmetries of the link. Next we take the product of that link with a single edge,
\begin{equation}
\label{eq:semi_tube_segment}
\includegraphics[valign=c]{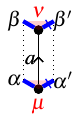}\;.
\end{equation}
It is again helpful to think of this as a 3-cell which looks like a little triangle wedge, by imagining another rectangle face between the two dotted lines at the back, such that there is a red shaded triangle on the bottom and top. Then an extended cellulation contains sequences of such ``semi-tube segments''. The red shaded triangles are attached to other semi-tube segments, whereas the two rectangular faces on the left and right are attached to the sides of nearby boundary triangles. At every red shaded triangle where two semi-tube segments meet,
\begin{equation}
\includegraphics[valign=c]{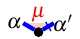}\;,
\end{equation}
there is one additional variable $\mu$. Those variables take values in some finite set $X$ which may depend on the value of the adjacent boundary vertex labels $\alpha$ and $\alpha'$. To each semi-tube segment with labels as in Eq.~\eqref{eq:semi_tube_segment}, we associate a weight
\begin{equation}
\kappa^{\alpha,\alpha'}(a)_\mu^\nu\;,
\end{equation}
noting that the values of $\beta$ and $\beta'$ are determined by the other labels.

Again, for stating the topological invariance of the boundary anyons and classifying them, we apply a compactification by pulling back the state-sum via the cartesian product with the interval link in Eq.~\eqref{eq:interval_link}. Each triangle is mapped to a sandwich of two boundary triangles,
\begin{equation}
\label{eq:boundary_anyon_compactification}
\includegraphics[valign=c]{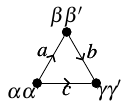}
\coloneqq
\includegraphics[valign=c]{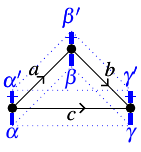}\;.
\end{equation}
We obtain a two-dimensional state-sum with one $G$-label at each edge and two $A$-labels at each vertex. The weight associated to each triangle is
\begin{equation}
\Psi^{\alpha,\alpha'}(a,b) \coloneqq \overline{\psi^\alpha(a,b)}\psi^{\alpha'}(a,b)\;.
\end{equation}
Note that one of the boundary triangle cells is reflected compared to Eq.~\eqref{eq:boundary_triangle} and thus complex conjugated. The resulting two-dimensional state-sum is a thick state-sum with vertex set $A\times A$ and right action given by
\begin{equation}
(\alpha,\alpha')\triangleleft g=(\alpha\triangleleft g, \alpha'\triangleleft g)\;.
\end{equation}
Accordingly $\Psi$ is a $G$ 2-cocycle with module $U(1)^{A\times A}$ and left action determined by the above.

As for bulk anyons, we extend the above to a mapping from 2-manifolds with boundary to 3-manifolds with boundary and boundary anyons, where the 1-dimensional boundary is mapped to a 1-dimensional boundary anyon world-line. Combinatorially, we map a boundary edge of a two-dimensional triangulation to a semi-tube segment,
\begin{equation}
\includegraphics[valign=c]{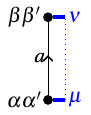}
\coloneqq
\includegraphics[valign=c]{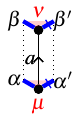}\;.
\end{equation}
Again, the topological invariance of boundary-anyon world-lines is equivalent to the topological invariance of the boundary above as in Eq.~\eqref{eq:2d_boundary_invariance}. So boundary anyons are determined by the equation
\begin{equation}
\Psi^{\alpha,\alpha'}(c,d) (\kappa^{\alpha,\alpha'}(cd))_\lambda^\nu = \sum_{\mu} (\kappa^{\alpha,\alpha'}(c))_\lambda^\mu (\kappa^{ \alpha\triangleleft g,\alpha'\triangleleft g}(d))_\mu^\nu\;.
\end{equation}
Again, what is commonly meant by boundary anyons are the irreducible $\kappa$. When we copy vertex labels onto edges in our two-dimensional state-sum we obtain the semi-tube algebra discussed in the main text. The boundary anyons are the irreducible representations of this semi-tube algebra.

\subsection{Ground states}
We can also consider 0-dimensional defects in space-time, i.e., embedded points. If the link of such a defect is not a 2-sphere, then the embedded points correspond to some kind of singularity. Such defects are in one-to-one correspondence with the ground states of the model on the 2-manifold which is the link. To represent such a defect combinatorially in the most concise way, we pick a minimal triangulation of the 2-manifold link. For a torus, for example,
we can choose a triangulation with two triangles,
\begin{equation}
\includegraphics[valign=c]{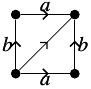}\;,
\end{equation}
with left/right and bottom/top identified. It might be intuitive to think of this triangulation as a solid-torus-like 3-cell, but we should keep in mind formally there is only the two-dimensional triangulation, and a more canonical filling would be given by a cone with a singularity in the center.  A combinatorial extended manifold is thus a 3-manifold with some torus 3-cells. At every torus 3-cell there is a weight
\begin{equation}
S^{a,b}\;.
\end{equation}

We want the 0-dimensional defects defined by the torus 3-cell to be topological. Roughly, this means that a layer of bulk padding the torus 3-cell can be absorbed into it. To concisely impose this topological invariance, we again use compactification. We pull back the state-sum along a mapping from 1-manifolds to 3-manifolds. Combinatorially, an edge is mapped to an edge times the torus link,
\begin{equation}
\label{eq:torus_link_mapping}
\includegraphics[valign=c]{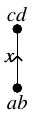}
\coloneqq
\includegraphics[valign=c]{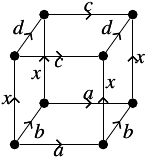}\;,
\end{equation}
where front/back and left/right are identified. We thus get a 1-dimensional state-sum with one $G$-variable at every edge and two $G$-variables at every vertex, subject to the constraints $d=x^{-1}bx$ and $c=x^{-1}ax$ at every edge with labels as above. The weight associated to every edge is
\begin{equation}
\begin{multlined}
P^{a,b}(x)\coloneqq
\overline{\omega(a,b,x)}\omega(a,x,x^{-1}bx)\overline{\omega(x,x^{-1}ax,x^{-1}bx)}\\\overline{\omega(b,a,x)\omega(b,x,x^{-1}ax)}\omega(x,x^{-1}bx,x^{-1}ax)\;,
\end{multlined}
\end{equation}
which can be obtain from a triangulation of Eq.~\eqref{eq:torus_link_mapping} with $6$ tetrahedra which is not shown. The topological invariance of this state-sum follows by construction from the topological invariance of the original state-sum. This is a thick 1-dimensional state-sum with vertex labels in $G\times G$ and right action
\begin{equation}
\label{eq:double_conjugation_action}
(a,b)\triangleleft x = (x^{-1}ax, x^{-1}bx)\;.
\end{equation}
Accordingly, $P$ is a $G$ 1-cocycle with module $U(1)^{G\times G}$, and left action determined by the above. We extend the mapping to a mapping from 1-manifolds with boundary to 3-manifolds with torus-link point defects, by mapping a boundary point to a torus 3-cell,
\begin{equation}
\includegraphics[valign=c]{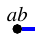}
\coloneqq
\includegraphics[valign=c]{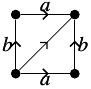}\;,
\end{equation}
such that topological invariance
\begin{equation}
\includegraphics[valign=c]{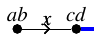}
=
\includegraphics[valign=c]{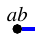}
\end{equation}
of the 1-dimensional state-sum boundary is equivalent to that of the torus-link point defects. Spelled out in letters, both are given by
\begin{equation}
\sum_{x} P^{a,b}(x) S^{x^{-1}ax,x^{-1}bx}=S^{a,b}\;,
\end{equation}
which can be rephrased as $dS=P$, if we interpret $S$ as a twisted 0-cochain. This equation is linear in $S$, thus the ground states form a vector space, which is a feature of all 0-dimensional defects. 1-dimensional state-sums with variables only on the vertices are given by projectors, and our state-sum can be brought into such a form by simply summing over $x$ in $P^{a,b}(x)$. Ground states $S^{a,b}$ are then vectors in the support of this projector.

\subsection{Bulk fusion events}
The link of a 0-dimensional defect might itself have defects. For example, take as link a 2-sphere with three embedded points,
\begin{equation}
\includegraphics[valign=c]{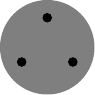}\;.
\end{equation}
Such defects correspond to points in space-time where three anyon world-lines meet. The points in the link are just the intersections of a sphere around the 0-dimensional defect with those three anyon world-lines. At the same time, those point-like defects are ground states on a sphere with three anyons
\footnote{Note that here we think of anyons as defects, so ``ground state with anyons'' means ground states of a Hamiltonian which is altered at some points to enforce the existance of anyons.}.
Here we will look at the fusion vertices between three different anyon world-lines which weights $\rho_0$, $\rho_1$ and $\rho_2$, in which case it would be natural to color the points differently.

This sphere link can be triangulated by one triangle (formed on the ``outside'' enclosed by all three edges in the following picture) with three red shaded 1-gons,
\begin{equation}
\label{eq:fusion_point_defect_volume}
\includegraphics[valign=c]{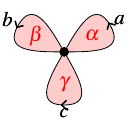}\;.
\end{equation}
Those three 1-gons are attached to three tube segments corresponding to the anyon world-lines, and the outside triangle is attached to a bulk tetrahedron. As usual we may think of this as a ``junction'' 3-cell. So extended cellulations are cellulations including sequences of tube segments (of three different colors) and junction 3-cells where three tube segments meet. To each junction 3-cell, we associate a weight
\begin{equation}
\label{eq:fusion_point_defect}
S^{a,b}_{\alpha\beta\gamma}\;.
\end{equation}

As usual, we consider the 1-dimensional state-sum arising from compactification with the link above. Combinatorially, we take the triangulation in Eq.~\eqref{eq:fusion_point_defect_volume} times an edge, and choose a 3-dimensional triangulation for the resulting volume,
\begin{equation}
\label{eq:fusion_vertex_compactification}
\includegraphics[valign=c]{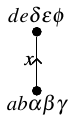}
\quad\leftrightarrow\quad
\includegraphics[valign=c]{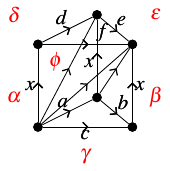}\;.
\end{equation}
On the right-hand side, the three edges labeled $x$ are identified, such that the edges $a$, $b$, $c$, $d$, $e$, and $f$ form loops, which are the start and end loops of three tube segments as in Eq.~\eqref{eq:anyon_segment}. The labels $\alpha$, $\beta$, $\gamma$, $\delta$, $\epsilon$, and $\phi$ are the labels at the bottom and top of the three tube segments, whose red 1-gons have not been drawn.. The resulting 1-dimensional state sum has one $G$-label at every edge, and two $G$-labels and three $X$-labels at every vertex, subject to the constraints $d=x^{-1}ax$, $e=x^{-1}bx$ at every edge. The weight associated to each edge is
\begin{equation}
\begin{multlined}
P^{a,b}(x)_{\alpha\beta\gamma}^{\delta\epsilon\phi}
\coloneqq
\omega(a,b,x)\overline{\omega(a,x,x^{-1}bx)}\omega(x,x^{-1}ax,x^{-1}bx)\\
\rho^a_0(x)_\alpha^\delta\rho^b_1(x)_\beta^\epsilon\overline{\rho^{ab}_2(x)_\gamma^\phi}\;,
\end{multlined}
\end{equation}
coming from the cellulation using three tetrahedra and three tube segments. The resulting 1-dimensional state-sum has vertex labels in $G\times G$ with action as in Eq.~\eqref{eq:double_conjugation_action}, as well as three free vertex labels. The fusion events are in one-to-one correspondence with boundaries of this 1-dimensional state-sum. The topological invariance is thus determined by
\begin{equation}
\label{eq:fusion_point_defect_axiom}
\sum_{\delta\epsilon\phi} P^{a,b}(x)_{\alpha\beta\gamma}^{\delta\epsilon\phi} S^{x^{-1}ax,x^{-1}bx}_{\delta\epsilon\phi}
=
S^{a,b}_{\alpha\beta\gamma}\;.
\end{equation}
As in the previous section, summing over $x$ yields a projector, and $S$ are vectors in the support of this projector. The dimension of this support vector space is the ground space dimension on the sphere with three anyons, or equivalently the fusion multiplicity $N^{\rho_0\rho_1}_{\rho_2}$. It can be calculated by taking the trace of the projector, or another compactification to a 0-dimensional state-sum via the cartesian product with the circle in Eq.~\eqref{eq:looping_edge}. Plugging Eq.~\eqref{eq:fusion_vertex_compactification} into this compactification means summing and identifying $a$ and $d$, $b$ and $e$, $c$ and $f$, $\alpha$ and $\delta$, $\beta$ and $\epsilon$, $\gamma$ and $\phi$, yielding,
\begin{equation}
\begin{multlined}
N^{\rho_0\rho_1}_{\rho_2} =
\sum_{\substack{a,b,x\in Z(a)\cap Z(b),\\\alpha,\beta,\gamma}} P^{a,b}(x)_{\alpha\beta\gamma}^{\alpha\beta\gamma}\\
=
\sum_{a,b,x\in Z(a)\cap Z(b)}
\omega(a,b,x)\overline{\omega(a,x,b)}\omega(x,a,b)\operatorname{Tr}(\rho^a_0(x))\operatorname{Tr}(\rho^b_1(x))\operatorname{Tr}(\overline{\rho^{ab}_2(x)})\;.
\end{multlined}
\end{equation}

\subsection{Bulk-boundary fusion events}
Next, let us consider fusion events at the boundary, between a bulk anyon and a boundary anyon. Those are 0-dimensional defects with link
\begin{equation}
\label{eq:bulk_boundary_fusion_link}
\includegraphics[valign=c]{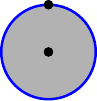}\;.
\end{equation}
A simple triangulation of this link consists of only one (thickened) boundary edge, an anyon 1-gon and a boundary-anyon triangle,
\begin{equation}
\label{eq:bulk_boundary_fusion_link_triangulation}
\includegraphics[valign=c]{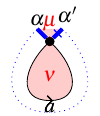}\;.
\end{equation}
Combinatorial extended manifolds are 3-manifolds with bulk tube segments and boundary wedge segments, meeting at ``junction'' 3-cells like above. At every such junction, we have a weight
\begin{equation}
S^{\alpha,a}_{\mu\nu}\;,
\end{equation}
noting that $\alpha'$ is determined by $\alpha'= \alpha\triangleleft a$.

As usual, we pull back the 3-dimensional state-sum along the cartesian product with the link in Eq.~\eqref{eq:bulk_boundary_fusion_link}. Concretely, an edge of a 1-dimensional triangulation is mapped to its cartesian product with Eq.~\eqref{eq:bulk_boundary_fusion_link_triangulation},
\begin{equation}
\label{eq:bulk_boundary_fusion_compactification}
\includegraphics[valign=c]{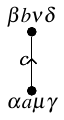}
\coloneqq
\includegraphics[valign=c]{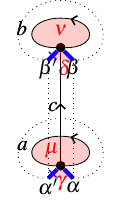}\;.
\end{equation}
The right hand side consists of one tube segment, one semi-tube segment, as well as one thickened boundary 4-gon, which can be triangulated with two boundary triangles. The resulting 1-dimensional state-sum has one $G$-label at each edge, and one $G$-label, one $A$-label, one anyon-space label, and one boundary-anyon-space label at each vertex. Those labels are subject to the constraints $\beta=\alpha\triangleleft c$ and $b=c^{-1}ac$. For an anyon $\rho$ and boundary anyon $\kappa$, the weight associated to an edge is
\begin{equation}
P^{a,\alpha}(c)_{\mu\nu}^{\psi\kappa} \coloneqq
\overline{\psi^{\alpha}(a,c)}\psi^{\alpha\triangleleft a}(c,c^{-1}ac) \rho^a(c)_\mu^\nu \kappa^{\alpha,\alpha\triangleleft a}(c)_\gamma^\delta\;.
\end{equation}
So bulk-to-boundary anyon fusion events are determined by
\begin{equation}
\sum_{\psi,\kappa} P^{a,\alpha}(c)_{\mu\nu}^{\psi\kappa} S^{\alpha\triangleleft c,c^{-1}ac}_{\psi\kappa} \coloneqq
S^{\alpha,a}_{\mu\nu}\;.
\end{equation}

If we want to know the fusion multiplicity for a bulk anyon $\rho$ and boundary anyon $\kappa$, we need to evaluate the 1-dimensional model on a circle. Equivalently, we identify the top and bottom in Eq.~\eqref{eq:bulk_boundary_fusion_compactification}, yielding
\begin{equation}
\sum_{\alpha\in A, a\in G, c\in S(\alpha)\cap Z(a)}
\overline{\psi^{\alpha}(a,c)}\psi^{\alpha\triangleleft a}(c,a) \operatorname{Tr}(\rho^a(c)) \operatorname{Tr}(\kappa^{\alpha,\alpha\triangleleft a}(c))\;,
\end{equation}
where $S(\alpha)$ is the stabilizer of $\alpha$, i.e., the set of $G$-elements $c$ such that $\alpha\triangleleft c=\alpha$. When both the boundary as well as $\rho$ and $\kappa$ are irreducible, we have $A=A\backslash G$, $\rho$ corresponding to a conjugacy class $C$ and $\kappa$ corresponding to a double-coset $x$, we get,
\begin{equation}
\label{eq:bulk_to_boundary_fusion}
\sum_{\substack{\alpha\in H\backslash G, a\in C\cap \alpha^{-1} x \alpha,\\c\in \alpha^{-1}\alpha\cap a^{-1}\alpha^{-1}\alpha a\cap Z(a)}}
\overline{\psi^{\alpha}(a,c)}\psi^{\alpha\triangleleft a}(c,a) \operatorname{Tr}(\rho^a(c)) \operatorname{Tr}(\kappa^{\alpha, \alpha\triangleleft a}(c))\;.
\end{equation}
Note that $x$ labels subsets of labels $(\alpha,\alpha')\in H\backslash G\times H\backslash G$ with transitive action via $x=\alpha'\alpha^{-1}$. The condition $a\in \alpha^{-1}x\alpha$ arises from this together with $\alpha'=\alpha a$. Moreover, $c\in Z(a)$ arises from $b=c^{-1}ac=a$, $c\in \alpha^{-1}\alpha$ from $\beta=\alpha c=\alpha$, and $c\in a^{-1}\alpha^{-1}\alpha a$ from $\beta'=\alpha' c=\alpha'$. Eq.~\eqref{eq:bulk_to_boundary_fusion} is equal to Eq.~\eqref{eq:fusion_mult_general} from the main text, apart from some choices of conventions.

\subsection{\texorpdfstring{Bulk anyon $F$-symbol}{Bulk anyon F-symbol}}
The $F$-symbol of the resulting anyon theory does not correspond to a defect like in all the previous sections, but simply to the evaluation of the state-sum on a specific extended cellulation. The corresponding extended manifold is a 3-sphere with 6 anyon world-lines meeting at 4 3-valent fusion point defects, forming the 1-skeleton of an embedded tetrahedron. Now, the $F$-symbol is not a single state-sum evaluation, but the collection of such evaluations for all different anyons and fusion point defects. For the anyons, we choose one irreducible representation of the tube algebra from each such isomorphism class. For the fusion point defects, we use an isomorphism from some abstract fusion vector space into the vector space of all point defects. I.e., instead of a single $S^{a,b}_{\alpha\beta\gamma}$ as in Eq.~\eqref{eq:fusion_point_defect} fulfilling Eq.~\eqref{eq:fusion_point_defect_axiom}, we consider
\begin{equation}
(S^{a,b}_{\alpha\beta\gamma})_\chi[\rho_a,\rho_b,\rho_c]\;
\end{equation}
fulfilling
\begin{equation}
\sum_{\chi} (S^{a,b}_{\alpha\beta\gamma})_\chi[\rho_a,\rho_b,\rho_c] \overline{(S^{c,d}_{\delta\epsilon\phi})_\chi[\rho_a,\rho_b,\rho_c]}
=
\sum_{\substack{x:x^{-1}ax=c,\\ x^{-1}bx=d}} P^{a,b}(x)_{\alpha\beta\gamma}^{\delta\epsilon\phi}[\rho_a,\rho_b,\rho_c]\;.
\end{equation}
More precisely, we can choose $S[\rho_a,\rho_b,\rho_c]$ to be an isometry (and we should do so in practice), such that the dimension of $\chi$ is the rank of $P$, namely $N^{\rho_a,\rho_b}_{\rho_c}$.

Now we choose an explicit extended cellulation. The smallest one consists of only four fusion point defects and one bulk tetrahedron,
\begin{equation}
\includegraphics[valign=c]{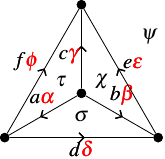}\;.
\end{equation}
Above we only drew the bulk tetrahedron. All four vertices are identified such that all 6 edges form loops. Each of the triangles is the triangle in one of the 4 junction 3-cells shown in Eq.~\eqref{eq:fusion_point_defect_volume}. With this, the evaluation is given by
\begin{equation}
\begin{multlined}
F^{\rho_a\rho_b\rho_c\rho_d\rho_e\rho_f}_{\sigma\chi\tau\psi}
= \sum_{\substack{a\in C_a,d\in C_d,e\in C_e,\\\alpha,\beta,\gamma,\delta,\epsilon,\phi}} \omega(a,d,e) (S^{a,d}_{\alpha\delta\beta})_\sigma[\rho_a,\rho_d,\rho_b]\\
\times (S^{ad,e}_{\beta\epsilon\gamma})_\chi[\rho_b,\rho_e,\rho_c] \overline{(S^{a,ade}_{\alpha\phi\gamma})_\tau[\rho_a,\rho_f,\rho_c]} \overline{(S^{d,e}_{\delta\epsilon\phi})_\psi[\rho_d,\rho_e,\rho_f]}\;.
\end{multlined}
\end{equation}

\section{Thin versus thick boundaries}
\label{app:thin_boundary}
In this appendix, we describe the space-time picture for a different microscopic way of defining a boundary. We will call this the \emph{thin boundary} state-sum as opposed to the \emph{thick boundary} described in the main text and in Appendix~\ref{app:state_sum}. We will first discuss the thin boundary in $1+1$ dimensions and then in $2+1$ dimensions, such that the generalization to arbitrary space-time dimensions will be straight-forward.

\subsection{1+1 dimensions}
Recall that a two-dimensional group-cohomology state-sum is defined on two-dimensional branching structure triangulations with one group label at every edge, and one constraint $ab=c$ and one weight $\omega(a,b)$ at every triangle as in Eq.~\eqref{eq:state_sum_triangle}. The thin-boundary state-sum is defined for a subgroup $H\subset G$ and a $H$ 1-cochain $\psi$ such that $d\psi=\omega|_H$, or more explicitly
\begin{equation}
\label{eq:2d_thin_boundary}
\psi(a)\psi(b)=\psi(ab) \omega(a,b)\;,
\end{equation}
for all $a,b\in H$. At every boundary edge,
\begin{equation}
\includegraphics[valign=c]{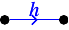}\;,
\end{equation}
the variables only take values in $H$ and not $G$, and we have an associated weight $\psi(h)$. The simplicity of the boundary comes at the expense of having a slightly more complicated topological invariance. A move with only one bulk triangle on one side as for the thick boundary in Eq.~\eqref{eq:2d_boundary_invariance} does not hold. Instead, the moves
\begin{equation}
\label{eq:thin_boundary_move1}
\includegraphics[valign=c]{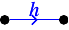}
=
\includegraphics[valign=c]{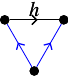}
\end{equation}
and
\begin{equation}
\label{eq:thin_boundary_move2}
\includegraphics[valign=c]{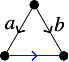}
=
\includegraphics[valign=c]{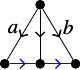}
\end{equation}
do follow from Eq.~\eqref{eq:2d_thin_boundary}.

Let us now discuss the relation between the thick and thin boundary. Note that in Appendix~\ref{app:state_sum}, a thick boundary was defined with values in an arbitrary right $G$-set $A$ at the vertices. In $1+1$ dimensions, there is a constraint $\beta=\alpha\triangleleft g$ and a weight $\psi^\alpha(a)$ at every thickened edge as in Eq.~\eqref{eq:boundary_edge}. Every set with right $G$ action is isomorphic to a disjoint union of left coset sets $H\backslash G$ for different $H$. Each $H$ is determined up to conjugation, and the right action on $H\backslash G$ is transitive and given by multiplication with the $G$-element from the right. Note that physically, boundaries with a transitive action are those which are irreducible, that is, robust to perturbations.

An irreducible thick boundary $\hat\psi$ (i.e., one with transitive action) can be obtained from a thin boundary $\psi$ with the same subgroup $H$ as follows. We start by padding each thin boundary edge with little bulk rectangle, and flatten those rectangles to get a generalized thick boundary,
\begin{equation}
\label{eq:2d_thin_to_thick_graph}
\includegraphics[valign=c]{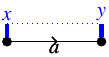}\coloneqq
\includegraphics[valign=c]{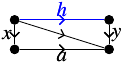}\;,
\end{equation}
which has labels in $G$ instead of $H\backslash G$ at the boundary vertices. It is easy to see that the moves for the thick boundary such as Eq.~\eqref{eq:2d_boundary_invariance} follow from the thin boundary moves such as in Eq.~\eqref{eq:thin_boundary_move1} and Eq.~\eqref{eq:thin_boundary_move2}, after we plug in Eq.~\eqref{eq:2d_thin_to_thick}. We say ``generalized'' because the label $y$ is not determined by $x$ and $a$ via some action, but only constrained through $h\coloneqq xay^{-1}\in H$.

In order to obtain a proper thick boundary $\hat\psi$, we realize that the state-sum on the right of Eq.~\eqref{eq:2d_thin_to_thick_graph} has a gauge symmetry acting on the edges adjacent to a boundary vertex,
\begin{equation}
\includegraphics[valign=c]{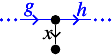}
=
\includegraphics[valign=c]{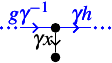}\;,
\end{equation}
for all $\gamma\in H$, where the equality is for the weights at the surrounding triangles and boundary edges. On the left of Eq.~\eqref{eq:2d_thin_to_thick_graph}, the same gauge symmetry is only acting on a single vertex label,
\begin{equation}
\includegraphics[valign=c]{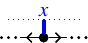}
=
\includegraphics[valign=c]{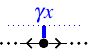}\;.
\end{equation}
Thus, we remove the gauge freedom by replacing the vertex labels $x,y\in G$ by left cosets in $\alpha,\beta\in H\backslash G$. The constraint $\beta=\alpha\triangleleft a$ of the thick boundary then directly follows from $x a= hy$ and $h\in H$ on the right. The weight $\hat\psi$ after gauge fixing can be obtained from the weight before gauge fixing by choosing a standard representative $R(\alpha)\in G$ for each coset $\alpha$, and using
\begin{equation}
\label{eq:representatives_shortcut}
x=R(\alpha),\qquad y=R(\beta)=R(\alpha\triangleleft a)\;.
\end{equation}
The weight can then be read off from the right-hand side of Eq.~\eqref{eq:2d_thin_to_thick_graph},
\begin{equation}
\label{eq:1d_thick_from_thin}
\hat\psi^\alpha(a)\coloneqq \overline{\omega(x,a)}\omega(xay^{-1},y) \psi(xay^{-1})\;,
\end{equation}
using the shortcut in Eq.~\eqref{eq:representatives_shortcut}.

Vice versa, a thin boundary $\tilde\psi$ can be obtained from an irreducible thick boundary $\psi$ by realizing that the thick boundary has the following gauge symmetry around a boundary vertex: Change every ingoing adjacent edge by $a\rightarrow a\gamma$, every outgoing adjacent edge by $a\rightarrow \gamma^{-1}a$, and the coset label at the vertex itself by $\alpha\rightarrow\alpha\triangleleft \gamma$, for any $\gamma\in G$. Since every such gauge symmetry only affects a single boundary vertex, we can use it to fix all the coset labels to the trivial coset $H$. So $\tilde\psi$ is obtained from $\psi$ by simply setting the coset label to $H$, which automatically restricts the group label to the subgroup $H$,
\begin{equation}
\label{eq:2d_thick_to_thin_mapping}
\tilde\psi(a)\coloneqq \psi^H(a)\;,
\end{equation}
for $a\in H$.

It is easy to see that if we first transform a thin boundary into a thick one, and go back to a thin one, we end up with the same thin boundary again,
\begin{equation}
\tildehat{\psi}^\alpha(g) =\psi^\alpha(g)\;.
\end{equation}
However, going from a thick to a thin and then back to a thick boundary does not yield the same boundary, but one that is in the same phase/cohomology class.
In order to see this, we consider the following domain wall $\eta$ between the thick boundaries $\psi$ and $\hattilde\psi$,
\begin{equation}
\label{eq:11d_thick_thick_domain}
\includegraphics[valign=c]{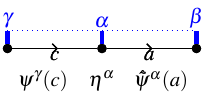}
=
\includegraphics[valign=c]{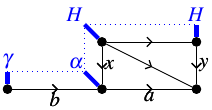}\;,
\end{equation}
using $x=R(\alpha)$ and $y=R(\beta)=R(\alpha\triangleleft a)$ as usual. On the right-hand side, we read off
\begin{equation}
\hattilde\psi^\alpha(a)\coloneqq \overline{\omega(x,a)}\omega(xay^{-1},y) \psi^H(xay^{-1})\;,
\end{equation}
in accordance with Eq.~\eqref{eq:2d_thick_to_thin_mapping} and Eq.~\eqref{eq:1d_thick_from_thin}. The weight
\begin{equation}
\label{eq:1d_thinthick_double_domainwall}
\eta^\alpha\coloneqq \overline{\psi^H(x)}
\end{equation}
associated to the vertical thick boundary edge on the right-hand side defines the weight of the domain wall on the left-hand side. With this domain wall, $\psi$ and $\hattilde\psi$ are related by
\begin{equation}
\label{eq:1d_thinthick_doublemapping_gauge}
\hattilde{\psi}^\alpha(g) = \psi^\alpha(g)\overline{\eta^\alpha}\eta^{\alpha\triangleleft g}\;.
\end{equation}
That is, $\psi$ and $\hattilde\psi$ are twisted 1-cochains which differ by a twisted 1-coboundary $d\eta$.

\subsection{2+1 dimensions}
Let us now look at the thin boundary in $2+1$ dimensions. Again, it is defined for $H\subset G$ and a $H$ 2-cochain $\psi$ with $d\psi=\omega|_H$, that is,
\begin{equation}
\label{eq:boundary_cocycle}
\psi(a,b)\psi(ab,c) \omega(a,b,c) = \psi(a,bc)\psi(b,c)\;.
\end{equation}
Again, we restrict the group labels at the boundary edges to $H$, and associate to every boundary triangle,
\begin{equation}
\includegraphics[valign=c]{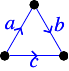}\;,
\end{equation}
the constraint $ab=c$ and the weight
\begin{equation}
\psi(a,b)\;.
\end{equation}
Again, the topological moves that hold are slightly more complicated. In fact, the move in Eq.~\eqref{eq:boundary_move} does still hold for the thin boundary,
\begin{equation}
\label{eq:thin_boundary_move}
\includegraphics[valign=c]{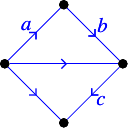}
=
\includegraphics[valign=c]{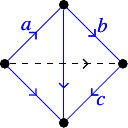}\;.
\end{equation}
This is because all involved edges labels are constrained to $H$ and so this is just Eq.~\eqref{eq:boundary_cocycle}. However, the thick boundary is invariant under additional moves, such as
\begin{equation}
\label{eq:additional_boundary_pachner_move}
\includegraphics[valign=c]{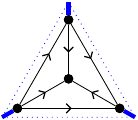}
=
\includegraphics[valign=c]{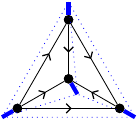}\;.
\end{equation}
This move consists of one boundary triangle and one bulk tetrahedron on the left, and three boundary triangles only on the right. This move does not hold for the thin boundary, since the three edges adjacent to the central vertex are constrained to $H$ on the right, but can take values in all of $G$ on the left. This can be fixed by padding the right-hand side with three bulk tetrahedra.

Another drawback of the thin boundary is that it is not directly compatible with the more general way of defining boundaries in terms of $F$ and $L$ symbols. That is, there is no analogue of a thin boundary for non-group-cocycle $F$-symbols or $L$-symbols.

Let us discuss the relation between the thin and thick boundary. We can construct a transitive thick boundary $\hat\psi$ from a thin boundary $\psi$ as follows. Analogous to the $1+1$-dimensional case, we start with a generalized thick boundary obtained by padding each thin boundary triangle with a triangle prism,
\begin{equation}
\label{eq:thin_thick_mapping}
\includegraphics[valign=c]{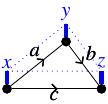}
\coloneqq
\includegraphics[valign=c]{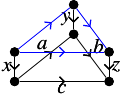}\;.
\end{equation}
It is easy to see that the moves for the thick boundary in Eq.~\eqref{eq:boundary_move} and Eq.~\eqref{eq:additional_boundary_pachner_move} follows form the thin boundary moves such as in Eq.~\eqref{eq:thin_boundary_move}, after we plug in Eq.~\eqref{eq:thin_thick_mapping}.

The $H$ boundary group labels on the right are determined by the other labels $\alpha,a,b\in G$ so they do not appear on the left. Just as in the $1+1$-dimensional case, we have a gauge symmetry on the right involving all edges adjacent to a fixed boundary vertex, which becomes a gauge symmetry acting on a single boundary vertex label on the left. Again, this allows us to replace the $G$-elements $x,y,z$ on the left by left cosets $\alpha,\beta,\gamma$. Then using a triangulation of the above prism similar to Eq.~\eqref{eq:anyon_compactification}, we find
\begin{equation}
\label{eq:2d_thin_to_thick}
\hat\psi^\alpha(a,b)\coloneqq \overline{\omega(x,a,b)} \omega(xay^{-1},y,b) \overline{\omega(xay^{-1},ybz^{-1},z)}\psi(xay^{-1},ybz^{-1})\;,
\end{equation}
using
\begin{equation}
\label{eq:coset_representatives}
x\coloneqq R(\alpha),\quad y\coloneqq R(\alpha\triangleleft a), \quad z\coloneqq R(\alpha\triangleleft ab)\;,
\end{equation}
where $R$ denotes a choice of standard representative of every coset as in the previous paragraph.

Vice versa, we can construct a thin boundary $\tilde\psi$ from a thick boundary $\psi$ by realizing that the thick boundary has a gauge symmetry acting on a single boundary vertex label. This gauge freedom can be fixed by setting all the cosets to the trivial coset $H$, yielding
\begin{equation}
\label{eq:3d_thick_to_thin}
\tilde\psi(a,b)\coloneqq \psi^H(a,b)\;,
\end{equation}
for $a,b\in H$. As in the $1+1$-dimensional case, we have
\begin{equation}
\tildehat{\psi}^\alpha(g,h) =\psi^\alpha(g,h)\;,
\end{equation}
but not vice versa. Similar to the $1+1$-dimensional case, $\hattilde\psi$ and $\psi$ are separated by a 1-dimensional domain wall which associates a weight $\eta$ to the according edges. This weight comes from flattening the vertical side of a ``step'' representing the domain wall. While in Eq.~\eqref{eq:11d_thick_thick_domain} the side step is an edge flattened to a point, here we have a rectangle flattened to an edge,
\begin{equation}
\includegraphics[valign=c]{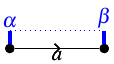}\coloneqq
\includegraphics[valign=c]{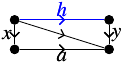}\;.
\end{equation}
So the weight is given by
\begin{equation}
\label{eq:2d_thinthick_double_domainwall}
\eta^\alpha(a)=\psi^H(x,a)\overline{\psi^H(xay^{-1},y)}\;,
\end{equation}
using Eq.~\eqref{eq:coset_representatives}. The relation between $\psi$ and $\hattilde\psi$ is now
\begin{equation}
\label{eq:2d_thinthick_doublemapping_gauge}
\hattilde{\psi}^\alpha(g,h) = \psi^\alpha(g,h)\eta^\alpha(g)\eta^{\alpha\triangleleft g}(h)\overline{\eta^\alpha(gh)}\;.
\end{equation}

The generalization of thick versus thin boundaries to higher dimensions is straight-forward. To map a thin to a thick boundary of an $n$-dimensional space-time bulk analogous to Eq.~\eqref{eq:2d_thin_to_thick_graph}, pad the $n-1$-simplex with the $n-1$-simplex times an edge. The resulting $n$-cell can be triangulated using $n$ $n$-simplices. The edge labels of the edges perpendicular to the boundary are set to standard representatives of the cosets of the thick boundary. Evaluation of the space-time volume yields an expression with $n$ bulk weigths $\omega$ and one thin-boundary weight $\psi$. Moreover, the domain wall $\eta$ between $\psi$ and $\hattilde\psi$ analogous to Eq.~\eqref{eq:2d_thinthick_double_domainwall} is obtained from triangulations of a boundary $n-2$-simplex times an edge, yielding a formula with $n-1$ times $\psi$.

\subsection{Thin versus thick bulk}
When the bulk $\omega$ is trivial, then the thick boundary $\psi$ gives rise to a state-sum on its own, which we will refer to as a \emph{thick} state-sum. Such a thick state-sum is a two-dimensional state-sum with vertex labels equipped with a right $G$-action, and $G$-elements on the edges. Examples for such state-sums-with-action arose in the compactification discussed around Eq.~\eqref{eq:anyon_compactification}. As discussed in the previous paragraph, the set of vertex labels is isomorphic to a direct sum left coset sets $H\backslash G$ on which $G$ acts transitively, for different subgroups $H$ determined up to conjugation. On the other hand, we will refer to the conventional state-sum with only $H$-elements on the edges as \emph{thin} state-sum.

The mapping in Eq.~\eqref{eq:2d_thin_to_thick} for a 1-dimensional boundary of a two-dimensional bulk can also be used to map a standalone 1-dimensional thin state-sum to a thick state-sum (with the same $H$),
\begin{equation}
\hat\psi^\alpha(a)\coloneqq \psi(xay^{-1})\;,
\end{equation}
using the short-hand notation $x\coloneqq R(\alpha)$ and $y\coloneqq R(\alpha\triangleleft a)$ as in Eq.~\eqref{eq:representatives_shortcut}. Up to some conventions, this is just precomposition with the cohomological isomorphism $m^{(1)}$ discussed in Appendix~\ref{app:sec:isomorphism}. Vice versa, Eq.~\eqref{eq:2d_thick_to_thin_mapping} can also be used to map a standalone thick state-sum $\psi$ to a thin state-sum $\tilde\psi$. Furthermore, the relation in Eq.~\eqref{eq:1d_thinthick_doublemapping_gauge} with Eq.~\eqref{eq:1d_thinthick_double_domainwall} still can be used to relate $\psi$ and $\hattilde\psi$.

Also the mapping in Eq.~\eqref{eq:thin_thick_mapping} for two-dimensional boundaries of 3-dimensional bulks can be used for two-dimensional standalone bulks,
\begin{equation}
\psi^\alpha(a,b)\coloneqq \psi(xay^{-1},ybz^{-1})\;,
\end{equation}
with $z\coloneqq R(\alpha\triangleleft ab)$ as in Eq.~\eqref{eq:coset_representatives}. This is $m^{(2)}$ up to conventions. Vice versa, the Eq.~\eqref{eq:3d_thick_to_thin} also maps from a standalone thick state-sum $\psi$ to a thin state-sum $\tilde\psi$, and Eq.~\eqref{eq:2d_thinthick_doublemapping_gauge} with Eq.~\eqref{eq:2d_thinthick_double_domainwall} relate $\psi$ and $\hattilde\psi$. The generalization to higher dimensions is straight-forward.

\subsection{Boundaries of thick bulk from boundaries of thin bulk}
Let us now consider the case where the bulk itself is a transitive thick state-sum in $1+1$ dimensions given by $\omega^\alpha(a,b)$, which we map to a thin state-sum given by
\begin{equation}
\omega(a,b)\coloneqq \omega^\alpha(a,b)
\end{equation}
for $a,b,c\in H$. Now, consider a generalized thick boundary of the thin state-sum. Here, generalized means that the set of labels at the vertices is not equipped with $G$-action. At a boundary edge,
\begin{equation}
\includegraphics[valign=c]{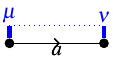}\;,
\end{equation}
the two vertex labels $\mu$, $\nu$ are independent and do not have to satisfy any constraints. The label set of $\mu$ and $\nu$ (which can also be understood as a vector space since there is a unitary gauge symmetry acting on $\mu$ and $\nu$) is allowed to depend on the value of the group label $a\in G$. The according weight is
\begin{equation}
\psi(a)_\mu^\nu\;.
\end{equation}

On the other hand, a generalized thick boundary of a transitive thick state-sum has edges,
\begin{equation}
\includegraphics[valign=c]{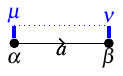}\;,
\end{equation}
with $a\in G$, two bulk coset labels $\alpha$ and $\beta$, and two free labels $\mu$, $\nu$ as before. The according weight is
\begin{equation}
\psi^\alpha(a)_\mu^\nu\;.
\end{equation}
Now we can obtain a thick boundary of the thick state-sum from the thick boundary of the thin state-sum by padding it with a layer of thick bulk. When doing so, we restrict the bulk coset labels at the boundary to the trivial coset $H$,
\begin{equation}
\includegraphics[valign=c]{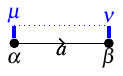}\coloneqq
\includegraphics[valign=c]{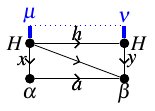}\;.
\end{equation}
As in the previous sections, $h$ on the right can be inferred from the other labels. Since $H\triangleleft x=\alpha$, $x$ is a representative of $\alpha$, and the same holds for $y$ and $\beta$. Evaluating the right-hand side above yields
\begin{equation}
\label{eq:boundary_of_thin_to_thick}
\psi^\alpha(a)_\mu^\nu
\coloneqq
\overline{\omega^H(x,a)} \omega^H(xay^{-1},y) \psi(xay^{-1})_\mu^\nu\;,
\end{equation}
where $x$ and $y$ are chosen representatives of $\alpha$ and $\beta$ as in Eq.~\eqref{eq:representatives_shortcut}.

Note that the two-dimensional state-sums arising from compactifications in Appendix~\ref{app:state_sum} are thick state-sums, as they have labels on the vertices that are acted on by $G$. The (boundary) anyons are in one-to-one correspondence with the (irreducible) generalized thick boundaries of those thick compactified state-sums. The formula above provides a way to obtain such boundaries from boundaries of a thin state-sum. The computation of the latter is simpler in practice as it takes place on smaller vector spaces. We follow the following steps, which are also discussed in a more algebraic way in Section~\ref{sec:twistedwithaction} in the main text.
\begin{itemize}
\item Decompose the thick compactified state-sum into transitive ones, with vertex label set $H\backslash G$ for different subgroups $H$.
\item For each transitive part, calculate the corresponding thin state-sum, that is, the corresponding $H$ 2-cocycle $\omega(a,b)$.
\item For each transitive part, find the irreducible generalized thick boundaries of this thin state-sum. This corresponds to finding the irreducible representations $\psi(h)_\mu^\nu$ of the $\omega$-twisted group algebra, or in other words, the projective irreducible representations of $H$ with 2-cocycle $\omega$.
\item Use Eq.~\eqref{eq:boundary_of_thin_to_thick} to obtain the irreducible generalized thick boundaries of the thick state-sum.
\end{itemize}

\bibliography{bibliography}

\end{document}